\documentclass{article}
\usepackage[utf8]{inputenc}
\usepackage[margin=1in]{geometry}
\usepackage{amsmath}
\usepackage[T1]{fontenc}
\usepackage{graphicx} % Required for inserting images
\usepackage{float}
\usepackage{subcaption}
\usepackage{parskip}
\usepackage{xcolor}
\usepackage{booktabs}
\usepackage{multirow}
\usepackage{tabularx}
\usepackage{pdflscape}
\usepackage{amssymb}
\graphicspath{{figures/cnn/architecture/}{figures/cnn/performance/}{figures/cnn/conformal/}{figures/h3n2/dpgr/}{figures/h3n2/landscape/}{figures/h3n2/shap/}{figures/h1n1/dpgr/}{figures/h1n1/landscape/}{figures/h1n1/shap/}}

\usepackage[hidelinks]{hyperref}
\usepackage[backend=biber,style=numeric,sorting=none]{biblatex}
\title{Inferring and Predicting Clade-Level Relative Transmission Fitness in Seasonal Influenza A Using Differential Population Growth Rate and Deep Learning}
\author{
    Ursula Senam Nkonu$^{1}$, Richard Annan$^{2}$, Letu Qingge$^{2}$, Hong Qin$^{3}$\\[0.5em]
    $^{1}$Biomedical Sciences PhD Program,\\
    Old Dominion University, Norfolk, VA, USA\\
    $^{2}$Department of Computer Science,\\
    North Carolina A\&T State University, Greensboro, NC, USA\\
    $^{3}$School of Data Science, Department of Computer Science,\\
    Old Dominion University, Norfolk, VA, USA
}
\date{}

\begin{document}
               
\maketitle
\begin{abstract}
Seasonal influenza A evolves rapidly, allowing newly emerged clades to
replace previously dominant lineages and complicate surveillance and
vaccine evaluation. Here, we applied the Differential Population Growth
Rate (DPGR) framework to GISAID-derived H3N2 and H1N1 surveillance data
collected from 1 January 2014 to 12 February 2026, including the 2025--2026
influenza season, to estimate clade-level relative
transmission fitness across continents and within the United States. We
identified windows of co-circulation with sliding-window regression,
reconstructed relative-fitness relationships among clades, and compared
inferred growth advantages with independent WHO and CDC surveillance
patterns. We further trained subtype-specific convolutional neural
networks on complete viral genomes to predict DPGR from sequence,
quantified predictive uncertainty with conformal prediction, and used
SHAP to localize genomic contributors to fitness. DPGR recovered
recurrent lineage turnover in both subtypes and consistently identified
the emerging H3N2 subclade K as fitter than the 2025-2026
vaccine-lineage background across multiple regions. Genome-based models
predicted DPGR accurately for H3N2 ($R^2 = 0.9577$) and H1N1
($R^2 = 0.9871$), while interpretation highlighted known
haemagglutinin antigenic sites together with contributions from internal
genes. These results support DPGR as an interpretable surveillance
signal and show that influenza fitness can be linked to genomic
prediction and biological interpretation in a unified framework.
\end{abstract}

\section{Introduction}

Seasonal influenza remains a persistent global public health challenge,
causing substantial annual respiratory morbidity and mortality worldwide
\cite{iuliano_global_2018}. A central reason influenza continues to
recur despite widespread immunity and vaccination is its rapid
evolution under transmission, host immune pressure, and vaccine-mediated
selection \cite{petrova_evolution_2018}. Within a single subtype,
newly emerged clades can replace previously dominant lineages, altering
epidemic intensity, geographic spread, and the degree of match between
circulating viruses and vaccine components. These evolutionary turnover
events are especially important for the major human influenza A
subtypes, A/H3N2 and A/H1N1, whose population dynamics can shift
substantially across regions and between seasons. Accurately
quantifying which clades are increasing in relative transmission
fitness is therefore important for surveillance, near-term forecasting,
and interpretation of vaccine performance.

The growth of global influenza sequence surveillance, particularly
through GISAID, has made it possible to study influenza evolution at
far greater temporal and geographic resolution than was previously
possible \cite{shu_gisaid_2017}. In parallel, peer-reviewed studies have
shown that influenza evolution can be forecast from genetic, antigenic,
and population-level data using sequence-based fitness models,
evolution-informed transmission models, phenotype-integrated forecast
frameworks, mutation-dynamics approaches, and relative lineage fitness
estimation \cite{luksza_predictive_2014,du_evolution_2017,
huddleston_genotypes_2020,lou_predictive_2024,zeller_forecasting_2025}. Machine-learning
methods have also improved sequence-based prediction of seasonal H3N2
antigenic change \cite{shah_seasonal_2024}. However, translating
sequence abundance data into an interpretable and comparable estimate
of relative transmission fitness remains challenging, especially when
the goal is to compare many co-circulating clades across regions and
seasons using a simple, reproducible quantitative metric. The
Differential Population Growth Rate (DPGR) framework offers one such
approach by estimating the relative growth advantage of one variant over
another during periods of co-circulation through the time-dependent
log-ratio of their observed counts \cite{pantho_data-driven_2025}.
Although this framework has been developed in other viral systems, its
application to human influenza across multiple clades, locations, and
flu seasons has not yet been systematically characterized.

In this study, we apply the DPGR framework to seasonal influenza A
viruses using GISAID-derived surveillance data spanning 1 January 2014
to 12 February 2026, including the 2025--2026 influenza season.
We focus on the major H3N2 and H1N1 clades that dominated or
co-circulated during recent flu seasons and estimate their pairwise
relative transmission fitness across multiple continents and within the
United States. By combining pairwise comparisons, sliding-window model
selection, and additive fitness relationships, we reconstruct a
population-level view of clade succession and competitive hierarchy
through time.

Beyond descriptive fitness estimation, this study also evaluates
whether DPGR-derived signals are informative for broader influenza
surveillance and prediction. We first compare inferred fitness
advantages with independent WHO and CDC surveillance trends to assess
whether the variants identified as fitter by DPGR correspond to those
that subsequently expanded or contributed to reduced vaccine
effectiveness. We then train subtype-specific convolutional neural
networks on complete influenza genomes to predict DPGR fitness directly
from sequence, use conformal prediction to quantify uncertainty, and
apply SHAP to identify genomic regions most strongly associated with
fitness differences. Together, these analyses aim to provide a unified
framework linking observed epidemiologic growth, sequence-based fitness
prediction, and biologically interpretable determinants of influenza
variant success.

\section{Methodology}
\subsection{Dataset}
The influenza surveillance dataset used for this was acquired from the Global Initiative on Sharing All Influenza Data (GISAID) database \cite{shu_gisaid_2017}. The initial metadata files were downloaded as .xls files and then converted to a .tsv format using a custom Python script. The .tsv file contained a total of 57 columns, which were then filtered down to the Location, Collection Date, and Subtype columns. The locations in the dataset were North America, South America, Europe, Asia, Africa, and Oceania. The dataset spanned a time period from 2010 to 2026 and contained 398,126 samples; the DPGR analysis was restricted to 2014 onwards, corresponding to the emergence of the earliest clades examined in this study.

\subsection{Data Preprocessing}
Raw metadata was obtained in tab-separated value (TSV) format 
from GISAID and subjected to a multi-stage cleaning pipeline 
prior to analysis. Records with missing values in any of the 
three columns required for downstream analysis were removed. 
Although the GISAID repository includes sequences from a broad 
range of host species, only records corresponding to the 
A/H1N1, A/H3N2, and influenza B subtypes were retained, as 
these constitute the principal seasonal influenza lineages 
circulating in human populations. Examination of host 
composition within each retained subtype revealed that 
non-human sequences (e.g., avian, swine origin) accounted 
for a negligible proportion of the total: 0.00\% for 
influenza B, 2.78\% for A/H3N2, and 7.24\% for A/H1N1. 
Given that over 92\% of sequences within each subtype 
originated from human hosts, no further host-based 
filtering was applied. Retaining the full subtype datasets 
maximised temporal continuity and sequence density while 
preserving a high degree of biological specificity for 
seasonal human influenza.

Following subtype filtering, the dataset comprised 324,956 
records spanning six continental regions. Geographic 
standardisation was achieved by mapping the \texttt{Location} 
column to predefined continental region lists, enabling 
consistent continent-level aggregation across the full 
analysis period. The \texttt{Collection\_date} field was 
parsed into Python \texttt{datetime} objects, and an 
additional column encoding the ISO collection week was 
derived to support weekly temporal aggregation. The dataset 
was subsequently restricted to the analysis window of 
1 January 2014 to 12 February 2026 and grouped by subtype, 
location, and collection week to compute weekly sequence 
frequency counts for each variant at each location. The 
weekly start date was extracted from each ISO week identifier 
to ensure unambiguous temporal alignment across records. 
These weekly counts were then summed within each 
subtype--location--week stratum to produce a compact 
summary dataset suitable for frequency-based modelling 
and visualisation.

This preprocessing pipeline produced a clean, temporally 
and geographically standardised dataset that served as 
the foundation for all subsequent DPGR fitness estimation 
and population-level analyses described in the following 
sections.

For linkage to vaccine efficacy, the major clades which showed dominance and co-circulated during the various flu seasons were used. For H3N2; 3C.2a, 3C.2a1, 3C.2a2, 3C.3a1, 3C.2a1b.1, 3C.2a1b.2a.2b, 3C.2a1b.2a.2a.1, 3C.2a1b.2a.2a.3, 3C.2a1b.2a.2a.3a.1, and 3C.2a1b.2a.2a.3a.1 / K.

For A/H1N1; 6B.1, 6B.1A, 6B.1A.1, 6B.1A.5a, 6B.1A.5b, 6B.1A.5a.1, 6B.1A.5a.2, 6B.1A.5a.2a, and 6B.1A.5a.2a.1.

\subsection{Differential Population Growth Rate (DPGR)}
We applied the Differential Population Growth Rate (DPGR) model \cite{pantho_data-driven_2025} to estimate relative transmission fitness between variant pairs. The DPGR model assumes exponential growth of co-circulating variants during applicable time windows:
\begin{equation}
\log_{10}\!\left(\frac{N_1(t)}{N_2(t)}\right) = \mathrm{DPGR}_{1,2} \cdot t + C
\label{eq:dpgr}
\end{equation}
where \(N_1(t)\) and \(N_2(t)\) represent weekly case counts of variants 1 and 2, \(t\) is time 
measured in days, and \(\mathrm{DPGR}_{1,2}\) represents the daily differential growth rate.
For variant pairs without sufficient temporal overlap, the additive property of DPGR was employed to account for intermediate variants that had intersecting periods of co-circulation: 
\begin{equation}
\log_{10}\!\left(\frac{a}{b}\right)
=
\log_{10}\!\left(\frac{a}{c}\right)
+
\log_{10}\!\left(\frac{c}{b}\right)
\label{eq:dpgr_additive}
\end{equation}

Resulting in the equation below as long as variant c can be co‐sampled with both variant a and b at the same location:
\begin{equation}
\mathrm{DPGR}_{a,b} = \mathrm{DPGR}_{a,c} + \mathrm{DPGR}_{c,b}
\label{eq:dpgr_chain}
\end{equation}

\subsection{Sliding Window}
As required for the DPGR model \cite{pantho_data-driven_2025}, a sliding window approach was implemented to pinpoint periods where the log-linear assumption held \cite{pantho_data-driven_2025}. For each variant pair and location, we tested windows ranging from 4 to 12 weeks, calculating the coefficient of determination ($R^2$) and $p$-value for each window. Windows were selected based on $R^2 > 0.90$ and $p < 0.05$. Weeks in which either variant recorded zero counts were excluded from the log-ratio computation, as the logarithm is undefined at zero. No additional minimum weekly submission count threshold was imposed beyond zero-count exclusion; the $R^2 > 0.90$ and $p < 0.05$ linearity criteria served as the effective quality filter, ensuring only windows with stable, well-supported log-linear trends were retained, consistent with the original DPGR methodology \cite{pantho_data-driven_2025}. These window parameters and selection criteria were established through sensitivity analysis in the original DPGR study \cite{pantho_data-driven_2025} and are adopted here without modification.

\subsection{Conversion of Pairwise DPGR to Absolute Fitness}
Pairwise DPGR values are relative by definition, describing the growth advantage of one clade over another within a specific region and time window. To construct a single interpretable fitness scale, pairwise DPGR values were converted to absolute fitness scores following Pantho et al.~\cite{pantho_data-driven_2025}. A neighbour-joining tree was constructed from the pairwise DPGR distance matrix, with the earliest circulating clade designated as the reference point at fitness zero. Absolute fitness values for all other clades were inferred from branch lengths in this tree. Each sequence in the CNN dataset was then assigned the absolute fitness value corresponding to its clade and geographic region. This absolute fitness score serves as the regression target for the deep learning models.

\subsection{Deep learning for prediction}

\subsubsection{Data Acquisition}
Raw genomic sequences of influenza A viruses were sourced from the GISAID EpiFlu database \cite{shu_gisaid_2017}. To capture the complete evolutionary landscape of the virus, data acquisition was strictly limited to isolates containing all eight viral RNA segments: PB2, PB1, PA, HA, NP, NA, MP, NS \cite{Bouvier2008InfluenzaBiology}. Due to significant disparities in global genomic surveillance and sequencing availability, the dataset was geographically restricted to regions with robust data density; specifically North America, Europe, and Asia. Isolates from Africa, Oceania, and South America were excluded to prevent statistical skewing from sparse geographical representation.

To standardize the alignment process, universal reference genomes were acquired: the WHO vaccine strain A/Hong Kong/4801/2014 (EPI\_ISL\_189814), obtained from the GISAID EpiFlu database (see Acknowledgments) \cite{GISAID_EPI_ISL_189814}
for the H3N2 cohort, and A/California/07/2009 (GCF\_001343785.1) \cite{NCBI_GCF_001343785} sourced from NCBI for the H1N1 cohort. Segment-specific FASTA files were downloaded globally by region and subsequently merged into a single unaligned file per segment.

\subsubsection{Multiple Sequence Alignment (MSA) and Trimming}
To establish a rigid, standardized coordinate grid for downstream machine learning, the universal reference sequences were anchored to the beginning of their respective global segment pools using bash concatenation. Multiple sequence alignment of the sequences to their reference genomes was carried out using the MAFFT multiple sequence alignment program \cite{Katoh2013MAFFT}.

The progressive FFT-NS-2 algorithm \cite{Katoh2013MAFFT} was employed with maximum CPU parallelization (\texttt{--thread -1}). To prevent the rejection of sequencing reads containing ambiguous bases, the \texttt{--anysymbol} flag was applied.

Following alignment, aggressive noise reduction was performed using \texttt{trimAl} \cite{CapellaGutierrez2009trimAl}. A strict gap threshold (\texttt{-gt 0.05}) was applied to the FASTA files, forcing the algorithm to delete any vertical alignment column consisting of more than 95\% gap characters (\texttt{-}). This methodology effectively eliminated "ragged ends" and sequencing artifacts while preserving statistically significant, biologically relevant insertions.

\subsubsection{Full-Genome Reconstruction and Target Integration}
Following trimming, continuous full-length viral genomes were reconstructed. A custom Python streaming algorithm matched the eight isolated segments by their unique EPI\_ISL accession identifiers, concatenating them in strict biological order (PB2 through NS). Isolates missing one or more segments were actively dropped, yielding an intermediate dataset of 88,621 complete genomes.

To integrate the evolutionary fitness metric, a custom matching algorithm was developed using a Python script. Because viral fitness varies by geographical environment, the script utilized a composite key comprising [Clade + Continent] to map isolates to their corresponding absolute Differential Population Growth Rate (DPGR) scores. Sequences lacking comprehensive metadata or corresponding DPGR values were systematically excluded.

The final datasets to be used for the deep learning comprised 55,638 unique sequences for H3N2 (standardized to exactly 13,629 nucleotides in length) and 30,535 unique sequences for H1N1 (standardized to exactly 13,632 nucleotides). The dataset was partitioned into training and test sets using a stratified random 80:20 split, stratified by clade and geographic region to preserve class balance, following the methodology established in Annan et al.~\cite{Annan_Nkonu_Hatami_Pantho_Qingge_Qin_2025}.

\subsubsection{Convolutional Neural Network (CNN) Architecture}
To predict continuous DPGR fitness scores, Convolutional Neural Networks (CNNs) adapted from \cite{Annan_Nkonu_Hatami_Pantho_Qingge_Qin_2025} were developed for each viral subtype. The models adapted an established baseline architecture, implementing critical optimizations including a shift from Stochastic Gradient Descent (SGD) to the Adam optimizer \cite{kingma2017adammethodstochasticoptimization}, an increased batch size from 32 to 128 to stabilize gradient updates, and Min-Max normalization applied to the target fitness values.

The genomic sequences were represented as 2D one-hot encoded arrays of shape $(L, 7)$, corresponding to sequence length ($L$) and the seven distinct sequence characters (A, C, G, T, N, I, \texttt{-}).

\textbf{H3N2 Model Architecture:} 
The H3N2 network accepted an input shape of $(13629, 7)$. Feature extraction was performed by three consecutive 1D Convolutional layers containing 92, 54, and 18 filters, respectively, utilizing kernel sizes of 3, 6, and 3, with a uniform stride of 2. Max-pooling layers (pool size 3, stride 2) were applied for spatial down-sampling. The flattened feature map was passed through four Dense hidden layers composed of 4280, 1756, 884, and 146 nodes utilizing ReLU activation. To mitigate overfitting, a 25\% dropout rate (0.25) was applied after the first and third Dense layers. The network was optimized with a learning rate of $9.926 \times 10^{-5}$.

\textbf{H1N1 Model Architecture:} 
The H1N1 network accepted an input shape of $(13632, 7)$. The three Conv1D layers utilized 92, 54, and 26 filters with kernel sizes of 7, 3, and 3, maintaining a stride of 2. Max-pooling was adjusted to a pool size of 2 (stride 2). The four Dense hidden layers contained 3512, 1756, 948, and 50 nodes (ReLU activation). A more aggressive dropout strategy of 25\% (0.25) was applied sequentially across the first, second, and third Dense layers. The network was optimized with a learning rate of $8.048 \times 10^{-5}$.

Both networks culminated in a single-node output layer utilizing a linear activation function to yield the final continuous DPGR prediction.

\begin{figure}[H]
    \centering
    \includegraphics[width=0.8\linewidth]{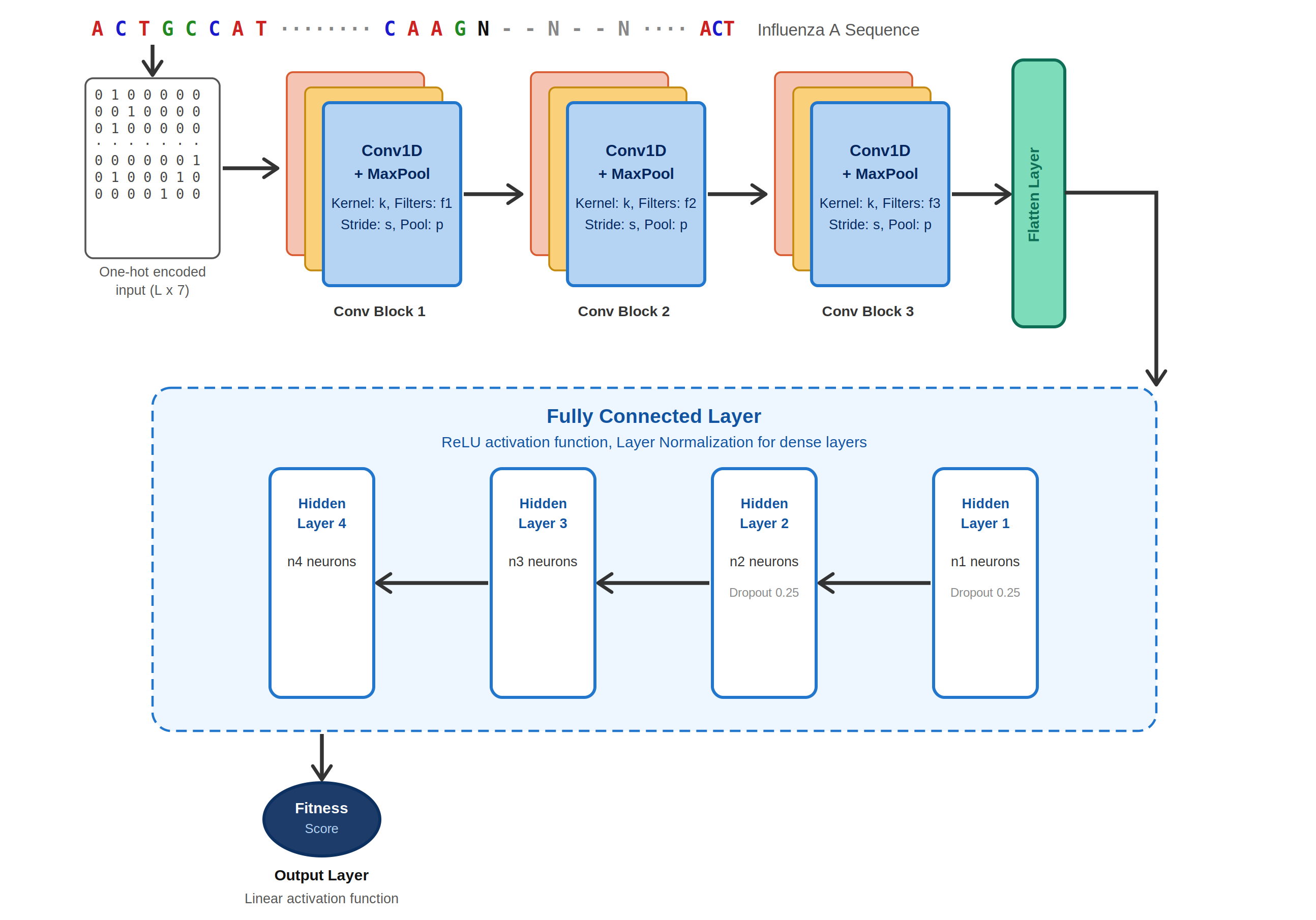} 
    \caption{Schematic representation of the subtype-specific 1D convolutional neural network (CNN) architecture. The model processes one-hot encoded full-genome sequences (input shape $L \times 7$) through three convolutional layers for hierarchical feature extraction, followed by four dense layers with ReLU activation and dropout regularization to predict the continuous DPGR fitness score.}
    \label{fig:cnn_architecture}
\end{figure}

\subsubsection{Conformal Prediction}
To provide statistically rigorous prediction intervals for CNN fitness estimates, we applied conformal prediction \cite{inbook} using a split conformal framework post-hoc. A held-out calibration set was partitioned from the test data after model convergence. For each calibration sample $i$, the nonconformity score was defined as the absolute residual:
\begin{equation}
    \alpha_i = |y_i - \hat{y}_i|
\end{equation}
where $y_i$ is the true DPGR score and $\hat{y}_i$ is the model's point prediction. For a desired coverage level $1 - \epsilon$ (here $0.95$), the conformal quantile $\hat{q}$ was computed as the $\lceil (n+1)(1-\epsilon) \rceil$-th smallest value among the $n$ calibration nonconformity scores. For each test sequence, the prediction interval was then constructed as:
\begin{equation}
    C(x) = [\hat{y} - \hat{q}, \hat{y} + \hat{q}]
\end{equation}
Conformal intervals were computed separately for the H3N2 and H1N1 models, using calibration sets drawn from held-out sequences not seen during training. All target values were transformed back to the original DPGR scale before interval reporting.

\subsubsection{Model Interpretability via SHAP}

To interpret the trained CNN models and identify genomic positions 
driving DPGR fitness predictions, we applied SHapley Additive 
exPlanations (SHAP) \cite{lundberg2017unifiedapproachinterpretingmodel}. SHAP assigns each 
feature an importance value $\phi_i$ for a particular prediction, 
where the explanation model takes the form of an additive linear 
function:

\begin{equation}
    g(z') = \phi_0 + \sum_{i=1}^{M} \phi_i z'_i
\end{equation}

\noindent where $z' \in \{0,1\}^M$, $M$ is the number of input 
features, and $\phi_i \in \mathbb{R}$ is the Shapley value assigned 
to feature $i$. The base value $\phi_0 = E[f(z)]$ represents the 
mean model output over the background distribution, and the sum of 
all $\phi_i$ values recovers the deviation of the prediction from 
this baseline: $f(x) = \phi_0 + \sum_{i=1}^{M} \phi_i$.

SHAP \cite{lundberg2017unifiedapproachinterpretingmodel} values were computed using the \texttt{GradientExplainer} \cite{lundberg2017unifiedapproachinterpretingmodel}
implementation in the \texttt{shap} Python library, which 
approximates Shapley values for differentiable models via 
gradient-based backpropagation through the network. A background 
reference distribution of 200 sequences was randomly sampled from 
the training set to serve as the baseline for computing expected 
gradients. SHAP values were computed for 100 representative 
sequences from the test set for each subtype (H3N2 and H1N1), 
yielding a SHAP matrix of shape $(100 \times L \times 7)$, where 
$L$ is the concatenated genome length (13,629 nt for H3N2; 13,632 
nt for H1N1) and 7 represents the one-hot nucleotide encoding 
channels.

For sequence-level interpretation, individual SHAP waterfall plots 
were generated for three representative sequences selected from the 
test set: one from the low-fitness tier, one from the mid-fitness 
tier, and one from the high-fitness tier. Feature names were constructed by comparing each position 
of the input sequence against the subtype-specific reference 
genome, labelling substitutions in standard mutation notation 
(e.g., A7460G), insertions (e.g., Ins11694A), and deletions 
(e.g., Del4644), with the corresponding viral segment appended in 
parentheses. The top six features by absolute SHAP value are 
displayed in each waterfall plot.

The base value $E[f(X)]$ used in all waterfall plots was computed 
as the mean model prediction over the 200-sequence background 
sample on the min--max normalised DPGR scale, corresponding to 
raw biological baselines of $0.0178$ $\log_{10}$/day for H3N2 and 
$0.0246$ $\log_{10}$/day for H1N1.

\section{Results}
Pairwise DPGR fitness estimates were computed for all H3N2 clade 
combinations with sufficient co-circulating weekly sequence data across 
the analysis period (2014-2025). Results are presented first at the 
continental level across North America, Europe, Asia, Oceania, and South 
America, followed by a dedicated United States analysis.

\subsection{A/H3N2}

\subsubsection{Continent-level clades A/H3N2 2025-2026 flu season}

DPGR analysis identified subclade~K (3C.2a1b.2a.2a.3a.1/K) as the 
faster-growing variant relative to the vaccine strain 
3C.2a1b.2a.2a.3a.1 across all continental regions with available data. In North America, the fitness 
advantage of subclade~K was detected over an 
11-week window (DPGR $= +0.02097$, window: 
Jul--Sep~2025)
In Europe, the fitness advantage of subclade~K was detected over a 
9-week window (DPGR $= +0.0172$, window: Oct--Dec~2025). In Asia, the 
signal appeared earlier over an 11-week window 
(DPGR $= +0.0151$, window: Aug--Oct~2025). In South America, the 
strongest fitness advantage was observed (DPGR $= +0.0354$, window: 
Oct--Dec~2025), and in Oceania a consistent signal was detected over 
12 weeks (DPGR $= +0.0175$, window: Sep--Nov~2025). The convergence 
of positive DPGR values for subclade~K across geographically distinct 
regions and non-overlapping time windows indicates a globally 
coordinated fitness shift, with subclade~K growing faster than the 
prevailing vaccine strain lineage across all monitored continents 

\begin{figure}[H]
  \centering
  \captionsetup[subfigure]{font=tiny, labelfont=tiny}

  % --- Top Row ---
  \begin{subfigure}{0.24\textwidth}
    \centering
    \includegraphics[width=\linewidth]{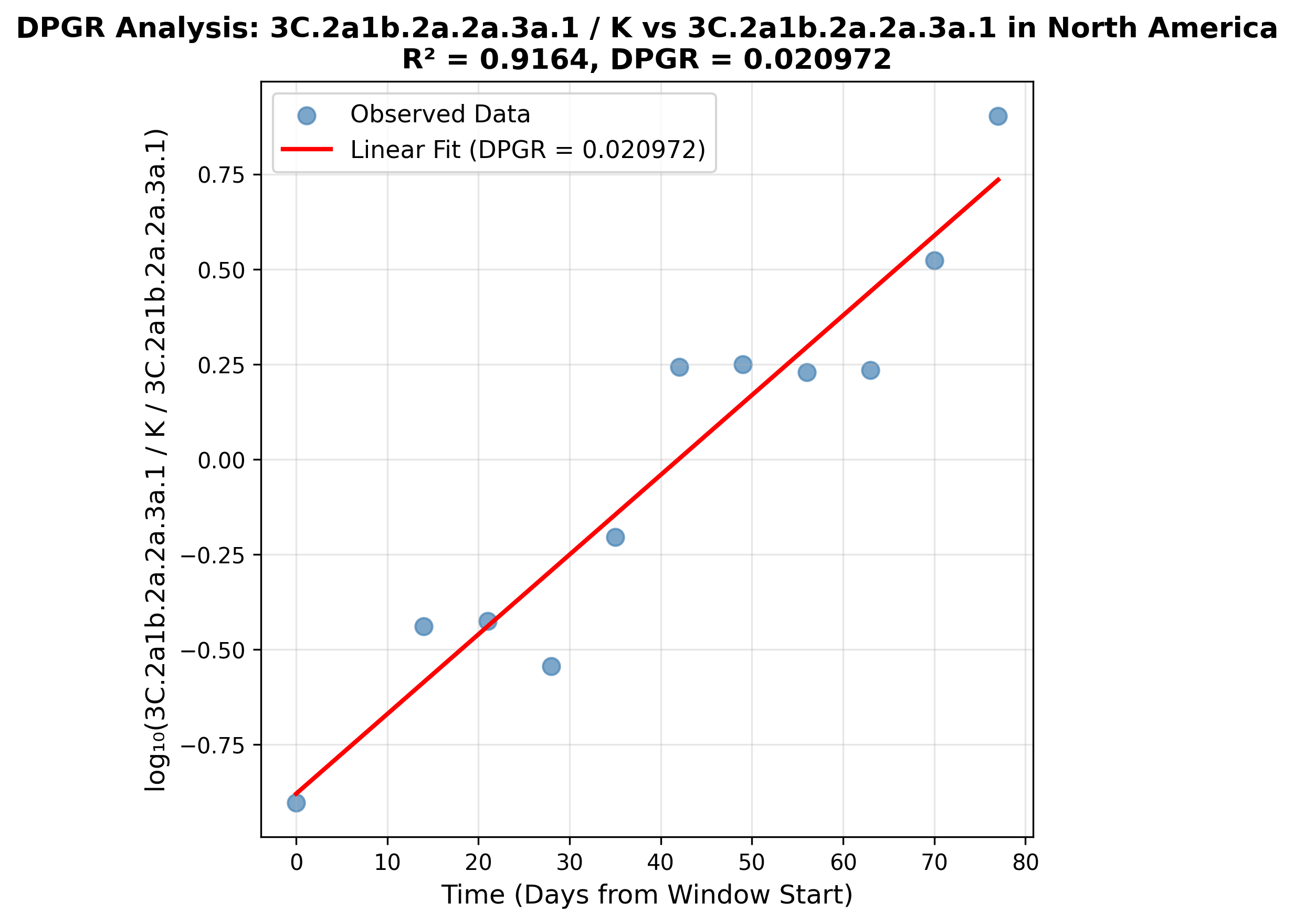}
    \caption{2025-07-14 - 2025-09-29}
  \end{subfigure}\hfill
  \begin{subfigure}{0.24\textwidth}
    \centering
    \includegraphics[width=\linewidth]{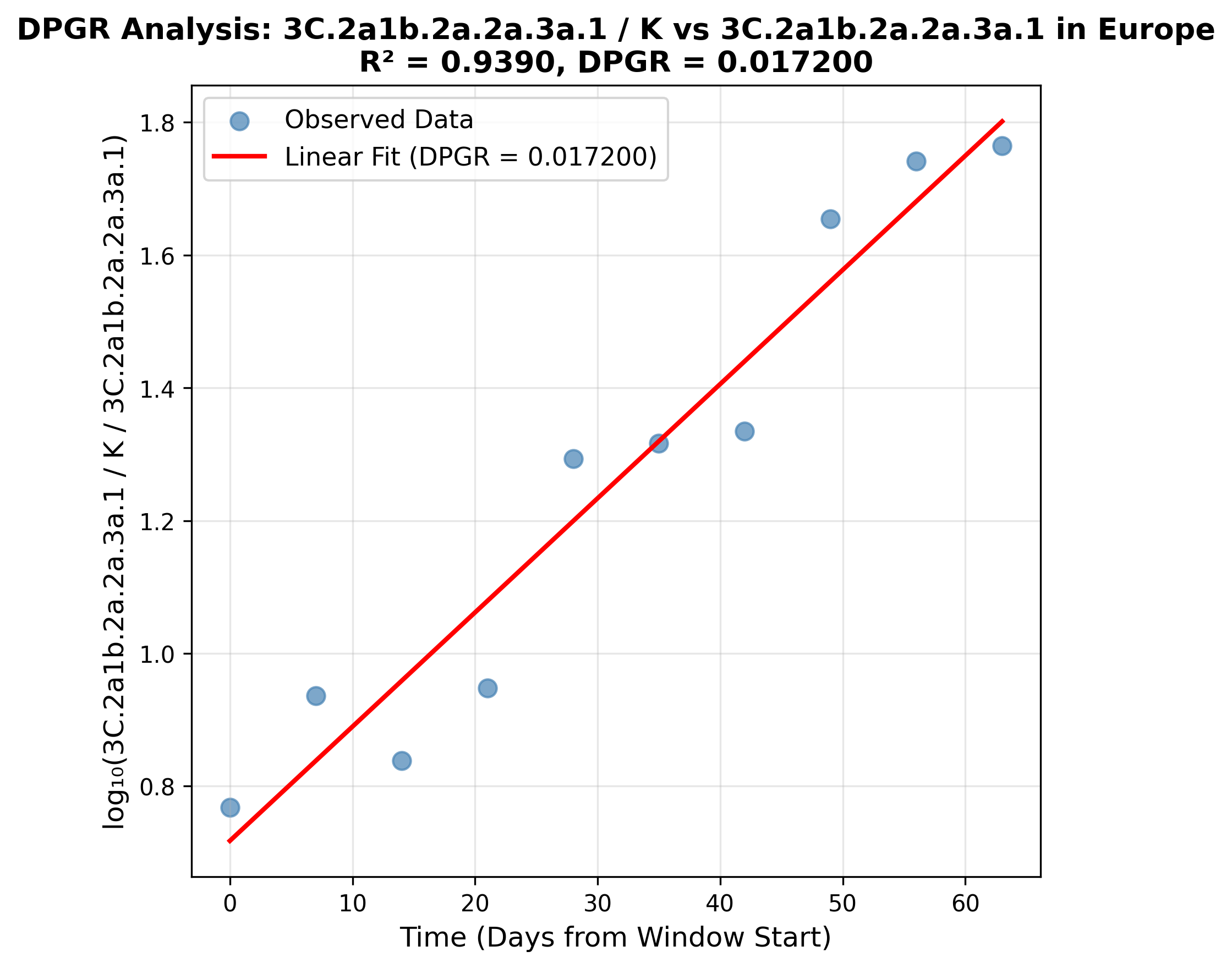}
    \caption{2025-10-13 - 2025-12-15}
  \end{subfigure}\hfill
  \begin{subfigure}{0.24\textwidth}
    \centering
    \includegraphics[width=\linewidth]{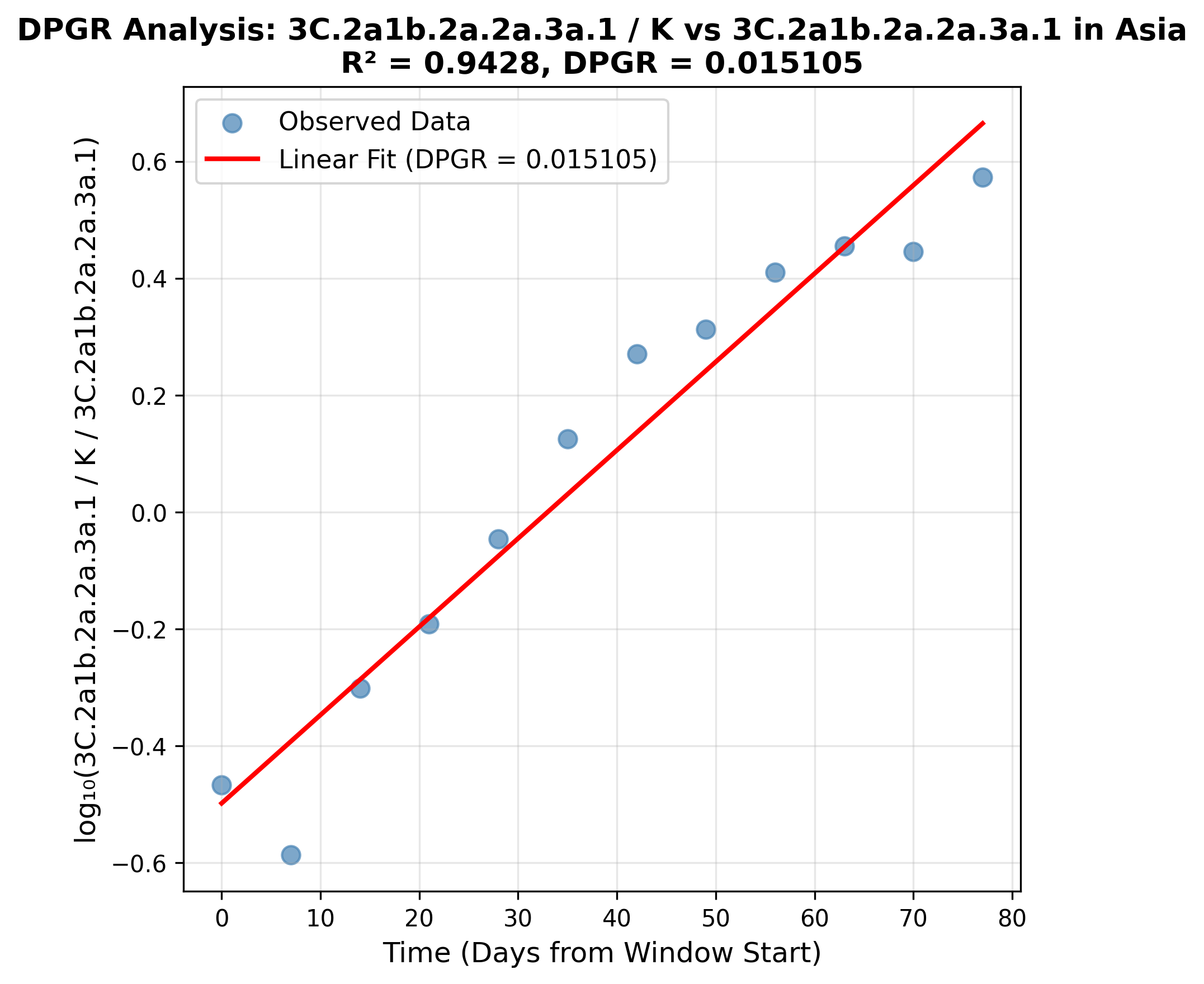}
    \caption{2025-08-11 - 2025-10-27}
  \end{subfigure}\hfill
  \begin{subfigure}{0.24\textwidth}
    \centering
    \includegraphics[width=\linewidth]{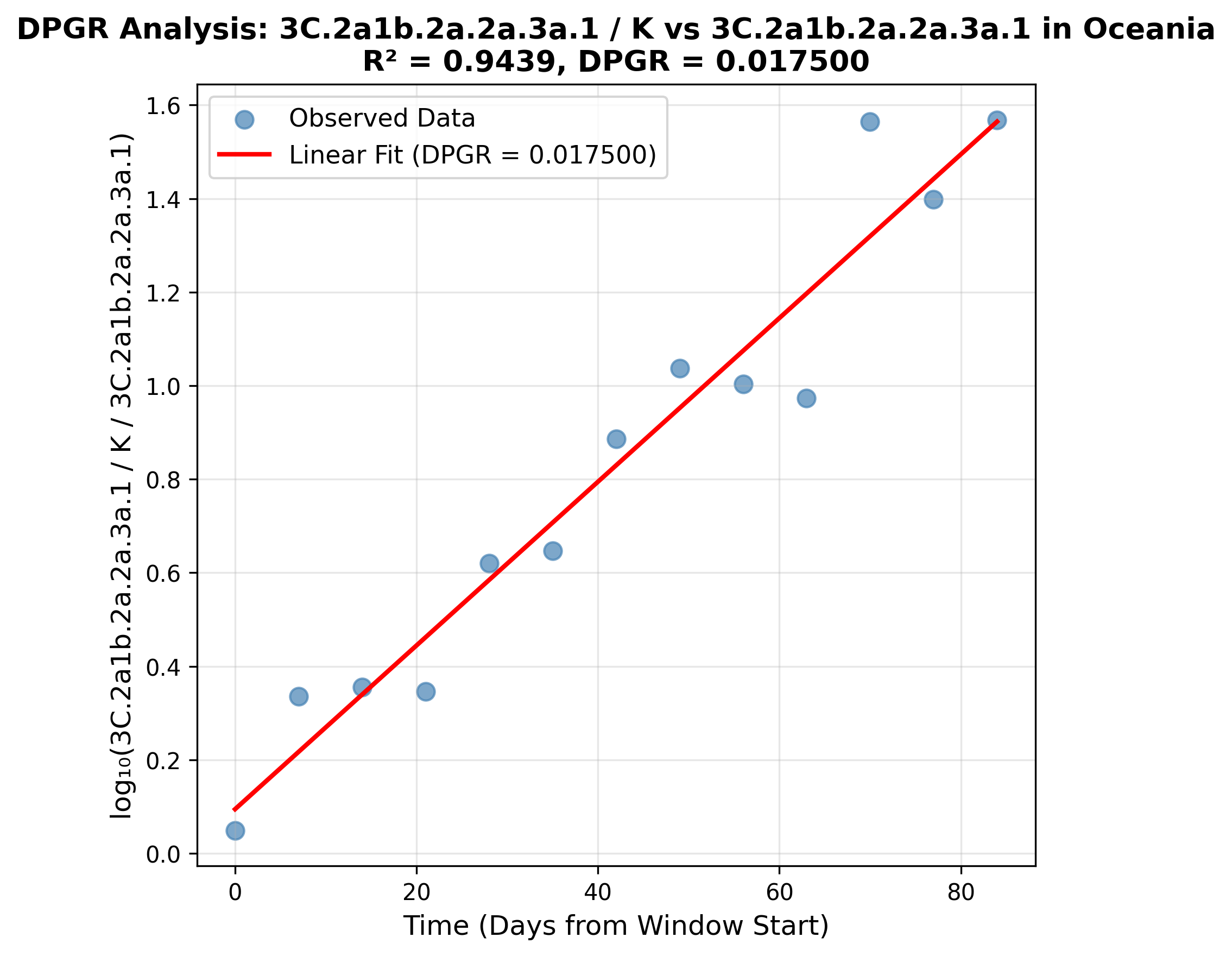}
    \caption{2025-09-01 - 2025-11-24}
  \end{subfigure}\hfill
  \begin{subfigure}{0.24\textwidth}
    \centering
    \includegraphics[width=\linewidth]{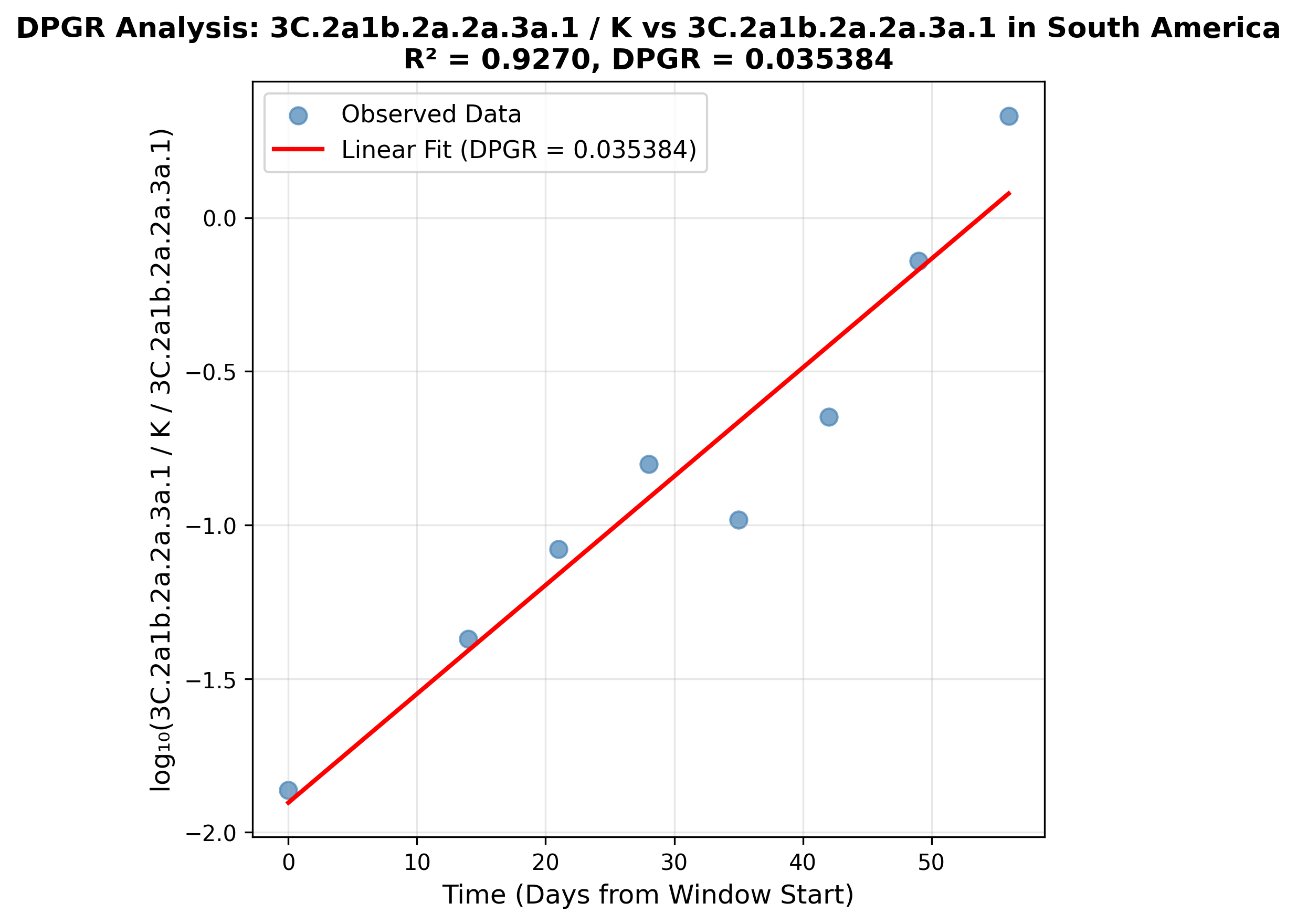}
    \caption{2025-10-13 - 2025-12-08}
  \end{subfigure}
  
  \caption{Clade-level analysis of A/H3N2 subtype in the 2025-2026 flu season per continent}
  \label{fig:continents_h3n2-25/26}
\end{figure}

\subsubsection{Continent-level clades A/H3N2 2022-2023 season}

DPGR analysis across continents consistently identified subclade 
3C.2a1b.2a.2a.3a.1 as the faster-growing variant relative to 
3C.2a1b.2a.2b. In North America, 3C.2a1b.2a.2a.3a.1 outgrew 
3C.2a1b.2a.2b (DPGR $= -0.0212$, window: Nov~2022--Feb~2023). In 
Europe, the same competitive relationship was observed 
(DPGR $= -0.0071$, window: Nov~2022--Jan~2023). In Asia, 
3C.2a1b.2a.2a.3a.1 similarly outgrew 3C.2a1b.2a.2b 
(DPGR $= -0.0221$, window: Dec~2022--Jan~2023). The negative slopes 
across all three continental regions reflect 3C.2a1b.2a.2a.3a.1 as the 
faster-growing denominator in each pair, indicating competitive 
dominance over 3C.2a1b.2a.2b across continents by early 2023.

\begin{figure}[H]
  \centering
  \captionsetup[subfigure]{font=tiny, labelfont=tiny}

  % --- Top Row ---
  \begin{subfigure}{0.24\textwidth}
    \centering
    \includegraphics[width=\linewidth]{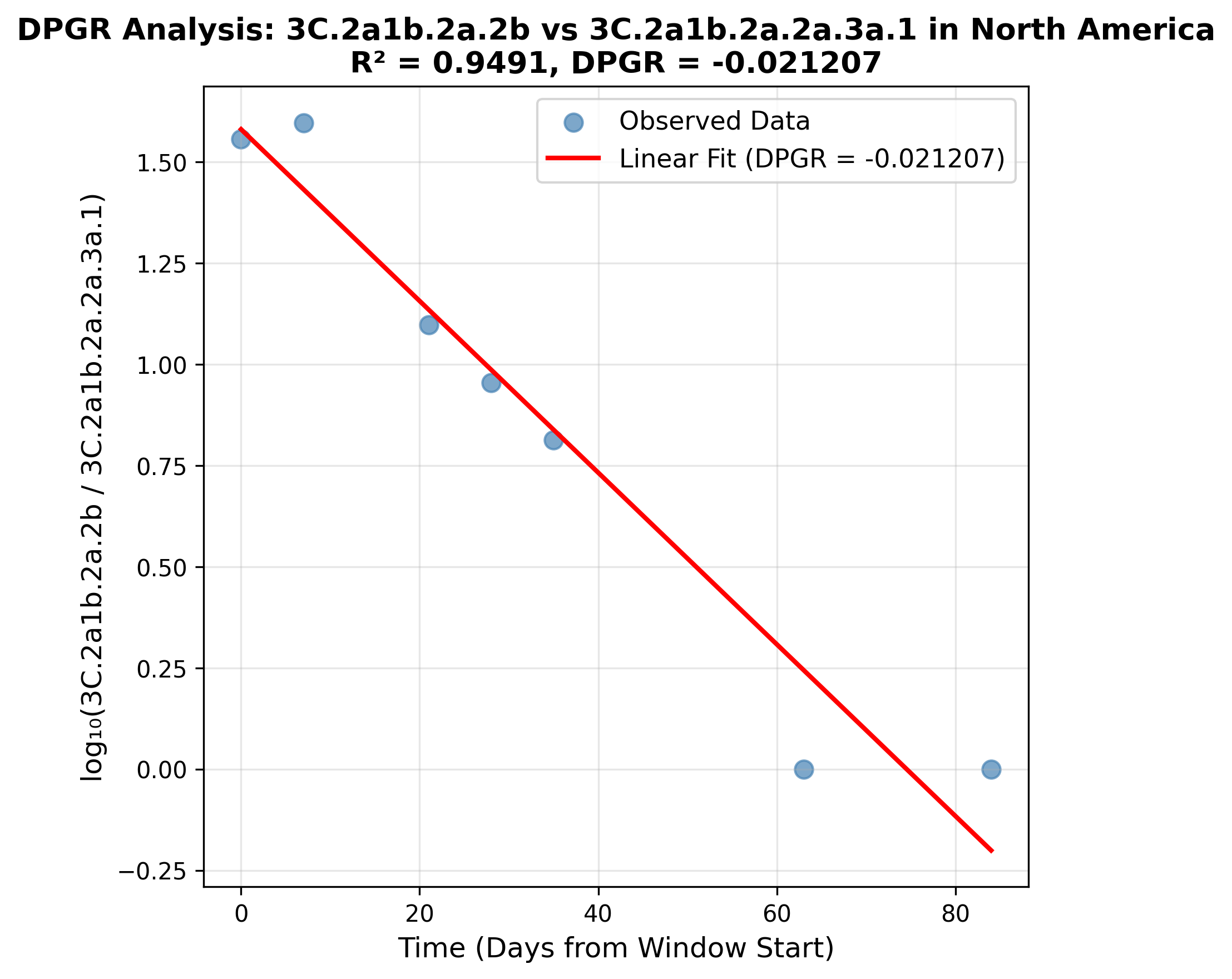}
    \caption{2022-11-28 - 2023-02-20}
  \end{subfigure}\hfill
  \begin{subfigure}{0.24\textwidth}
    \centering
    \includegraphics[width=\linewidth]{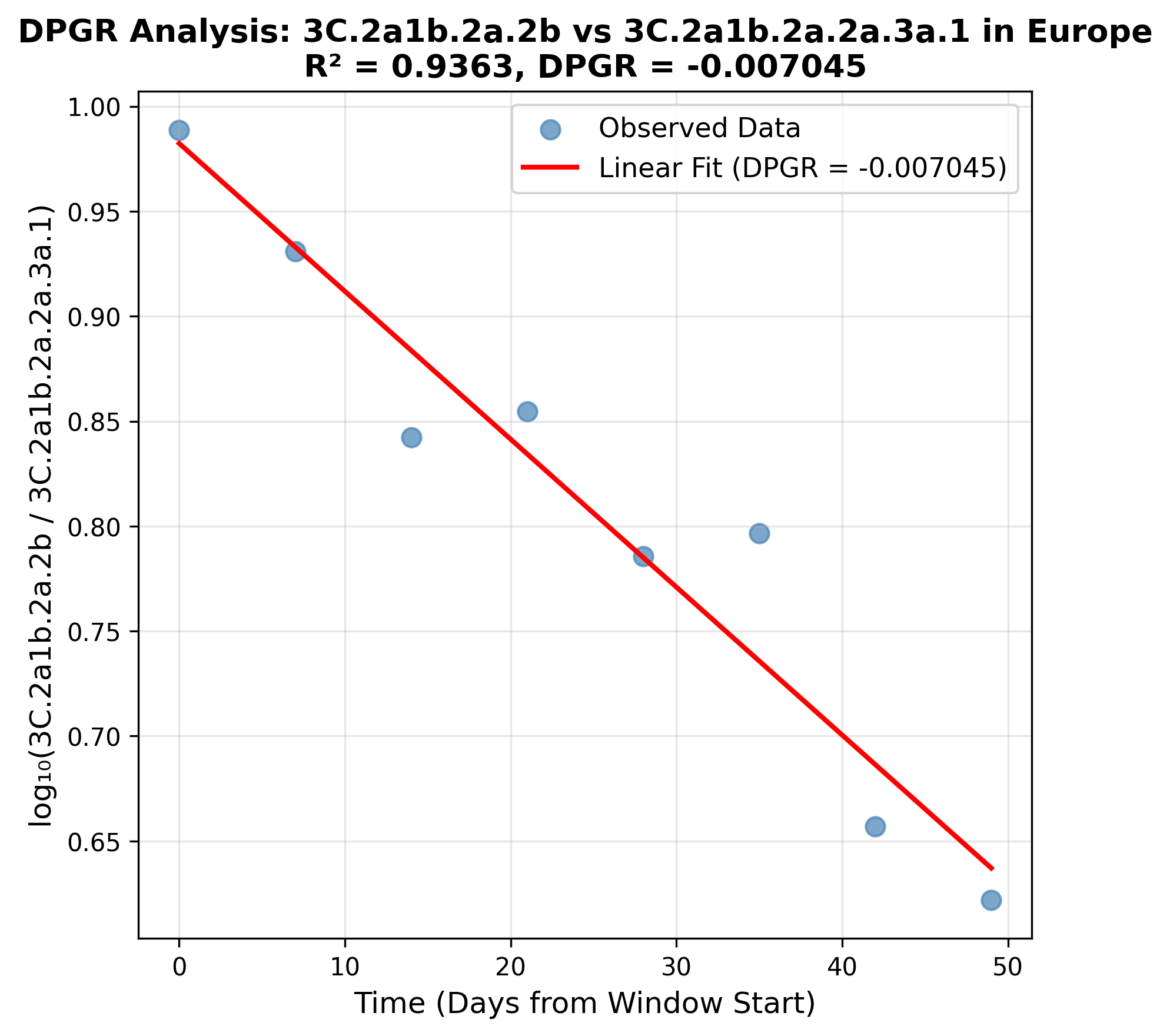}
    \caption{2022-11-21 - 2023-01-09}
  \end{subfigure}\hfill
  \begin{subfigure}{0.24\textwidth}
    \centering
    \includegraphics[width=\linewidth]{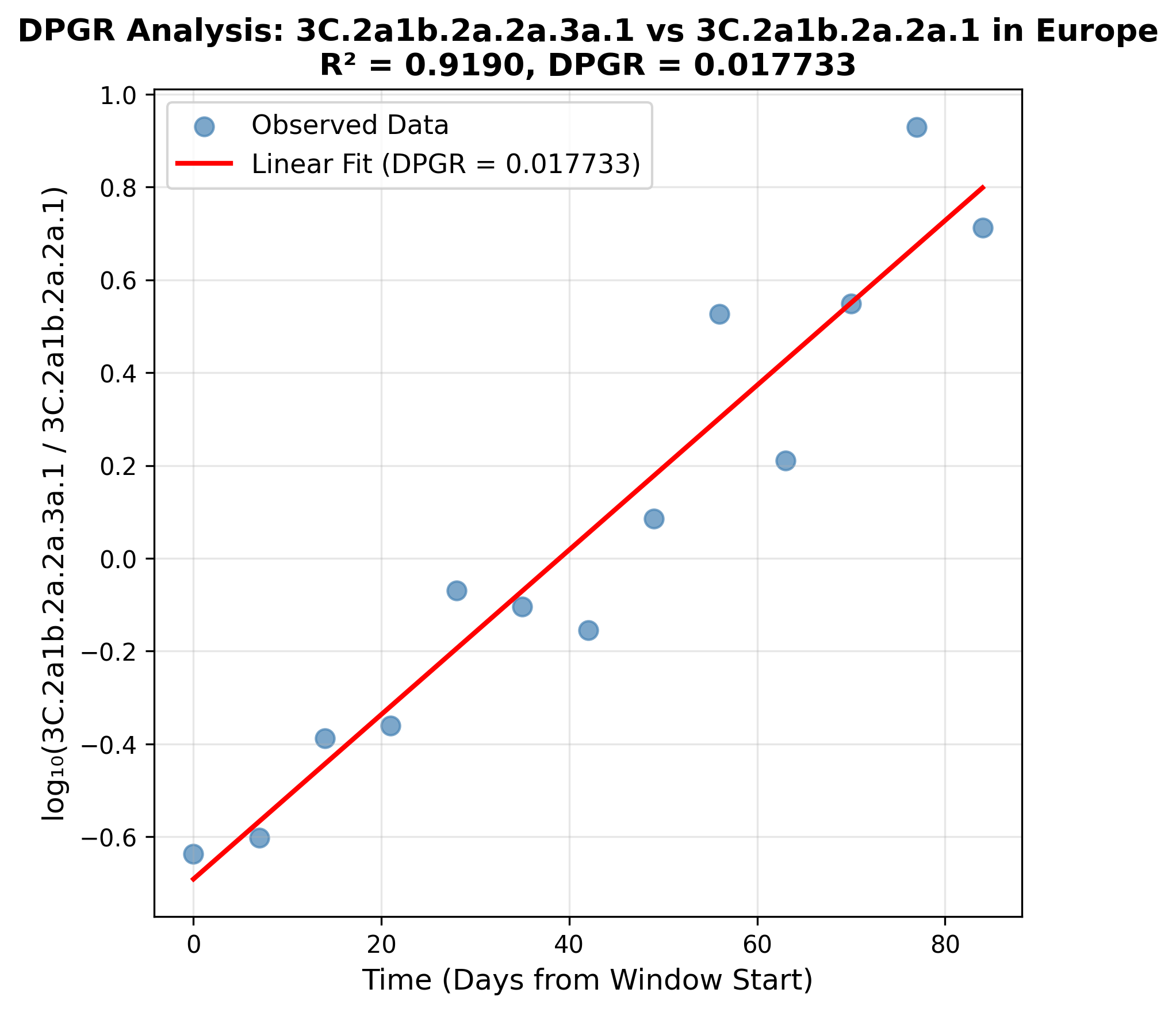}
    \caption{2022-09-19 - 2022-12-12}
  \end{subfigure}\hfill
  \begin{subfigure}{0.24\textwidth}
    \centering
    \includegraphics[width=\linewidth]{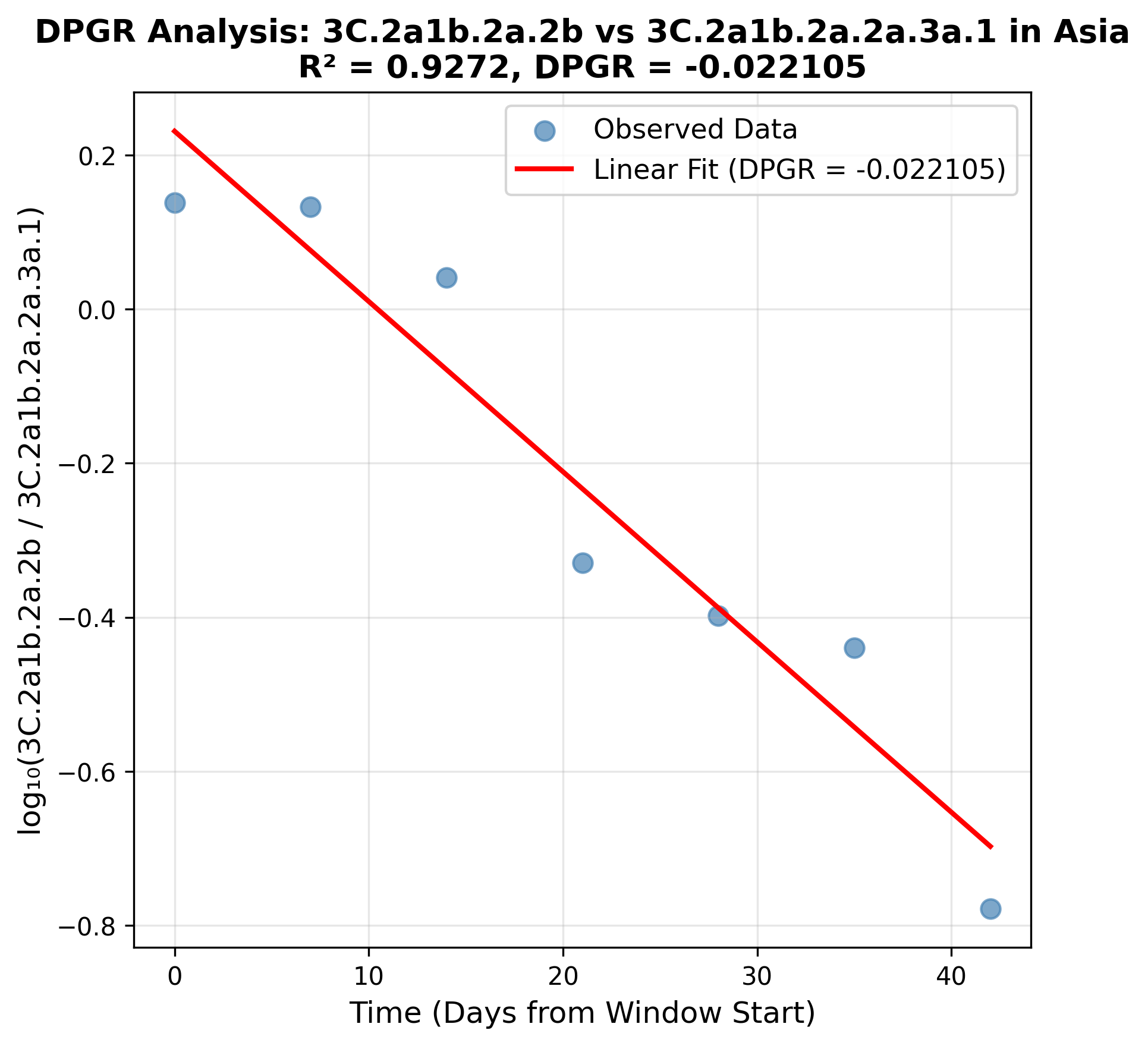}
    \caption{2022-12-12 - 2023-01-23}
  \end{subfigure}

  \caption{Clade-level analysis of A/H3N2 subtype in the 2022-2023 flu season per continent}
  \label{fig:continents_h3n2-2022/23}
\end{figure}

\paragraph{Supplementary seasonal panels.}
Additional historical H3N2 regression panels, the multi-season United
States panel collections are provided in Appendix~\ref{app:dpgr_appendix} to
keep the main text focused on the most current cross-regional results.

\subsubsection{Evolution Landscape of Relative Fitness for A/H3N2 in the USA}
To complement the pairwise regression results, we visualised the
relative-fitness landscape of contemporary H3N2 viruses in the United
States using a heatmap, fitness stair, and neighbour-joining tree.

\begin{figure}[H]
  \centering
  \captionsetup[subfigure]{font=tiny, labelfont=tiny}

  \begin{subfigure}[b]{0.48\textwidth}
    \centering
    \includegraphics[width=\linewidth]{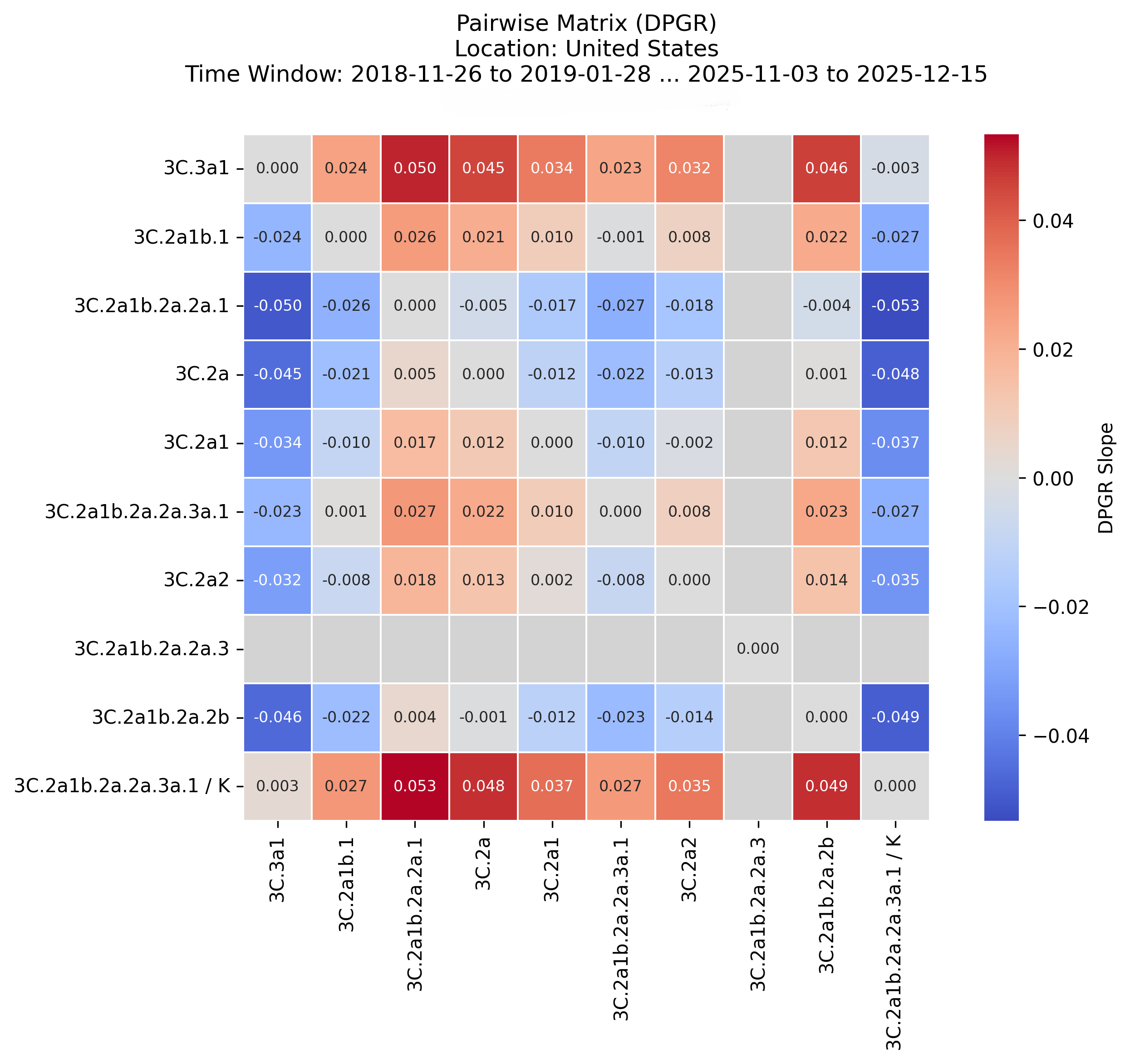}
    \caption{Heatmap Analysis}
    \label{fig:h3n2-heatmap}
  \end{subfigure}\hfill
  \begin{subfigure}[b]{0.48\textwidth}
    \centering
    \includegraphics[width=\linewidth, height=5cm]{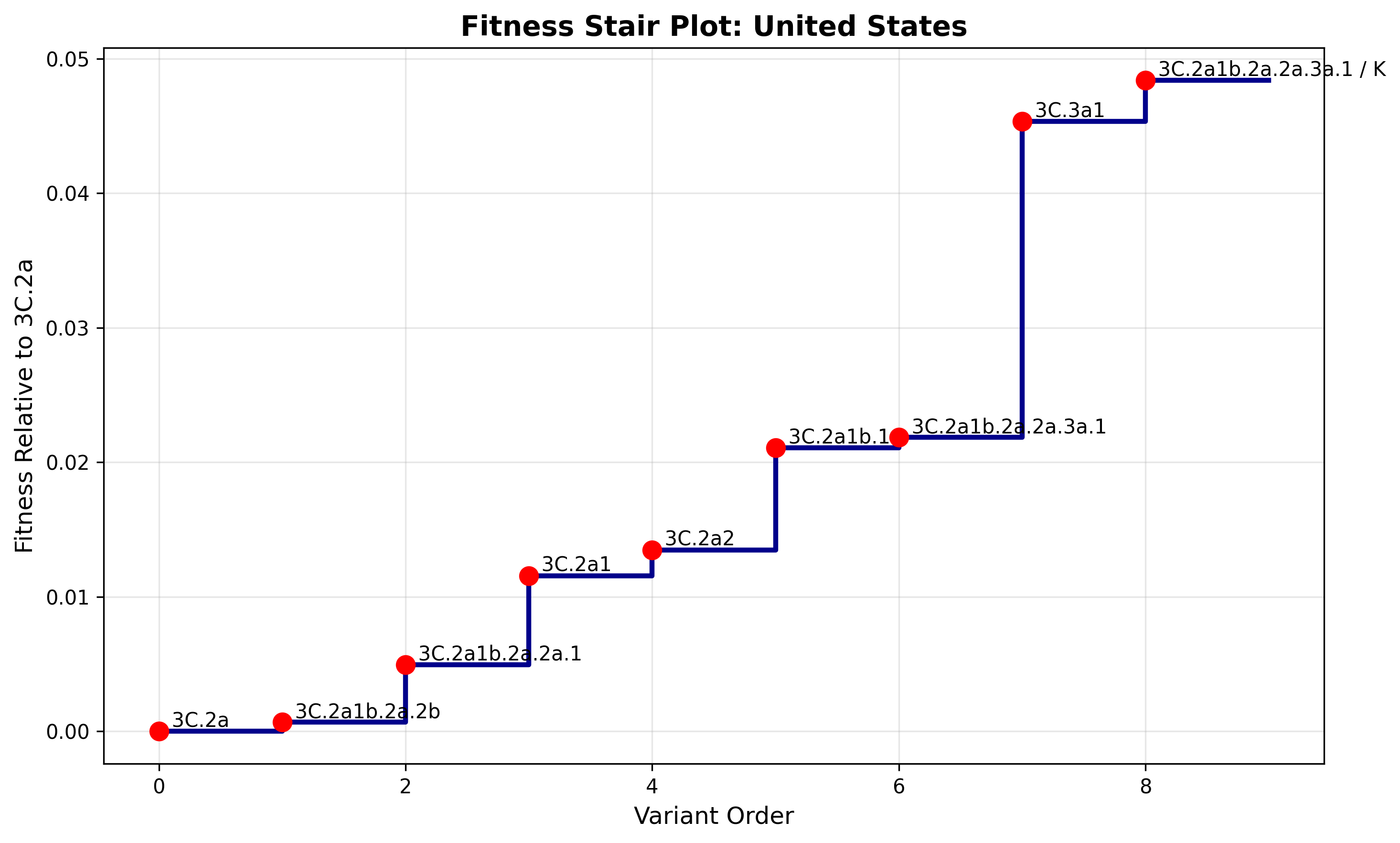}
    \caption{Fitness Stair}
    \label{fig:h3n2-fitness}
  \end{subfigure}

  \vspace{1em}

  \begin{subfigure}[b]{0.48\textwidth}
    \centering
    \includegraphics[width=\linewidth]{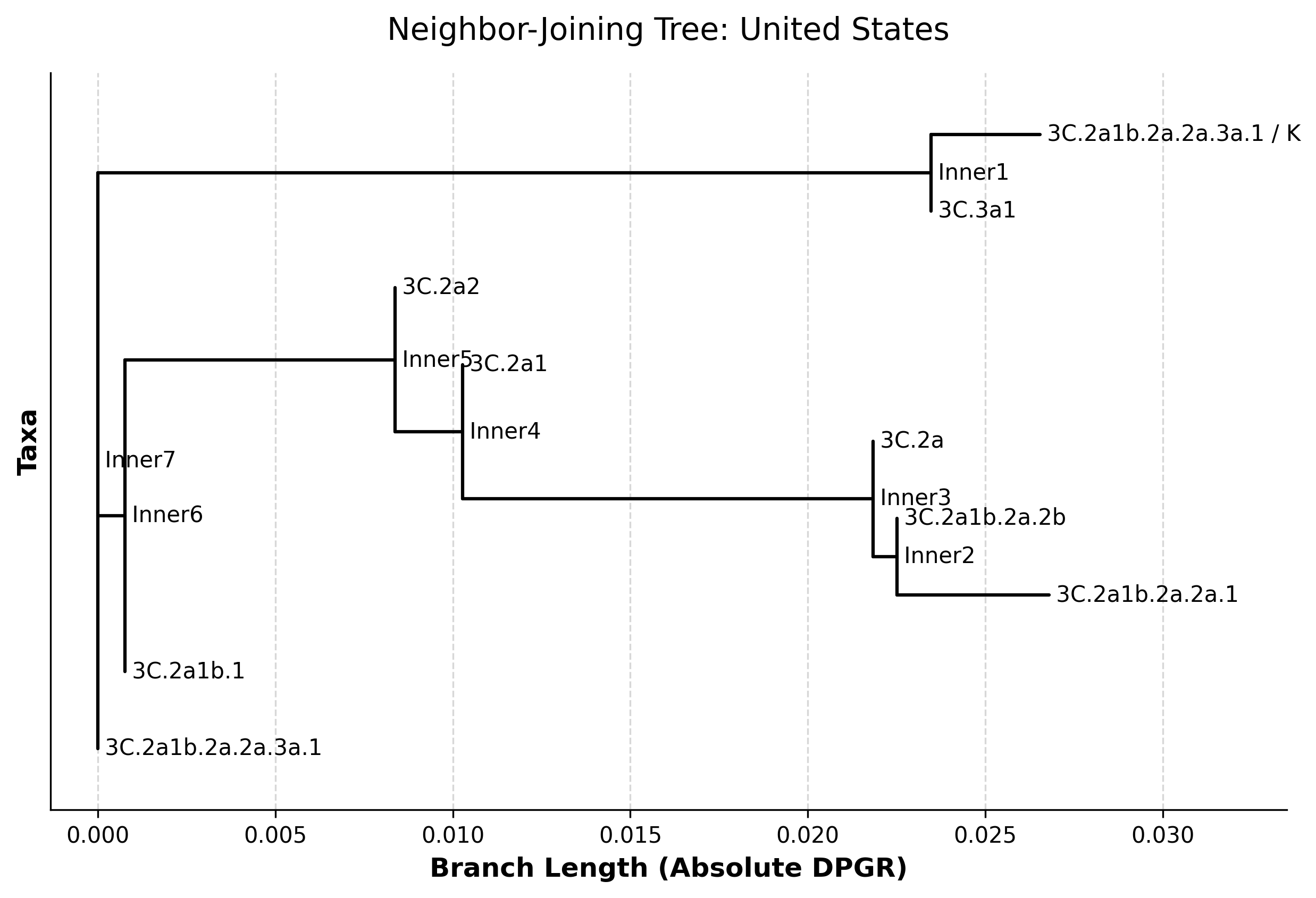}
    \caption{Neighbour Joining Tree}
    \label{fig:h3n2-tree}
  \end{subfigure}

  \caption{A/H3N2 evolution landscape analysis for the United States.}
  \label{fig:h3n2-landscape}
\end{figure}

\subsection{A/H1N1}
DPGR likewise captured recurrent H1N1 clade turnover across continents
and within the United States. The full continental and U.S. regression
panels are provided in Appendix~\ref{app:dpgr_appendix}.

\subsubsection{Evolution Landscape of Relative Fitness for A/H1N1 in the USA}
We likewise summarised the United States H1N1 fitness structure with a
heatmap, fitness stair, and neighbour-joining tree to contextualise the
pairwise DPGR comparisons.

\begin{figure}[H]
  \centering
  \captionsetup[subfigure]{font=tiny, labelfont=tiny}

  \begin{subfigure}[b]{0.48\textwidth}
    \centering
    \includegraphics[width=\linewidth]{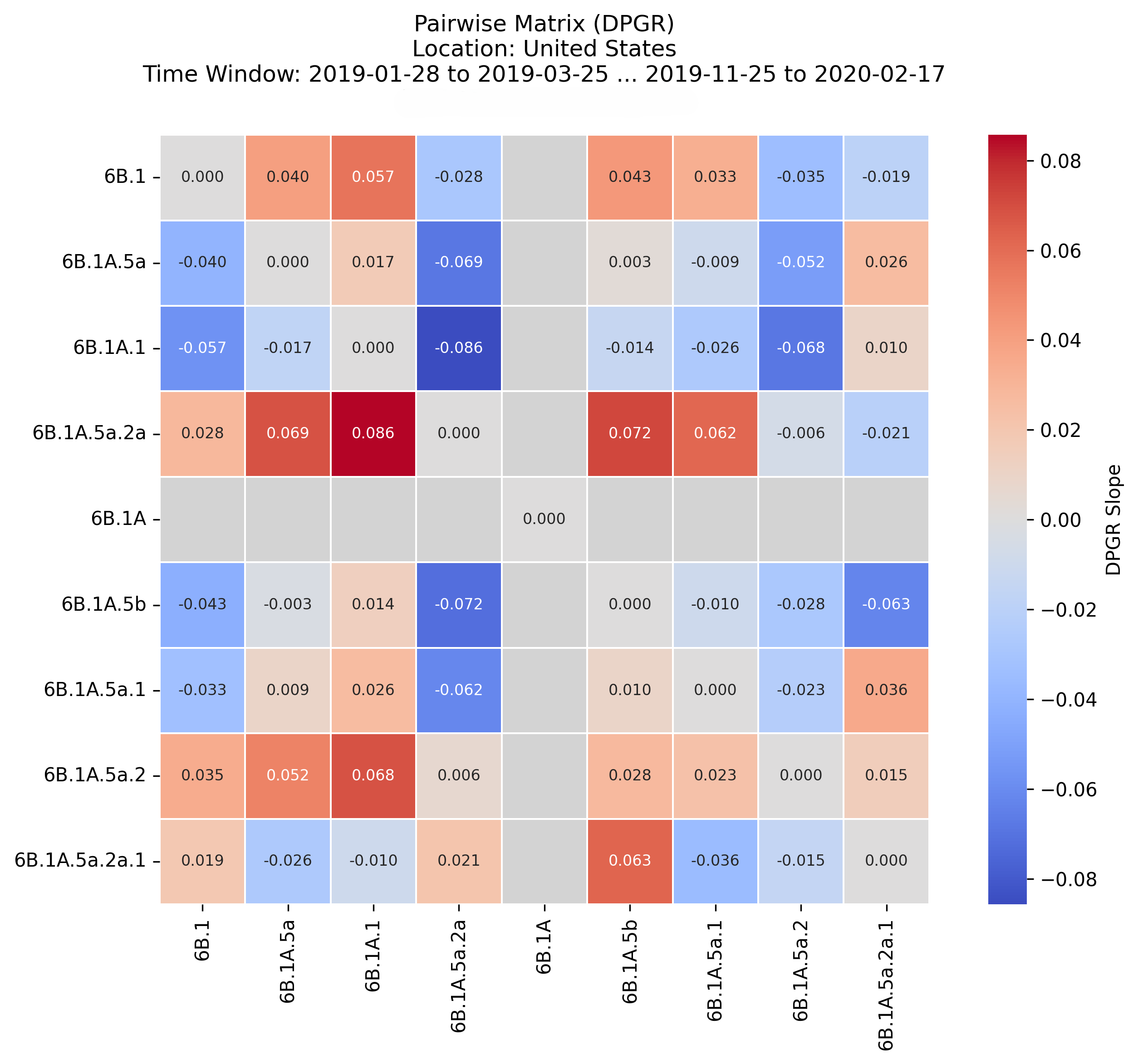}
    \caption{Heatmap Analysis}
    \label{subfig:h1n1-heatmap}
  \end{subfigure}\hfill
  \begin{subfigure}[b]{0.48\textwidth}
    \centering
    \includegraphics[width=\linewidth, height=5cm]{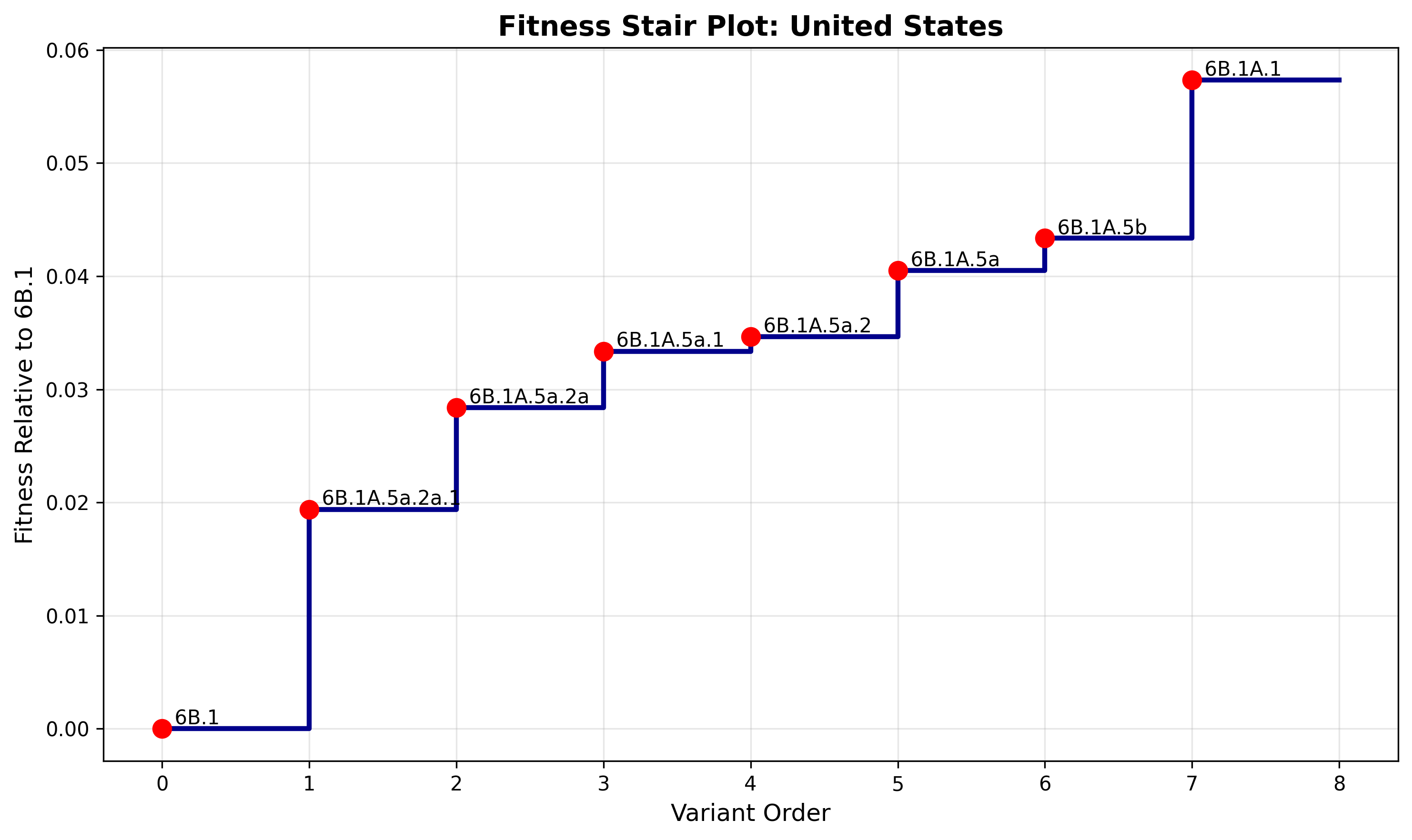}
    \caption{Fitness Stair}
    \label{subfig:h1n1-fitness}
  \end{subfigure}

  \vspace{1em}

  \begin{subfigure}[b]{0.48\textwidth}
    \centering
    \includegraphics[width=\linewidth]{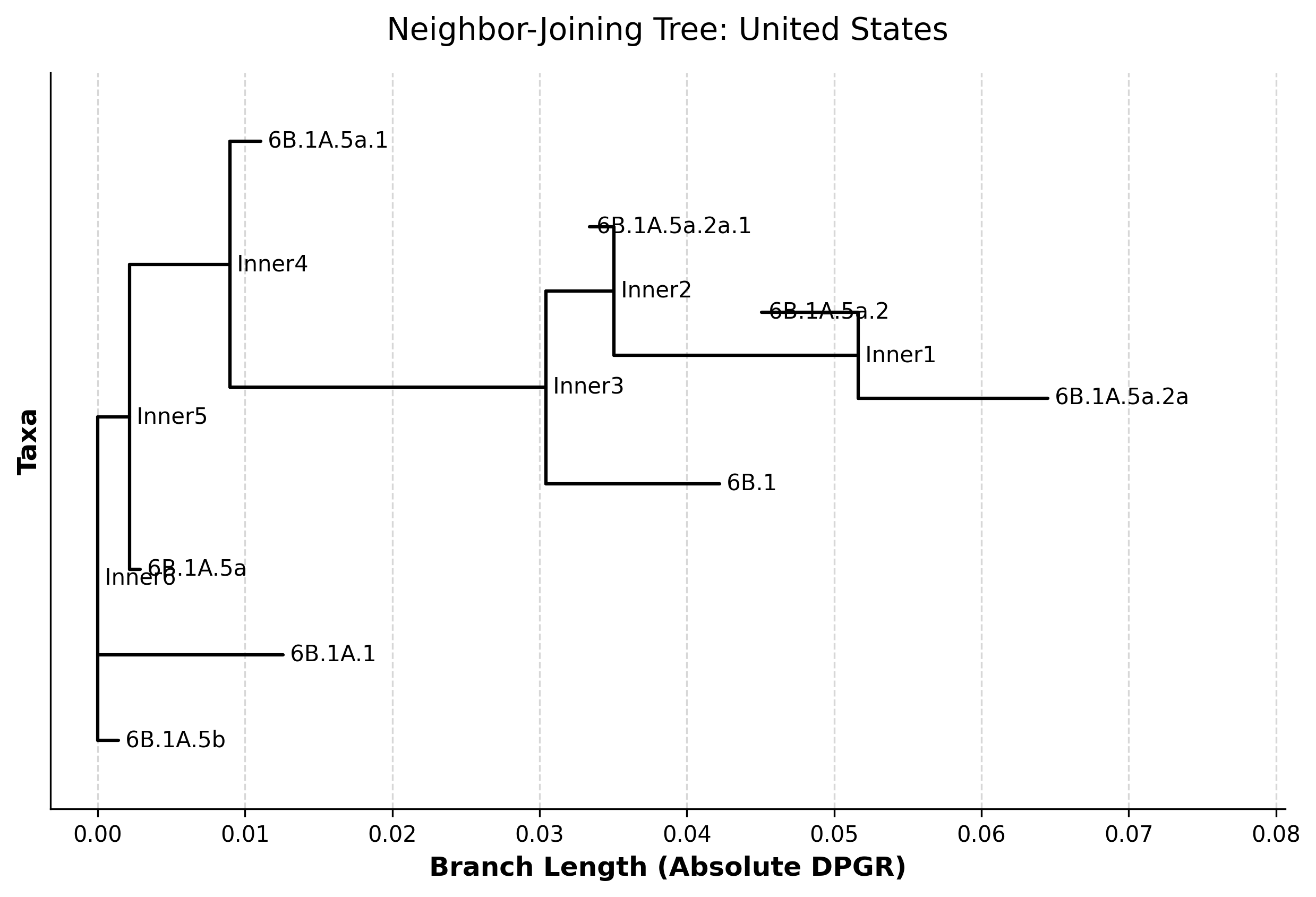}
    \caption{Neighbour Joining Tree}
    \label{subfig:h1n1-tree}
  \end{subfigure}

  \caption{A/H1N1 evolution landscape analysis for the United States.}
  \label{fig:h1n1-landscape}
\end{figure}

\subsection{Comparison With WHO and CDC Surveillance Trends}
To assess whether DPGR-derived fitness estimates are consistent with real-world
transmission dynamics, we compared our results against published 
WHO global influenza surveillance reports and CDC vaccine effectiveness 
(VE) data across five H3N2 seasons and 3 H1N1 seasons. In all pairwise 
comparisons, DPGR$_{1,2}$ is defined such that a positive slope indicates 
Variant~1 (numerator) growing faster than Variant~2 (denominator), and a 
negative slope indicates the opposite, consistent with 
Equation~\ref{eq:dpgr}. It should be noted that GISAID clade annotations 
follow WHO/FluNet phylogenetic nomenclature; the colloquial subclade 
``J'' labels used in some surveillance reports are not present in GISAID 
metadata, and full phylogenetic designations are used throughout.
 
\subsubsection{A/H3N2 Seasonal surveillance comparison}
 
\paragraph{2025-2026 season.}
The vaccine component was updated to A/Croatia/10136RV/2023 and 
A/District~of~Columbia/27/2023, both representing subclade 
3C.2a1b.2a.2a.3a.1. DPGR analysis identified subclade~K 
(3C.2a1b.2a.2a.3a.1/K) as having a fitness advantage over the vaccine 
strain across all five regions with available data 
(Table~\ref{tab:k_dpgr}).
 
\begin{table}[ht]
\centering
\caption{DPGR estimates for subclade~K (3C.2a1b.2a.2a.3a.1/K) versus 
the 2025--2026 vaccine strain (3C.2a1b.2a.2a.3a.1) by region.}
\label{tab:k_dpgr}
\begin{tabular}{llc}
\toprule
\textbf{Region} & \textbf{Time window} & \textbf{DPGR} \\
\midrule
United States  & 2025-11-03 to 2025-12-15 & $+0.0265$ \\
Europe         & 2025-10-13 to 2025-12-15 & $+0.0172$ \\
Asia           & 2025-08-11 to 2025-10-27 & $+0.0151$ \\
South America  & 2025-10-13 to 2025-12-08 & $+0.0354$ \\
Oceania        & 2025-09-01 to 2025-11-24 & $+0.0175$ \\
\bottomrule
\end{tabular}
\end{table}
 
The globally consistent positive DPGR values for subclade~K indicate 
faster growth relative to the vaccine strain across all monitored 
regions. This is corroborated by BC~CDC surveillance reporting that a 
substantial proportion of J.2.4 viruses, including the recently emerged 
subclade~K (formerly J.2.4.1), show immune escape potential from the 
2025--2026 J.2 vaccine strain~\cite{Sabaiduc2025}. Reports from 
Australia and New Zealand further indicate that subclade~K viruses were 
both genetically and antigenically distinct from the 2025 vaccine strain 
and from previously circulating subclade~J viruses, suggesting likely 
further expansion during the 2025--2026 northern hemisphere winter 
season~\cite{Dapat2025}.
 
\paragraph{2022-2023 season.}
The vaccine component represented clade 3C.2a1b.2a.2, with 
egg-based vaccines using A/Darwin/9/2021 and cell culture-based vaccines 
using A/Darwin/6/2021. DPGR analysis identified subclade 
3C.2a1b.2a.2a.3a.1 as having a consistent fitness advantage over the 
co-circulating subclade 3C.2a1b.2a.2b across all analysed regions: 
United States (DPGR $= -0.0225$, window: Dec~2022--Mar~2023), North 
America (DPGR $= -0.0212$, window: Nov~2022--Feb~2023), Europe 
(DPGR $= -0.0071$, window: Nov~2022--Jan~2023), and Asia 
(DPGR $= -0.0221$, window: Dec~2022--Jan~2023), with all negative slopes 
reflecting 3C.2a1b.2a.2a.3a.1 as the faster-growing denominator. 
However, since 3C.2a1b.2a.2a.3a.1 is a direct descendant of the vaccine 
clade 3C.2a1b.2a.2, WHO serological data showed that post-vaccination 
geometric mean titres against A(H3N2) viruses representing subclades 2a 
and 2b were not significantly reduced compared to titres against the 
cell culture-propagated A/Darwin/6/2021 vaccine virus~\cite{WHO2023}. 
This is consistent with CDC estimates that flu vaccination contributed 
to meaningful reductions in influenza impact during the 2022--2023 
season~\cite{CDC2023}. This case illustrates an important distinction: 
a positive DPGR fitness differential reflects competitive growth within 
a season but does not necessarily indicate vaccine failure when sufficient 
antigenic cross-reactivity is maintained within the same clade lineage.
 
\paragraph{2018-2019 season.}
The northern hemisphere vaccine contained 
A/Singapore/INFIMH-16-0019/2016 (clade 3C.2a1) and the southern 
hemisphere vaccine contained A/Switzerland/8060/2017 (clade 3C.2a1b). 
DPGR analysis continued to identify clade 3C.3a1 as the faster-growing 
variant, with 3C.2a1b.1 declining relative to 3C.3a1 in the United States 
(DPGR $= -0.0242$, window: Nov~2018--Jan~2019), and 3C.2a1 similarly 
outpaced by 3C.3a1 (DPGR $= -0.0338$, window: Oct--Dec~2018). WHO 
surveillance reported that the majority of A(H3N2) viruses collected from 
September~2018 to January~2019 belonged to subclade 3C.2a1b, but that 
the number of clade 3C.3a viruses increased substantially from 
November~2018 onward across multiple geographic regions~\cite{WHO2019}. 
The predominance of 3C.3a viruses in the latter part of the season was 
associated with decreased VE, and WHO subsequently updated the A(H3N2) 
vaccine component to a clade 3C.3a virus for the 2019--2020 northern 
hemisphere season~\cite{Flannery2019}.
 
\paragraph{2017-2018 season.}
The vaccine strain remained A/Hong~Kong/4801/2014 (clade 3C.2a). DPGR 
analysis detected a consistent fitness advantage for clade 3C.3a1 over 
all co-circulating 3C.2a subclades in North America. Specifically, 3C.3a1 
outgrew 3C.2a1b.1 (DPGR $= -0.0181$, window: Dec~2017--Feb~2018), 
3C.2a1 (DPGR $= -0.0320$, window: Dec~2017--Feb~2018), and 3C.2a2 
(DPGR $= -0.0240$), with all negative slopes reflecting 3C.3a1 as the 
faster-growing denominator variant. Within the 3C.2a lineage, 3C.2a2 also 
outgrew 3C.2a1 in North America (DPGR $= +0.0152$, window: 
Oct~2017--Jan~2018), and in Oceania, 3C.2a1 outgrew 3C.2a1b.1 
(DPGR $= +0.0262$, window: Sep--Oct~2017). WHO surveillance reported 
the emergence of clade 3C.3a viruses predominantly in the Americas during 
this period~\cite{WHO2018}. Overall VE against A(H3N2) was estimated at 
22\%, with reduced protection attributed to multifactorial causes 
including the same egg-adaptation effects observed in the preceding 
season~\cite{Rolfes2019}.
 
\paragraph{2016-2017 season.}
The H3N2 vaccine component was A/Hong~Kong/4801/2014 (clade 3C.2a). 
DPGR analysis identified clade 3C.2a1 as growing faster than the vaccine 
strain 3C.2a in the United States (DPGR $= +0.0116$, window: 
Nov~2016-Feb~2017) and in Europe (DPGR $= +0.0211$, window: 
Jan--Mar~2017), with clade 3C.2a2 also outcompeting 3C.2a in the 
United States (DPGR $= +0.0135$) and Europe (DPGR $= +0.0324$). 
Additional regional comparisons showed 3C.2a1 outgrowing 3C.2a in South 
America (DPGR $= -0.0357$, where 3C.2a is the faster-growing denominator 
in that pair), while in Europe and Asia, 3C.2a1b.1 showed a fitness 
advantage over 3C.2a1 (DPGR $= -0.0162$ and $-0.0249$ respectively), 
reflecting active subclade competition within the broader 3C.2a lineage. 
These results are consistent with WHO surveillance data reporting that 
the large majority of A(H3N2) viruses collected from September~2016 to 
February~2017 belonged to clades 3C.2a and 3C.2a1~\cite{WHO2017}. 
Although the vaccine was antigenically well-matched to circulating viruses 
in standard laboratory assays using cell-propagated reference strains, 
real-world VE against H3N2 was approximately 33\%, largely because 
egg-based vaccine manufacturing introduced a key glycosylation-site loss 
via the T160K mutation in the vaccine haemagglutinin, causing elicited 
antibodies to poorly recognise the glycosylated surface of the actual 
circulating viruses~\cite{Flannery2019}.

\subsubsection{A/H1N1 Seasonal surveillance comparison}

\paragraph{2024-2025 season.}
The vaccine component for 2024--2025 was 
6B.1A.5a.2a.1 (A/Victoria/4897/2022 egg-based; 
A/Wisconsin/67/2022 cell culture-based). DPGR 
analysis identified 6B.1A.5a.2a.1 as growing 
faster than its parent subclade 6B.1A.5a.2a 
across all five regions analysed, confirming 
the vaccine strain clade as the globally 
dominant circulating lineage. This is 
consistent with interim VE estimates from four 
US surveillance networks (IVY, NVSN, US Flu VE, 
VISION), which reported that the 2024--2025 
vaccines provided meaningful protection against 
A(H1N1)pdm09, with outpatient VE up to 72\% 
and hospitalization VE of 63\% among 
individuals under 18 years \cite{ivy2025interim}. 
The positive DPGR values in this season 
therefore represent confirmation of vaccine 
strain relevance rather than a mismatch signal: 
when DPGR identifies the vaccine strain clade 
as the competitively dominant lineage, vaccine 
effectiveness is preserved.

\paragraph{2019-2020 season.}
The vaccine component was A/Brisbane/02/2018 
(H1N1)pdm09-like virus (6B.1A.1). DPGR analysis 
identified 6B.1A.5a.2 as having substantial 
fitness advantages over all co-circulating 
subclades including the vaccine-matched 6B.1A.1, 
with the highest DPGR value in the entire H1N1 
dataset recorded for 6B.1A.5a.2 vs 6B.1A.5a 
in the United States (DPGR $= +0.0516$). This 
fitness signal is consistent with documented 
antigenic drift during the 2019--2020 season. 
Early-season VE against A(H1N1)pdm09 was 
estimated at 37\% (95\% CI: 19--52\%) 
\cite{dawood2020interim}, reflecting moderate 
protection while 6B.1A.5a.2 was still emerging. 
In a subsequent analysis covering the full 
season, VE against A(H1N1)pdm09 declined to 
30\% overall (95\% CI: 21--39\%), with 
vaccination offering no meaningful protection 
(VE 7\%, 95\% CI: $-14$ to 23\%) against the 
antigenically drifted 6B.1A 183P-5A+156K 
subclades that predominated after January 2020 
\cite{tenforde2021antigenic}. The strong and 
consistent positive DPGR values for 6B.1A.5a.2 
across the US and continental analyses are 
directly concordant with this documented 
vaccine failure driven by antigenic drift 
within the 6B.1A lineage.

\paragraph{2018--2019 season.}
The vaccine component was A/Michigan/45/2015 
(H1N1)pdm09-like virus (6B.1). DPGR analysis 
identified faster growth of emerging 6B.1A 
subclades over the vaccine-matched 6B.1 clade 
and older 6B.1A.1 lineage across the United 
States, North America, Europe, and Asia. 
Despite this competitive displacement signal, 
interim VE estimates for 2018--2019 were 
46\% (95\% CI: 30--58\%) against 
A(H1N1)pdm09-associated medically attended 
illness across all ages, with notably stronger 
protection in children aged 6~months-17~years 
(VE = 62\%, 95\% CI: 40--75\%)
\cite{doyle2019interim}. This indicates 
that while 6B.1A subclades were gaining 
competitive fitness over the 6B.1 vaccine 
clade, sufficient antigenic cross-reactivity 
was maintained early in the season to preserve 
moderate overall effectiveness, a 
similar situation to the H3N2 2022-2023 
season, in which a DPGR-detected fitness 
advantage did not immediately translate into 
vaccine failure when cross-reactive antibodies 
were preserved within the same broad clade 
lineage.

Across the influenza seasons examined the variant identified by DPGR as having a transmission fitness 
advantage was subsequently found to be consistent with independent WHO and CDC
surveillance reports, aligning with the dominant emerging clade or the driver of
reduced vaccine effectiveness. The 2022-2023 H3N2 season further provides evidence consistent with vaccine strain relevance when the selected component is correctly 
matched to the dominant circulating lineage. Importantly, the 
2022--2023 season illustrates that DPGR captures relative competitive 
fitness rather than antigenic distance, which is a distinction with direct 
practical implications for interpreting DPGR signals in the context 
of vaccine strain selection. These results collectively support the 
utility of DPGR as an early quantitative signal of fitness shifts in 
circulating influenza~A variants.

\subsection{CNN Model Performance}
Both 1D-CNN models were trained to convergence on their respective labelled
genome datasets. Model performance was evaluated on a held-out test set using mean square error (MSE) and coefficient of determination coefficient ($R^2$) between predicted and true DPGR scores. Results are summarised in Table~\ref{tab:cnn}.
 
\begin{table}[H]
\centering
\caption{CNN model performance on held-out test sets.}
\label{tab:cnn}
\begin{tabular}{lccc}
\toprule
\textbf{Model} & \textbf{Total sequence count} & 
  \textbf{Test MSE} & \textbf{$R^2$} \\
\midrule
H3N2 CNN & 55{,}638 & 0.0038 & 0.9577 \\
H1N1 CNN & 30{,}535 & 0.0007 & 0.9871 \\
\bottomrule
\end{tabular}
\end{table}

\begin{figure}[H]
  \centering
  % Use a slightly smaller width so two fit on one line (e.g., 0.48 each)
  \begin{subfigure}[b]{0.48\textwidth}
    \centering
    \includegraphics[width=\linewidth]{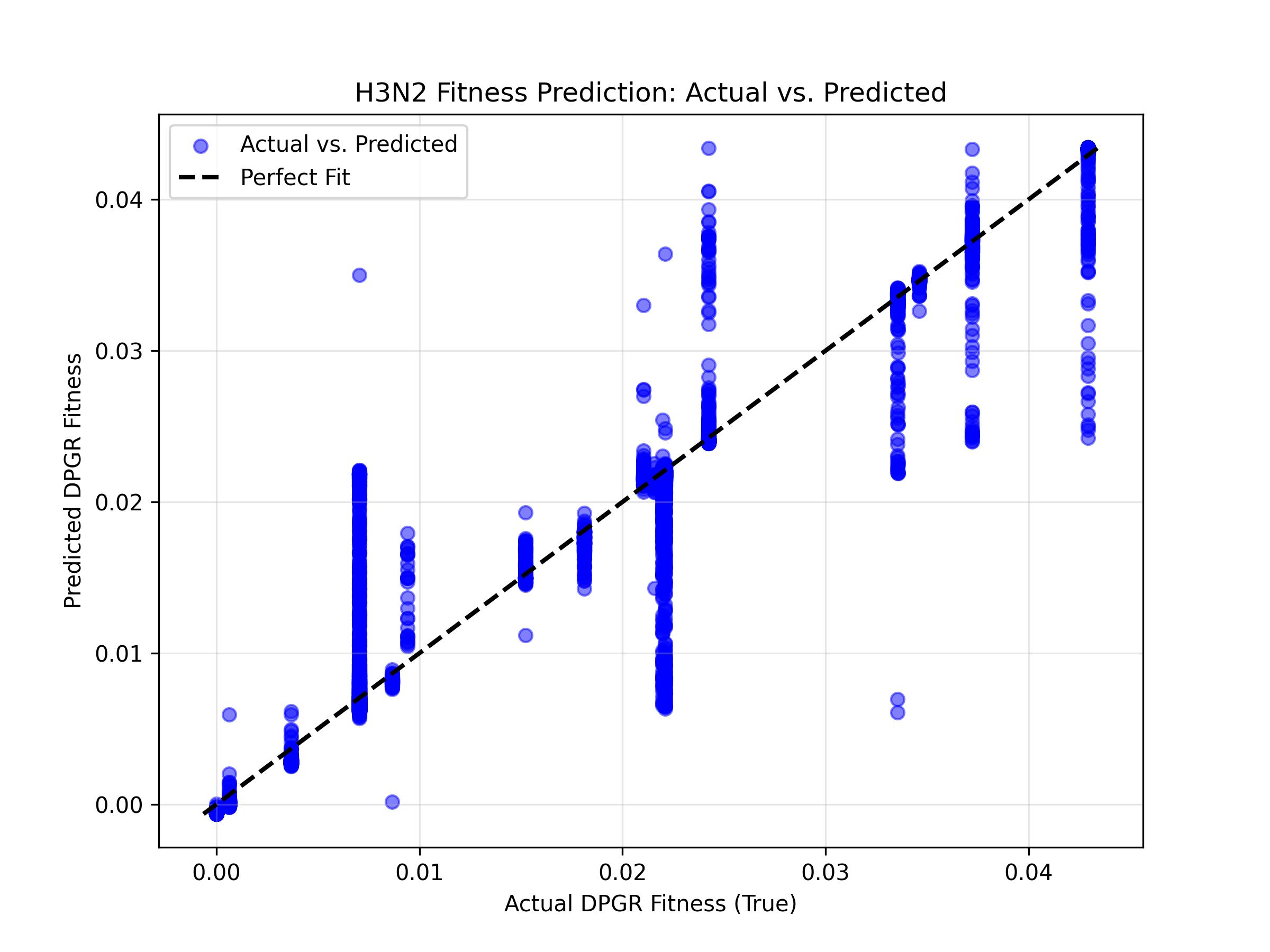}
    \caption{H3N2 ($R^2 = 0.9577$)}
  \end{subfigure}\hfill
  \begin{subfigure}[b]{0.48\textwidth}
    \centering
    \includegraphics[width=\linewidth]{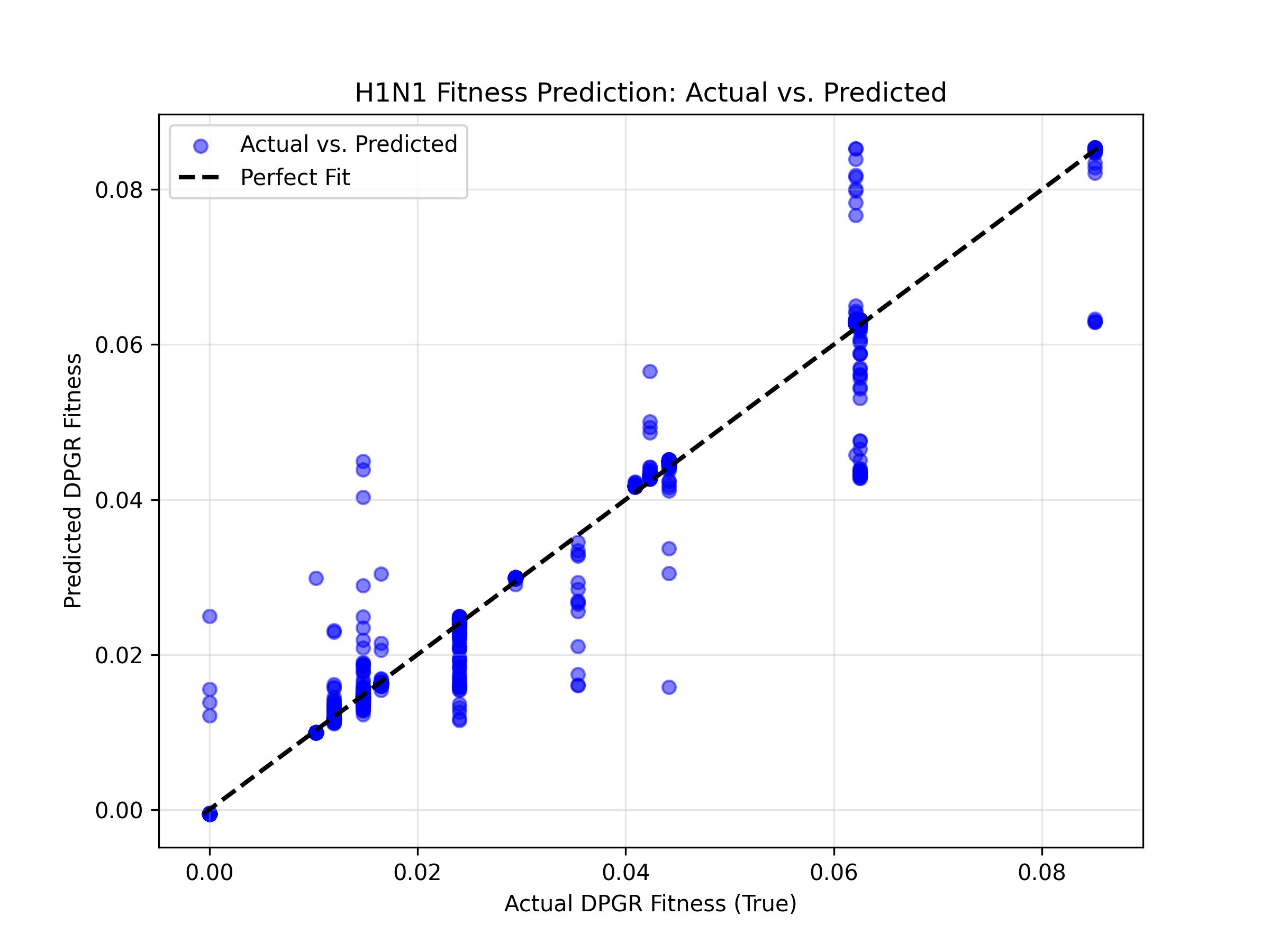}
    \caption{H1N1 ($R^2 = 0.9871$)}
  \end{subfigure}

  \caption{Scatter plots of predicted versus true DPGR fitness scores on the held-out test set for H3N2 (left) and H1N1 (right). The dashed line represents perfect prediction (slope $= 1$).}
  \label{fig:scatter}
\end{figure}

\begin{table}[H]
\centering
\caption{Comparison of CNN performance against non-sequence baselines for H3N2 
(n = 54,859 sequences; 779 excluded due to missing collection dates) and H1N1 
(n = 30,535 sequences), using a stratified 80:20 train/test split for baselines 
and 80:20 for CNN models. R\textsuperscript{2} 
values are reported on the held-out test set. The Clade+Continent lookup represents 
the theoretical ceiling since fitness labels are assigned at this level.}
\begin{tabular}{lcc}
\hline
\textbf{Model} & \textbf{H3N2 R\textsuperscript{2}} & \textbf{H1N1 R\textsuperscript{2}} \\
\hline
Continent-only lookup   & 0.130 & 0.061 \\
Clade-only lookup       & 0.753 & 0.955 \\
CNN                     & 0.958 & 0.987 \\
Clade+Continent ceiling & 1.000 & 1.000 \\
\hline
\end{tabular}
\label{tab:baselines}
\end{table}
The H3N2 model achieved an $R^2$ of 0.9577 between predicted and true 
DPGR scores on the test set ($\text{MSE} = 0.0038$), indicating that the 
model explains 95.77\% of the variance in H3N2 fitness scores. The H1N1 
model achieved $R^2 = 0.9871$ ($\text{MSE} = 0.0007$), explaining 98.71\% 
of variance in H1N1 fitness scores. This higher performance is consistent with 
the narrower and more structured range of normalised fitness values in the 
H1N1 dataset. Both scatter plots (Figure~\ref{fig:scatter}) show close 
alignment with the perfect-fit diagonal, with visible vertical banding 
reflecting the discrete nature of DPGR labels.
The results are shown in Figure~\ref{fig:scatter}. The H3N2 model converged at 
epoch 131 and H1N1 at epoch 166, both under early stopping with batch size 
128 and the Adam optimiser. Training and validation loss curves are presented 
in Figure~\ref{fig:cnn_training} in Appendix~\ref{app:dpgr_appendix}.

Both models converged without evidence of overfitting, as training and validation MSE curves remained closely aligned throughout training (Figure~\ref{fig:cnn_training}). Both plots track mean squared error on the min--max normalised DPGR scale; the left panel as the compiled loss function being optimised, and the right panel as the same metric evaluated independently at the end of each epoch. 

The H3N2 model exhibited rapid early convergence, with the epoch-level loss dropping sharply from an initial value of approximately 5.0 to near zero within the first 10 epochs, after which both training and validation curves plateaued stably through epoch 130. The first-epoch spike reflects unstable early batch gradients prior to optimizer stabilization rather than a true error on the normalized scale; the end-of-epoch MSE, a cleaner evaluation, showed a more representative initial value of approximately 0.35 and converged to near zero over the same initial phase, with training and validation MSE tracking indistinguishably thereafter. 

The H1N1 model followed a more gradual convergence trajectory, with both loss and MSE beginning at approximately 0.25 and 0.13 respectively and descending smoothly over roughly 75--100 epochs before reaching a stable minimum near zero. Validation curves tracked training curves closely throughout with no divergence observed. The absence of a train--validation gap in both models indicates that the dropout regularisation strategy and the scale of the training corpus were sufficient to prevent overfitting.

\subsection{Conformal Prediction Intervals}
Conformal prediction intervals were constructed at a nominal coverage 
level of $1 - \varepsilon = 0.95$ for both the H3N2 and H1N1 CNN 
models using an inductive split conformal framework. The full test set 
(20\% of each dataset, held out from training) was further partitioned 
into a calibration split (20\% of the test set, approximately 2,226 
sequences for H3N2 and 1,221 for H1N1) and a final evaluation split 
(80\% of the test set, approximately 8,902 sequences for H3N2 and 
4,886 for H1N1). All splits used \texttt{random\_state=42} and are 
fully reproducible. Nonconformity scores were computed as absolute 
residuals $\alpha_i = |y_i - \hat{y}_i|$ on the calibration set, and 
the conformal quantile $\hat{q}$ was calculated as the 
$\lceil (n+1)(1-\varepsilon)/n \rceil$-th order statistic of those 
residuals. Results are summarised in Table~\ref{tab:conformal}.
 
\begin{table}[H]
\centering
\caption{Conformal prediction results at 95\% nominal coverage for 
H3N2 and H1N1. All values on the min-max normalised DPGR scale.}
\label{tab:conformal}
\begin{tabular}{lcccc}
\toprule
\textbf{Model} & \textbf{Calibration $n$} & \textbf{$\hat{q}$}
  & \textbf{Empirical coverage} & \textbf{Interval width} \\
\midrule
H3N2 & $\approx$2,226 & 0.1058 & 95.2\% & 0.2116 \\
H1N1 & $\approx$1,221 & 0.0106 & 93.8\% & 0.0212 \\
\bottomrule
\end{tabular}
\end{table}
 
\begin{figure}[H]
  \centering
  % Use a slightly smaller width so two fit on one line (e.g., 0.48 each)
  \begin{subfigure}[b]{0.48\textwidth}
    \centering
    \includegraphics[width=\linewidth]{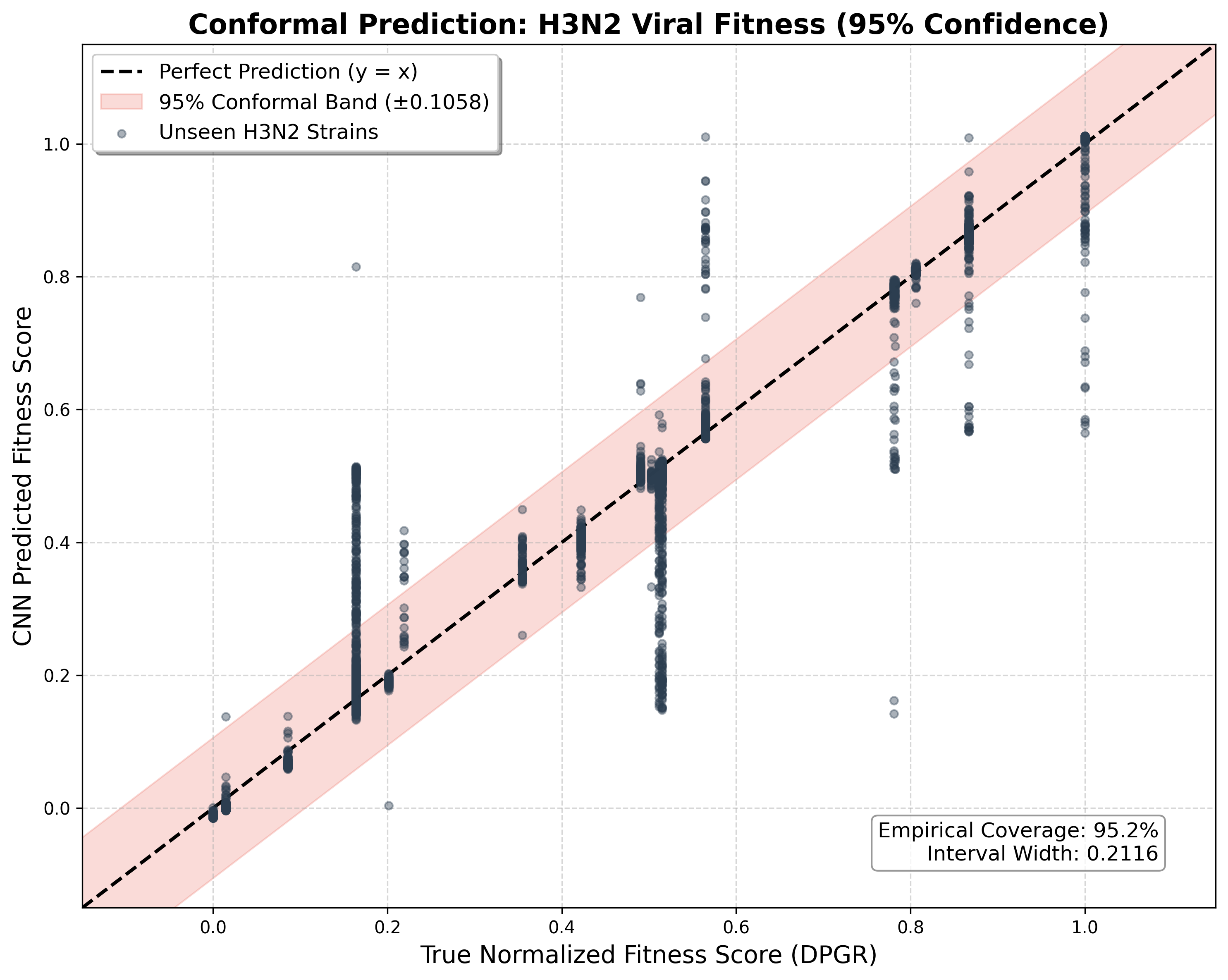}
    \caption{Conformal prediction intervals for unseen A/H3N2 test sequences. The shaded band denotes the 95\% conformal prediction region ($\hat{q} = 0.1058$, interval width $= 0.2116$), with empirical coverage of 95.2\%}
  \end{subfigure}\hfill
  \begin{subfigure}[b]{0.48\textwidth}
    \centering
    \includegraphics[width=\linewidth]{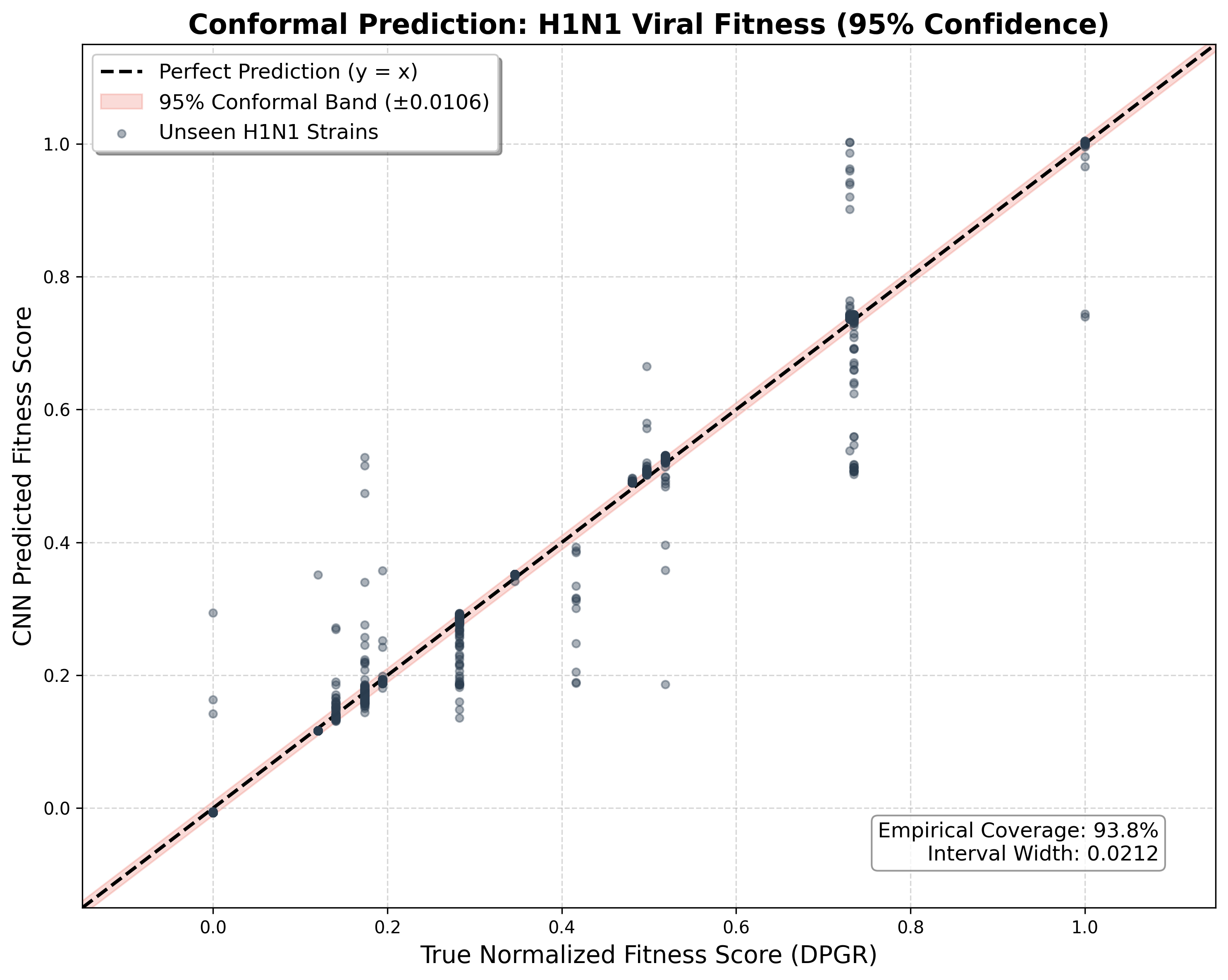}
    \caption{Conformal prediction intervals for unseen A/H1N1 test sequences. The shaded band denotes the 95\% conformal prediction region ($\hat{q} = 0.0106$, interval width $= 0.0212$), with empirical coverage of 93.8\%}
  \end{subfigure}

  \caption{Conformal prediction of DPGR-derived viral fitness scores for held-out A/H3N2 and A/H1N1 genome sequences. Each point represents a test sequence plotted against the diagonal perfect-prediction line (dashed). Vertical bars indicate individual prediction intervals; the shaded band shows the global 95\% conformal band. The narrower interval width and tighter clustering around the diagonal for H1N1 reflect its more constrained fitness distribution relative to H3N2, consistent with the stronger purifying selection and antigenic stability characteristic of A(H1N1)pdm09 in the post-pandemic period.}
  \label{fig:conformal}
\end{figure}

\paragraph{H3N2}
The conformal quantile for the H3N2 model was $\hat{q} = 0.1058$ on 
the normalised DPGR scale, yielding symmetric prediction intervals of 
the form $[\hat{y} - 0.1058,\; \hat{y} + 0.1058]$ with a total width 
of 0.2116. Empirical coverage on the final evaluation split was 
95.2\%, meeting the nominal 95\% guarantee 
(Figure~\ref{fig:conformal}a). The prediction band is visible as 
a consistent envelope around the perfect-fit diagonal, with the 
majority of unseen H3N2 strains falling within the interval. The 
relatively wide interval width reflects the broader range of fitness 
values across H3N2 clades and the greater diversity of genomic 
configurations present in the H3N2 training corpus. A small number 
of outliers, primarily at mid-range true fitness scores 
($\approx 0.5$--$0.6$), fell above the band with predicted values 
substantially exceeding the true value, likely corresponding to 
sequences from clades with atypical genomic compositions relative 
to the training distribution.
 
\paragraph{H1N1}
The conformal quantile for the H1N1 model was $\hat{q} = 0.0106$, 
yielding a substantially narrower interval width of 0.0212 --- 
approximately ten times tighter than the H3N2 intervals. This is 
consistent with the lower MSE (0.0007 vs 0.0038) and higher $R^2$ 
(0.9871 vs 0.9577) observed for H1N1 in the CNN performance analysis, 
reflecting the model's greater point-prediction precision on the 
H1N1 fitness task. Empirical coverage on the final evaluation split 
was 93.8\%, marginally below the nominal 95\% target 
(Figure~\ref{fig:conformal}b). This minor shortfall is 
attributable to the small calibration set size: with approximately 
1,221 calibration sequences, the finite-sample coverage guarantee is 
weaker than for H3N2, and the estimated $\hat{q}$ may slightly 
underestimate the true required quantile. Despite this, the 93.8\% 
empirical coverage remains close to nominal and does not indicate 
systematic model miscalibration. Rather, it reflects a 
statistical artefact of the calibration sample size. The narrow 
interval width confirms high model confidence on the H1N1 fitness 
prediction task.
 
Taken together, the conformal prediction results demonstrate that 
both models produce well-calibrated uncertainty estimates with 
statistically grounded coverage guarantees. The H3N2 model achieves 
the formal 95\% coverage guarantee with a moderate interval width 
that reflects the inherent variability of H3N2 fitness across a 
decade of clade succession. The H1N1 model's substantially tighter 
intervals reflect its higher point-prediction accuracy, with the 
marginal coverage shortfall attributable to calibration set size 
rather than model error. Together, these results support the use of 
both models for practical fitness prediction with quantified 
uncertainty, a property that is particularly relevant for 
public health applications where decision-making benefits from 
knowing not just a predicted fitness score but a statistically 
guaranteed range within which the true value is expected to fall.
 
\subsection{SHAP Feature Importance}

\subsubsection{H3N2}

\begin{figure}[H]
  \centering
  \captionsetup[subfigure]{font=tiny, labelfont=tiny}
  \begin{subfigure}[b]{0.31\textwidth}
    \centering
    \includegraphics[width=\linewidth]{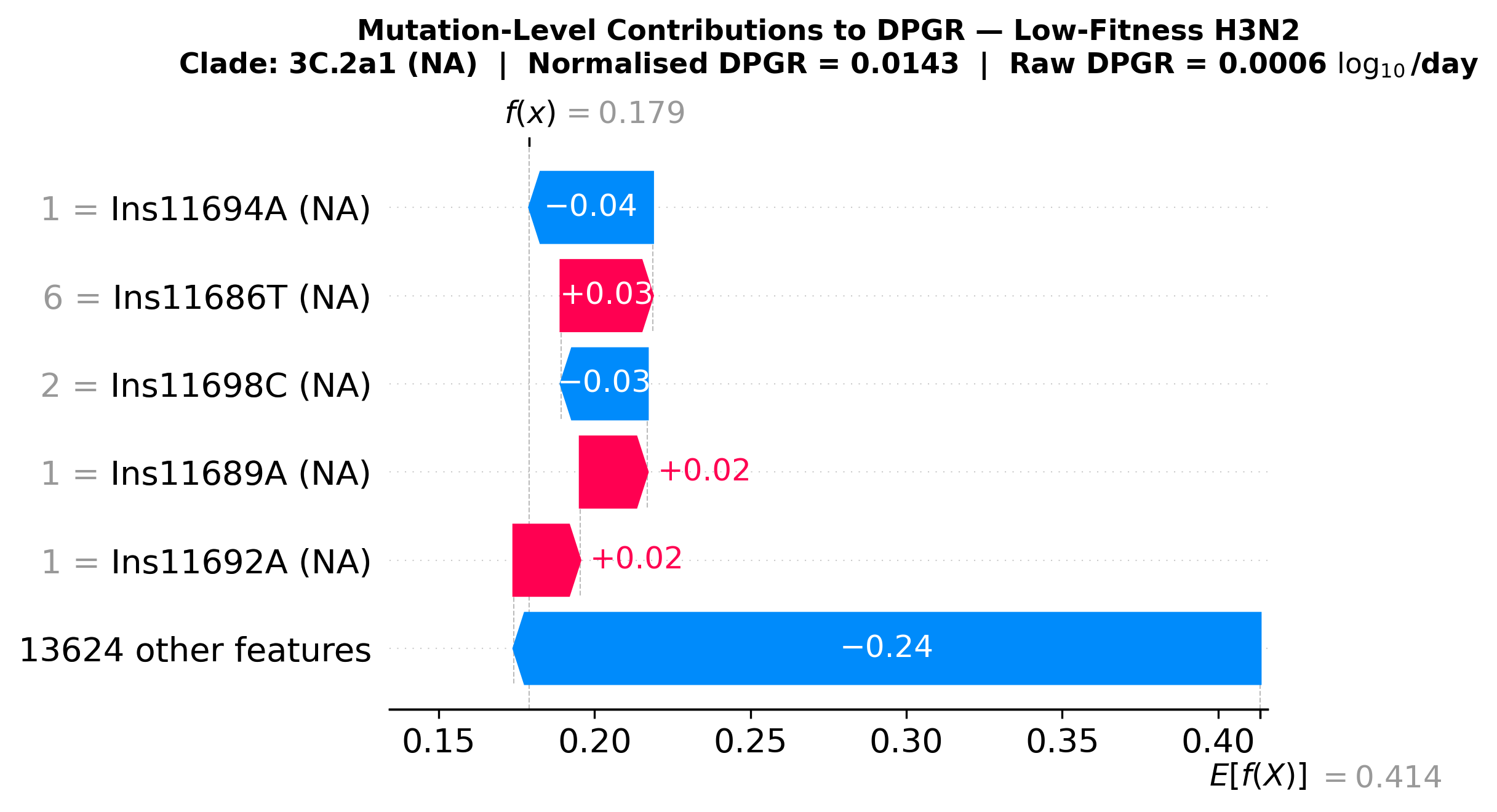}
    \caption{SHAP waterfall plot for a low-fitness
    A/H3N2 sequence (clade 3C.2a1, North America;
    normalised DPGR $= 0.0143$). Top-ranked positions
    are concentrated in the neuraminidase segment,
    with predominantly suppressive contributions
    relative to the model baseline ($f(x) = 0.414$).}
    \label{subfig:h3n2_waterfall_low}
  \end{subfigure}\hfill
  \begin{subfigure}[b]{0.31\textwidth}
    \centering
    \includegraphics[width=\linewidth]{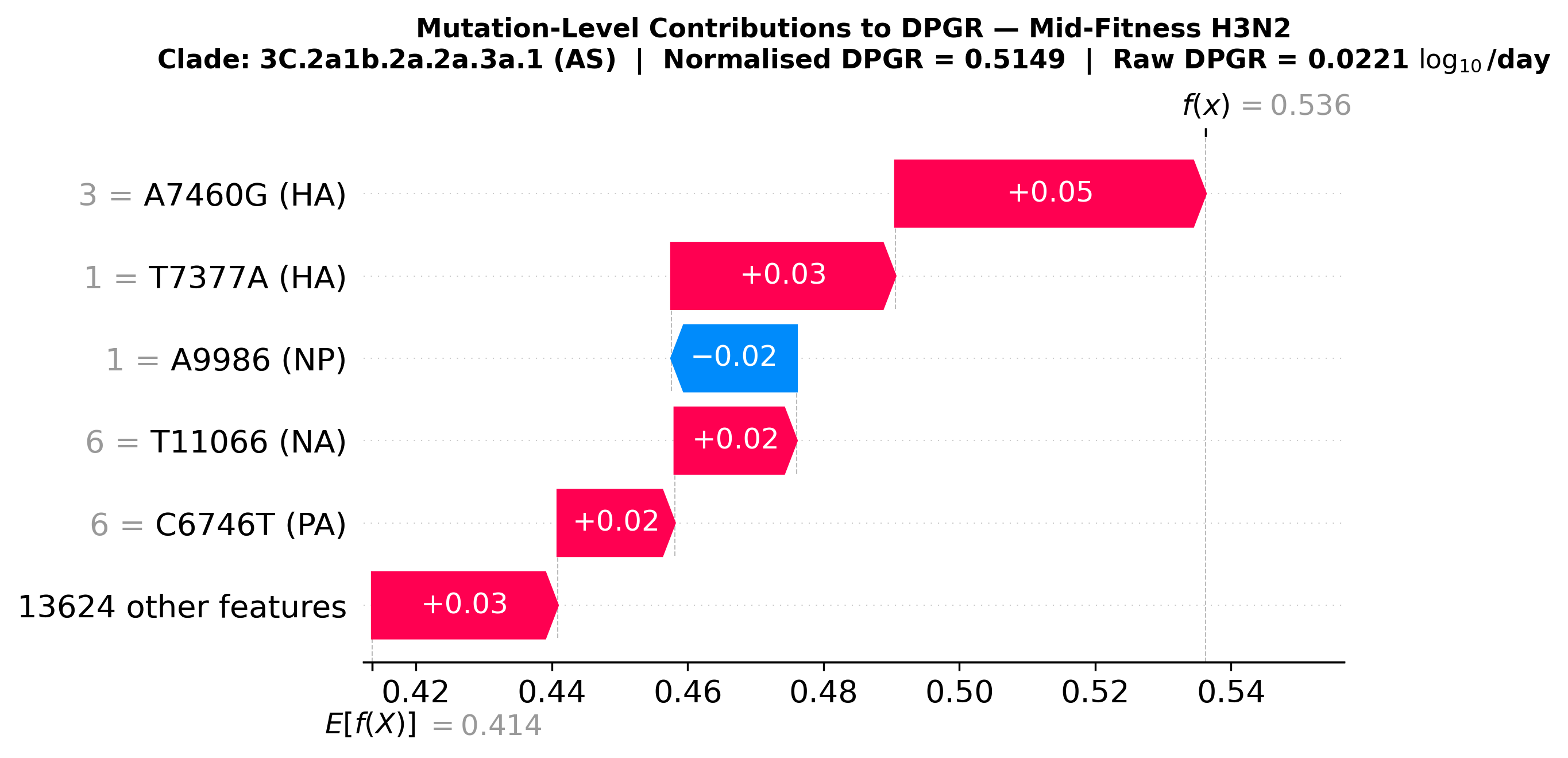}
    \caption{SHAP waterfall plot for a mid-fitness
    A/H3N2 sequence (clade 3C.2a1b.2a.2a.3a.1,
    Asia; normalised DPGR $= 0.5149$). Top-ranked
    positions span the NA and NP segments, with
    mixed contributions reflecting the transitional
    competitive position of this clade.}
    \label{subfig:h3n2_waterfall_mid}
  \end{subfigure}\hfill
  \begin{subfigure}[b]{0.31\textwidth}
    \centering
    \includegraphics[width=\linewidth]{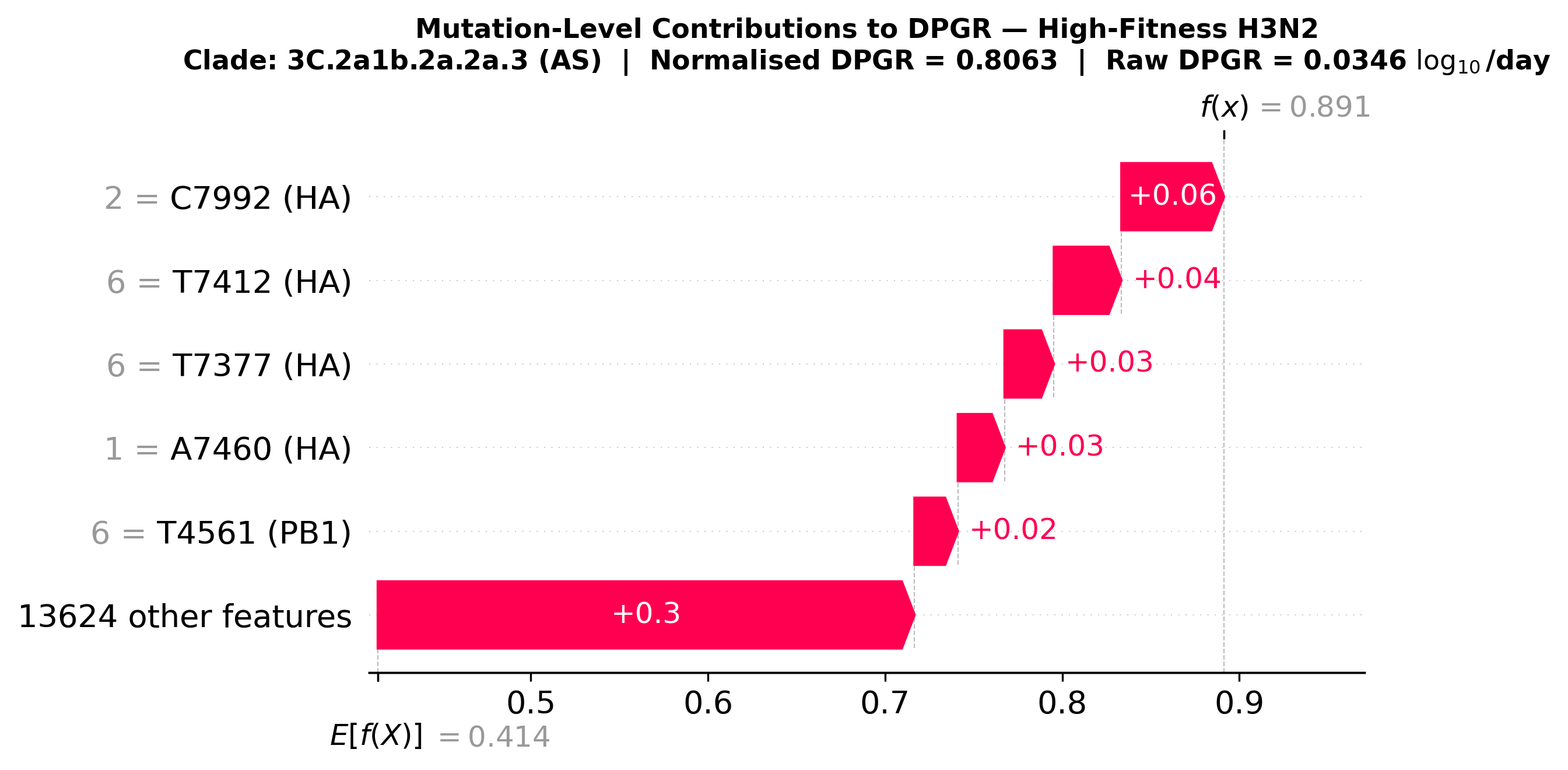}
    \caption{SHAP waterfall plot for a high-fitness
    A/H3N2 sequence (clade 3C.2a1b.2a.2a.3, Asia;
    normalised DPGR $= 0.8063$). Top-ranked positions
    shift toward the haemagglutinin segment, with
    strongly positive contributions at known
    antigenic sites and an additional PB1
    contribution.}
    \label{subfig:h3n2_waterfall_high}
  \end{subfigure}
  \vspace{1.5em}
  \begin{subfigure}[b]{0.7\textwidth}
    \centering
    \includegraphics[width=\linewidth]{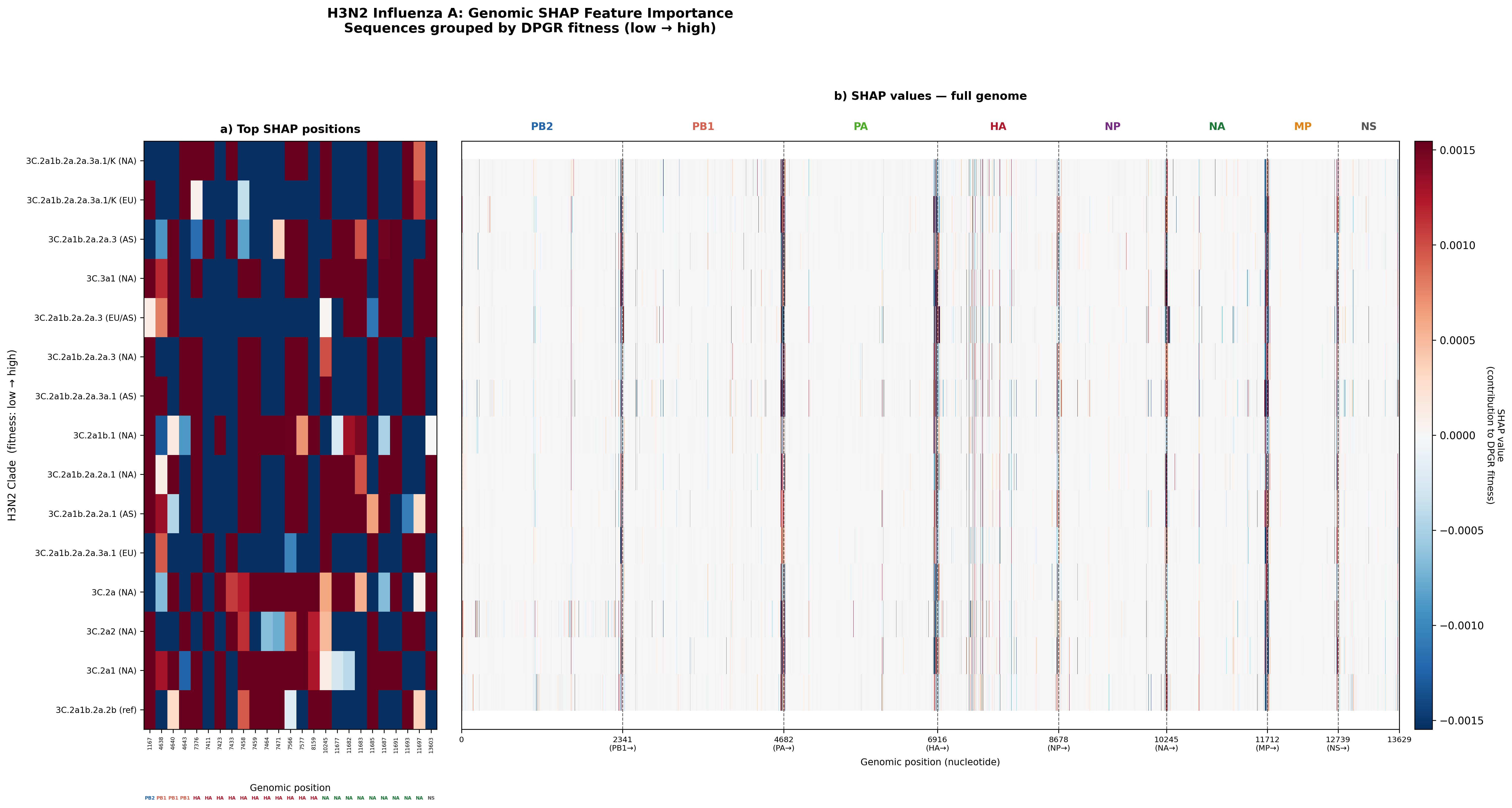}
    \caption{Genome-wide mean absolute SHAP contribution
    heatmap across all 15 DPGR fitness groups for
    100 representative H3N2 test sequences. Each row
    corresponds to a fitness group (low to high) and
    each column to a nucleotide position in the
    13,629-nt concatenated genome. Segment boundaries
    for all eight influenza gene segments are
    annotated. Warmer colours indicate positive
    contributions; cooler colours indicate suppressive
    contributions.}
    \label{subfig:h3n2_heatmap}
  \end{subfigure}
  \caption{SHAP-based feature importance analysis for
  the A/H3N2 convolutional neural network. Figures
  (a)-(c) show SHAP waterfall plots for
  representative low- (normalised DPGR $= 0.0143$),
  mid- (normalised DPGR $= 0.5149$), and high-fitness
  (normalised DPGR $= 0.8063$) H3N2 sequences,
  illustrating the systematic shift in top-ranked
  genomic contributions from neuraminidase-dominated
  suppression at low fitness to
  haemagglutinin-centred positive contributions at
  high fitness. Figure (d) shows the genome-wide
  heatmap of mean absolute SHAP values across all
  15 fitness groups, revealing that predictive signal
  is concentrated in the haemagglutinin and
  nucleoprotein segments, with polymerase and surface
  protein contributions varying across fitness tiers.}
  \label{fig:shap_heatmap_h3n2}
\end{figure}

SHAP GradientExplainer values were computed for 100 representative 
H3N2 test sequences to identify the genomic positions and viral 
segments most predictive of DPGR fitness. Two complementary 
visualisations are presented: a genome-wide heatmap summarising 
mean absolute SHAP contributions across all 15 DPGR fitness groups 
(Figure~\ref{fig:shap_heatmap_h3n2}d), and individual waterfall plots 
for three sequences representing low, mid, and high fitness clades 
(Figure~\ref{fig:shap_heatmap_h3n2}a--c).

\paragraph{Genome-wide SHAP heatmap.}
The full-genome SHAP heatmap (Figure~\ref{fig:shap_heatmap_h3n2}d) 
reveals that predictive signal is not uniformly distributed across 
the 13,629-nucleotide concatenated genome. The haemagglutinin (HA) 
segment (positions 6,916-8,678) exhibits the strongest and most 
consistent SHAP signal across fitness groups, with prominent positive 
contributions (red) in higher-fitness clades and predominantly 
negative contributions (blue) in lower-fitness groups. The 
nucleoprotein (NP) segment (positions 8,678-10,245) shows the 
second most prominent signal, with a concentrated hotspot visible 
in the full-genome panel. The neuraminidase (NA) segment 
(positions 10,245--11,712) contributes a distinct signal pattern, 
particularly in lower-fitness clades. The polymerase segments PB2, 
PB1, and PA show sparse but consistent contributions across fitness 
groups, while MP and NS display minimal signal.Panel~(a) of (Figure~\ref{fig:shap_heatmap_h3n2}d) confirms these 
findings at the level of individual top-ranked positions, showing 
that the highest-importance genomic positions are predominantly 
located in the HA, NP, and NA segments, with the pattern of 
positive versus negative SHAP values shifting systematically 
from low-fitness (predominantly blue) to high-fitness 
(predominantly red) groups.

\paragraph{Low-fitness sequence — 3C.2a1 (North America).}
The waterfall plot for the lowest-fitness representative sequence 
(clade 3C.2a1, North America; normalised DPGR $= 0.0143$, raw DPGR 
$= 0.0006$ $\log_{10}$/day) shows a predicted fitness of $f(x) = 
0.179$, correctly placed well below the model baseline of $E[f(X)] 
= 0.414$ (Figure~\ref{fig:shap_heatmap_h3n2}a). The dominant 
individual features are a cluster of NA segment insertions near 
positions 11,686-11,698, with alternating positive and negative 
contributions (Ins11694A: $-0.04$; Ins11686T: $+0.03$; Ins11698C: 
$-0.03$), reflecting complex insertion-deletion variation in the NA 
stalk region. The largest single contribution is the aggregate of 
13,624 remaining positions ($-0.24$), indicating that the 
genome-wide background signature of this early clade is the primary 
driver of its low predicted fitness rather than any single dominant 
mutation.

\paragraph{Mid-fitness sequence — 3C.2a1b.2a.2a.3a.1 (Asia).}
The mid-fitness representative (clade 3C.2a1b.2a.2a.3a.1, Asia; 
normalised DPGR $= 0.5149$, raw DPGR $= 0.0221$ $\log_{10}$/day) 
was also the most accurately predicted sequence in the test set 
(predicted $f(x) = 0.536$, absolute error $= 1.4 \times 10^{-4}$; 
(Figure~\ref{fig:shap_heatmap_h3n2}b). The two dominant 
fitness-increasing features are both located in HA: A7460G 
($+0.05$) and T7377A ($+0.03$). A single NP position, A9986, 
exerts a modest fitness-suppressing effect ($-0.02$), while NA 
(T11066: $+0.02$) and PA (C6746T: $+0.02$) contribute smaller 
positive signals. This multi-segment contribution pattern — 
centred on HA but with secondary inputs from NP, NA, and PA — 
is consistent with the known multi-genic basis of influenza 
transmission fitness.

\paragraph{High-fitness sequence — 3C.2a1b.2a.2a.3 (Asia).}
The high-fitness representative (clade 3C.2a1b.2a.2a.3, Asia; 
normalised DPGR $= 0.8063$, raw DPGR $= 0.0346$ $\log_{10}$/day) 
shows a predicted fitness of $f(x) = 0.891$, well above baseline 
(Figure~\ref{fig:shap_heatmap_h3n2}c). Strikingly, four 
of the five individually named top features are located in HA: 
C7992 ($+0.06$), T7412 ($+0.04$), T7377 ($+0.03$), and A7460 
($+0.03$), with a single PB1 contribution T4561 ($+0.02$). The 
aggregate genome-wide signal is strongly positive ($+0.30$), 
reflecting a coherent high-fitness genomic background. The 
near-exclusive HA dominance in the high-fitness waterfall, 
contrasting with the NA-dominated pattern in the low-fitness 
sequence, suggests a clear fitness-dependent shift in the 
genomic determinants identified by the model.

\paragraph{Summary of H3N2 SHAP findings.}
Across the three fitness tiers, a consistent pattern emerges: 
as DPGR fitness increases, the dominant SHAP contributions shift 
from NA-segment insertions and genome-wide suppressive signals 
(low fitness) through a mixed HA--NP--NA--PA pattern (mid fitness) 
to near-exclusive HA dominance with a strongly positive genome-wide 
background (high fitness). This gradient is biologically coherent 
with the established role of HA as the primary determinant of 
influenza A antigenic variation and transmission advantage. The 
model has learned this segment-level hierarchy from raw sequence 
data alone, without explicit annotation of segment boundaries or 
biological prior knowledge, supporting the validity of the CNN's 
learned sequence-fitness mapping.

\subsubsection{H1N1}

\begin{figure}[H]
  \centering
  \captionsetup[subfigure]{font=tiny, labelfont=tiny}

  \begin{subfigure}[b]{0.31\textwidth}
    \centering
    \includegraphics[width=\linewidth]{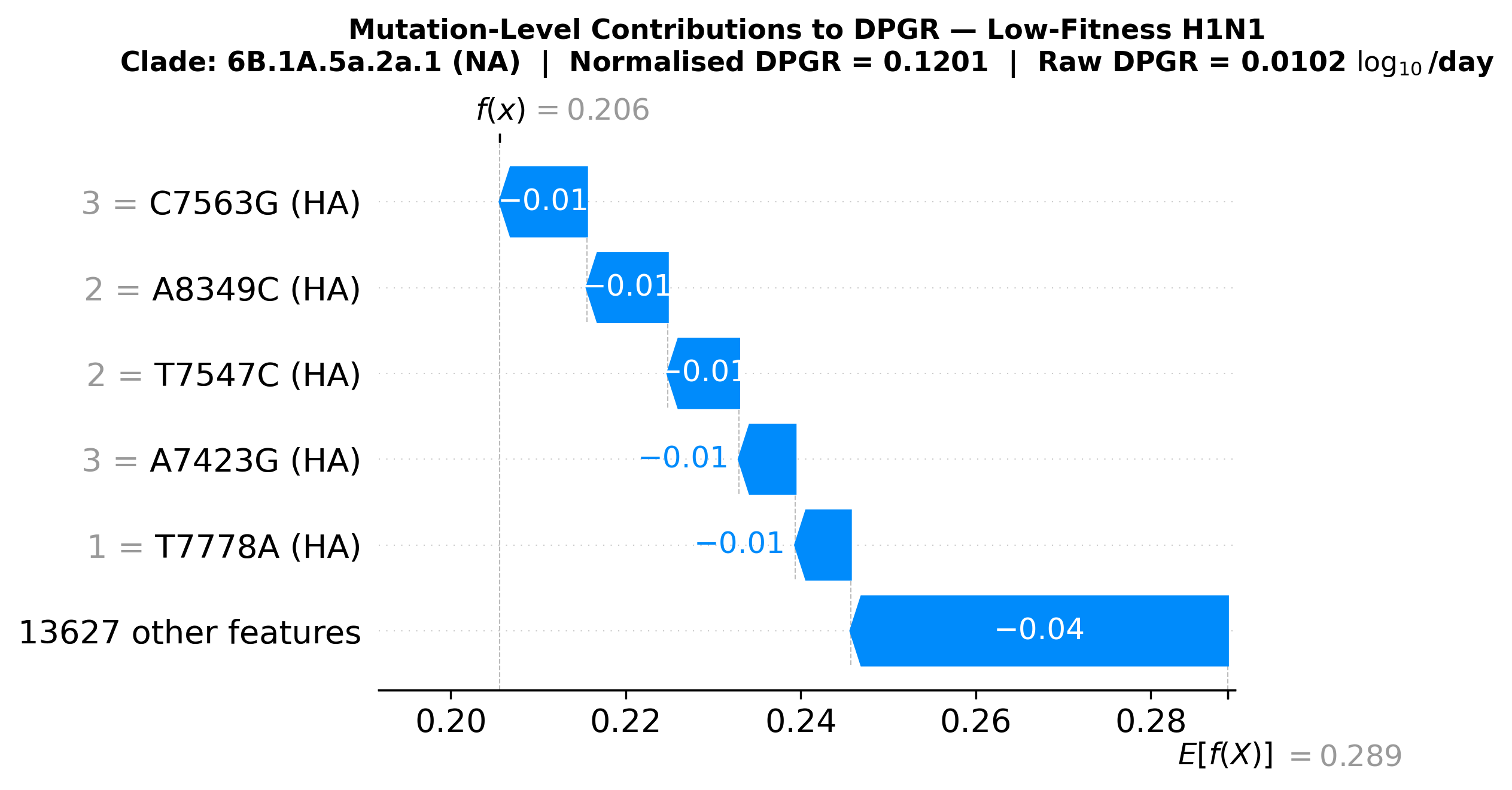}
    \caption{SHAP waterfall plot for a low-fitness
    A/H1N1 sequence (clade 6B.1A.5a.2a.1, North
    America; normalised DPGR $= 0.1201$). All five
    top-ranked positions are in the haemagglutinin
    segment and are uniformly suppressive, pushing
    the prediction below the model baseline
    ($E[f(X)] = 0.289$).}
    \label{subfig:h1n1_waterfall_low}
  \end{subfigure}\hfill
  \begin{subfigure}[b]{0.31\textwidth}
    \centering
    \includegraphics[width=\linewidth]{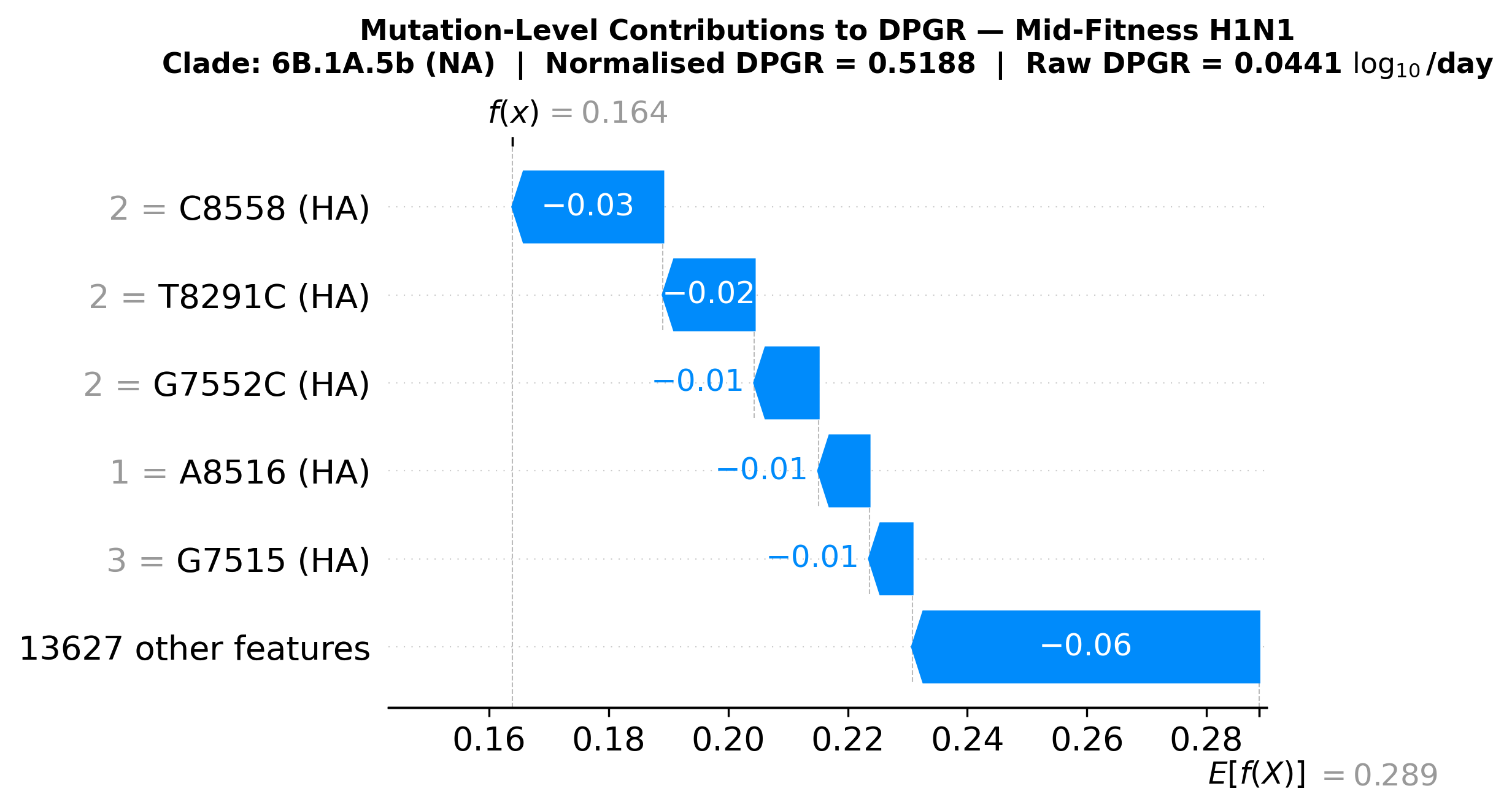}
    \caption{SHAP waterfall plot for a mid-fitness
    A/H1N1 sequence (clade 6B.1A.5b, North America;
    normalised DPGR $= 0.5188$). All five top-ranked
    positions are in the haemagglutinin segment and
    remain suppressive, with an additional negative
    genome-wide background contribution of $-0.06$.}
    \label{subfig:h1n1_waterfall_mid}
  \end{subfigure}\hfill
  \begin{subfigure}[b]{0.31\textwidth}
    \centering
    \includegraphics[width=\linewidth]{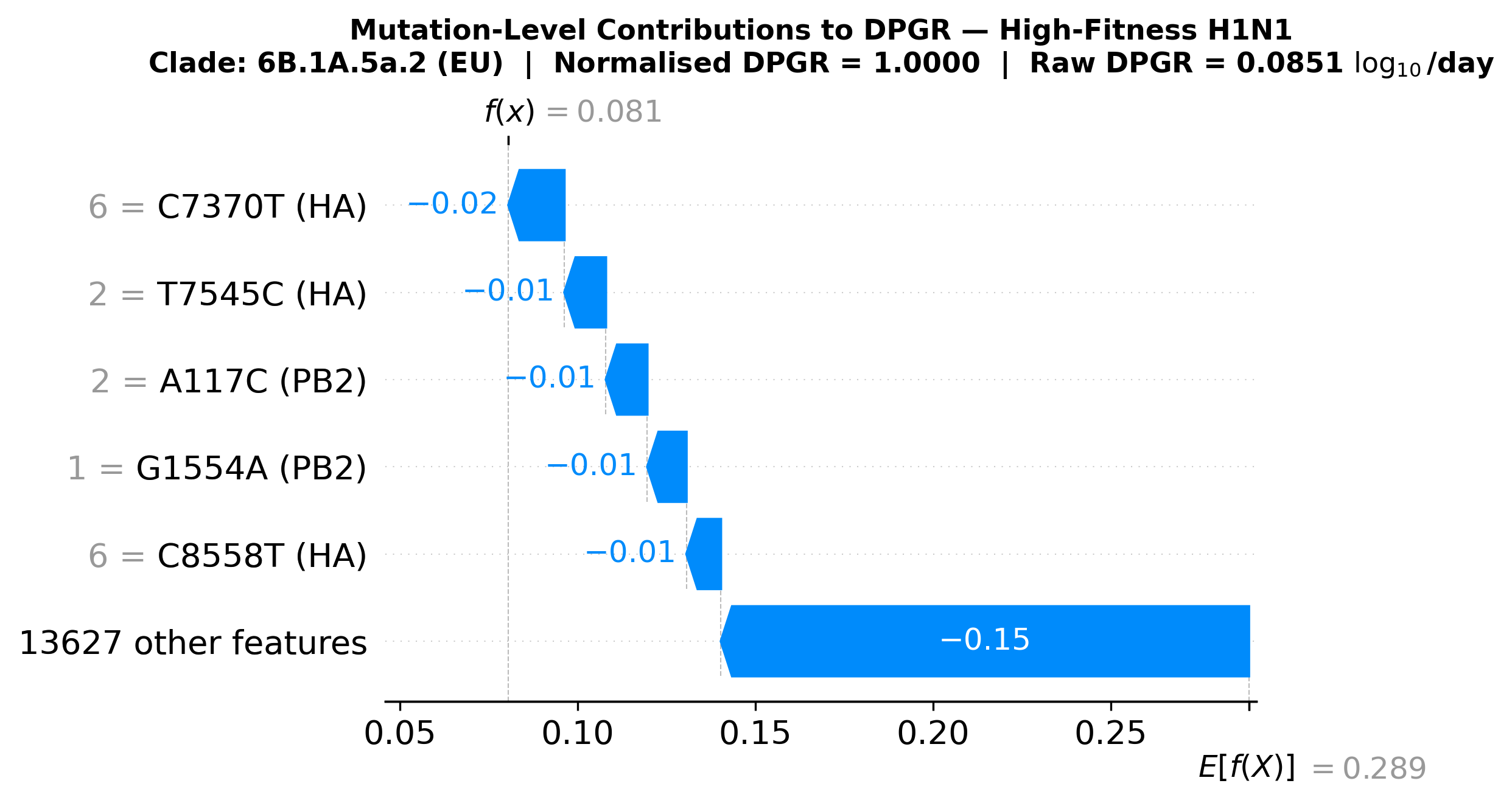}
    \caption{SHAP waterfall plot for a high-fitness
    A/H1N1 sequence (clade 6B.1A.5a.2, Europe;
    normalised DPGR $= 1.0000$). Top-ranked positions
    include haemagglutinin and PB2 contributions,
    all suppressive, with a strongly negative
    genome-wide background ($-0.15$), reflecting
    underrepresentation of high-fitness sequences
    in the training distribution.}
    \label{subfig:h1n1_waterfall_high}
  \end{subfigure}

  \vspace{1.5em}

  \begin{subfigure}[b]{0.7\textwidth}
    \centering
    \includegraphics[width=\linewidth]{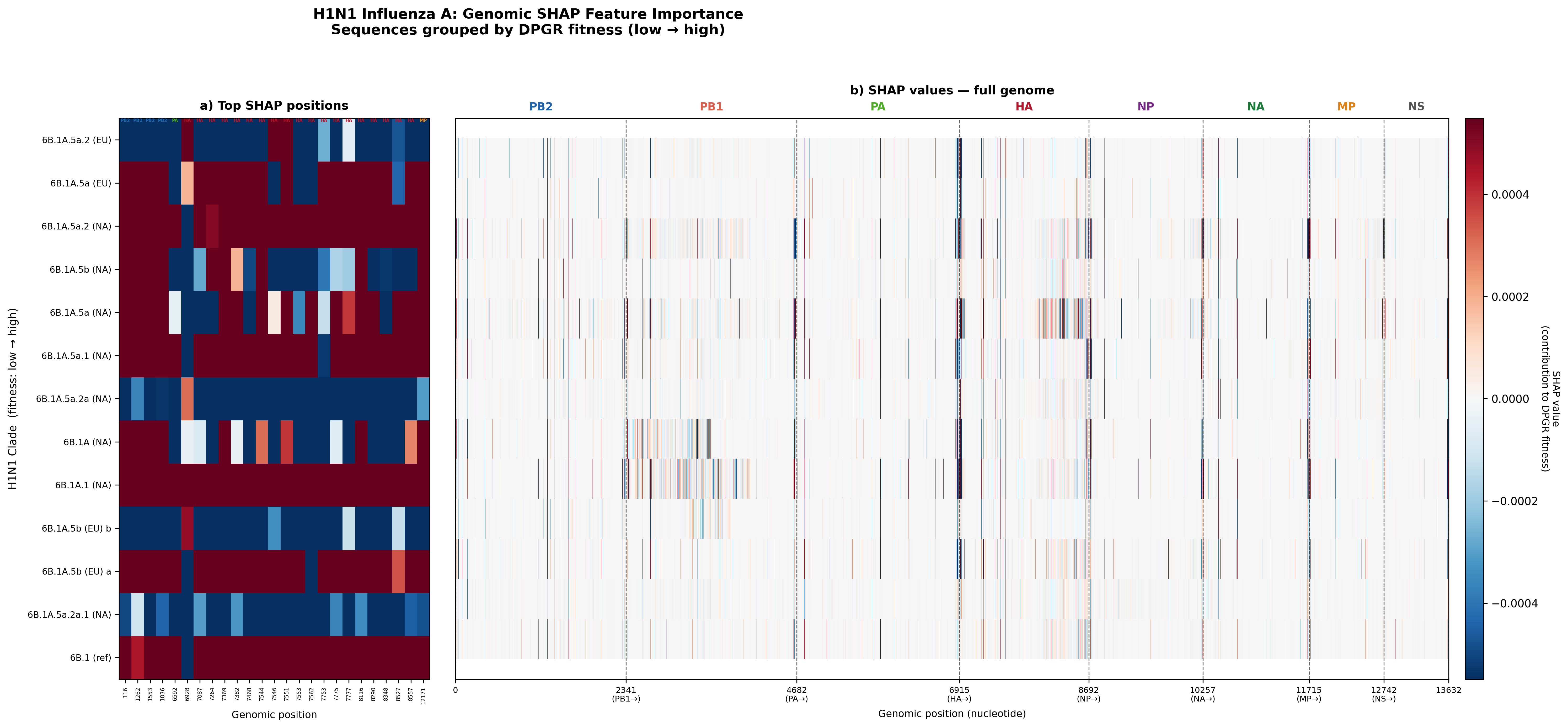}
    \caption{Genome-wide mean absolute SHAP contribution
    heatmap across 13 A/H1N1 DPGR fitness groups for
    100 representative H1N1 test sequences. Each row
    corresponds to a fitness group (low to high) and
    each column to a nucleotide position in the
    13,632-nt concatenated genome. Segment boundaries
    for all eight influenza gene segments are
    annotated. Contributions are uniformly suppressive
    across fitness tiers, with the strongest signal
    concentrated in the haemagglutinin, PB1, and PA
    segments.}
    \label{subfig:h1n1_heatmap}
  \end{subfigure}

  \caption{SHAP-based feature importance analysis for
  the A/H1N1 convolutional neural network. Figures
  (a)--(c) show SHAP waterfall plots for
  representative low- (normalised DPGR $= 0.1201$),
  mid- (normalised DPGR $= 0.5188$), and high-fitness
  (normalised DPGR $= 1.0000$) H1N1 sequences.
  Unlike H3N2, contributions remain uniformly
  suppressive across all fitness tiers, reflecting
  the skewed fitness distribution and stronger
  purifying selection characteristic of
  A(H1N1)pdm09 in the post-pandemic period.
  Figure (d) shows the genome-wide heatmap of mean
  absolute SHAP values across all 13 fitness groups,
  revealing that predictive signal is distributed
  across the haemagglutinin, PB1, and PA segments,
  with a more diffuse multi-segment architecture
  compared to H3N2.}
  \label{fig:shap_heatmap_h1n1}
\end{figure}

SHAP GradientExplainer values were computed for 100 representative 
H1N1 test sequences using the same framework applied to H3N2. The 
H1N1 test set fitness distribution is heavily skewed toward lower 
values (median normalised DPGR $= 0.120$, mean $= 0.241$), with 
only 33 of 6,107 test sequences ($0.5\%$) exhibiting normalised 
DPGR $> 0.75$. This reflects the biological reality of H1N1 
post-2009 pandemic evolution, in which the majority of circulating 
clades occupy a narrow low-to-mid fitness band with relatively few 
high-fitness outliers. The genome-wide SHAP heatmap 
(Figure~\ref{fig:shap_heatmap_h1n1}d) and three representative 
waterfall plots (Figure~\ref{fig:shap_heatmap_h1n1}a--c) are 
interpreted in this context.

\paragraph{Genome-wide SHAP heatmap.}
The H1N1 full-genome SHAP heatmap 
(Figure~\ref{fig:shap_heatmap_h1n1}d) reveals a distinct pattern 
from H3N2. While HA (positions 6,915--8,692) remains a prominent 
source of predictive signal, the PB1 segment (positions 
2,341--4,682) and PA segment (positions 4,682--6,915) exhibit 
stronger relative contributions compared to the H3N2 heatmap, 
with visible hotspots in both segments across multiple fitness 
groups. The NP segment (positions 8,692--10,257) shows a 
concentrated signal band particularly prominent in mid-to-high 
fitness groups. The NA segment (positions 10,257--11,715) 
contributes a moderate signal, while MP and NS show minimal 
contribution. Panel a of (Figure~\ref{fig:shap_heatmap_h1n1}d)
confirms that the highest-importance individual positions are 
distributed across HA, PB1, and PA, with the pattern of positive 
and negative contributions varying across fitness groups in a 
less clearly monotonic fashion than observed for H3N2, reflecting 
the more complex multi-segment fitness architecture of H1N1 
in the post-pandemic period.

\paragraph{Low-fitness sequence — 6B.1A.5a.2a.1 (North America).}
The low-fitness representative sequence (clade 6B.1A.5a.2a.1, 
North America; normalised DPGR $= 0.1201$, raw DPGR $= 0.0102$ 
$\log_{10}$/day) yields a predicted fitness of $f(x) = 0.206$, 
correctly placed below the model baseline of $E[f(X)] = 0.289$ 
(Figure~\ref{fig:shap_heatmap_h1n1}a). All five individually 
named top features are located in HA and are fitness-suppressing: 
C7563G ($-0.01$), A8349C ($-0.01$), T7547C ($-0.01$), A7423G 
($-0.01$), and T7778A ($-0.01$). The aggregate genome-wide signal 
is also negative ($-0.04$). The uniformly suppressive HA signal 
is consistent with this clade representing an early post-pandemic 
H1N1 lineage whose HA has not accumulated the antigenic changes 
associated with higher transmission fitness.

\paragraph{Mid-fitness sequence — 6B.1A.5b (North America).}
The mid-fitness representative (clade 6B.1A.5b, North America; 
normalised DPGR $= 0.5188$, raw DPGR $= 0.0441$ $\log_{10}$/day) 
yields a predicted fitness of $f(x) = 0.164$, below the model 
baseline (Figure~\ref{fig:shap_heatmap_h1n1}b). All five 
named top features are again located in HA and all are 
fitness-suppressing: C8558 ($-0.03$), T8291C ($-0.02$), G7552C 
($-0.01$), A8516 ($-0.01$), and G7515 ($-0.01$), with the 
genome-wide background contributing an additional $-0.06$. The 
model underestimates the true fitness of this sequence, which is 
attributable to the severe underrepresentation of mid-to-high 
fitness H1N1 sequences in the training distribution rather than 
a systematic model failure — the overall prediction distribution 
matches the true label distribution closely at the population 
level (mean predicted $= 0.240$, mean true $= 0.241$).

\paragraph{High-fitness sequence — 6B.1A.5a.2 (Europe).}
The high-fitness representative (clade 6B.1A.5a.2, Europe; 
normalised DPGR $= 1.0000$, raw DPGR $= 0.0851$ $\log_{10}$/day) 
yields a predicted fitness of $f(x) = 0.081$, substantially below 
the true label (Figure~\ref{fig:shap_heatmap_h1n1}c). 
The top features include two HA positions (C7370T: $-0.02$; 
C8558T: $-0.01$) and two PB2 positions (A117C: $-0.01$; G1554A: 
$-0.01$), with a strongly suppressive genome-wide background 
($-0.15$). The emergence of PB2 contributions alongside HA in 
the high-fitness waterfall is notable and consistent with the 
heatmap signal observed in the polymerase segments for higher 
fitness groups. The model's underestimation of this sequence 
reflects the rarity of truly high-fitness H1N1 sequences in 
the dataset — only 33 of 6,107 test sequences ($0.5\%$) 
exceed a normalised DPGR of 0.75 — and is expected given the 
skewed training distribution.

\paragraph{Summary of H1N1 SHAP findings.}
Across all three fitness tiers, the H1N1 SHAP analysis reveals 
that HA is the dominant segment driving model predictions, 
consistent with H3N2. However, several important differences 
emerge. First, PB1, PA, and PB2 show stronger relative 
contributions in H1N1 than in H3N2, particularly at higher 
fitness levels, suggesting a more distributed multi-segment 
fitness architecture for H1N1. Second, unlike H3N2 where SHAP 
contributions shift from negative to positive as fitness 
increases, H1N1 SHAP contributions remain predominantly 
suppressive across all three tiers, reflecting the skewed 
fitness distribution in which the vast majority of H1N1 
sequences occupy the low-fitness range. Third, the genome-wide 
background signal is consistently negative across fitness tiers 
for H1N1, in contrast to the strongly positive background 
observed for high-fitness H3N2 sequences. Together, these 
findings suggest that the CNN has learned distinct 
sequence--fitness mappings for the two subtypes, capturing 
the different evolutionary dynamics and fitness landscapes of 
H3N2 and H1N1 in the post-2009 pandemic period.

\subsection{Summary mapping of genomic importance to viral biology}

To facilitate biological interpretation of the deep learning models, the top-ranked nucleotide positions identified by SHAP were mapped to their corresponding viral proteins and amino acid residues (Table~\ref{tab:nt_to_aa_mapping}). This mapping bridges the raw genomic feature importance discovered by the CNNs with the experimentally validated antigenic sites and immune epitopes discussed in the following biological interpretation section.

\begin{table}[H]
\centering
\caption{Summary mapping of top-ranked SHAP nucleotide positions to viral protein residues for H3N2 and H1N1.}
\label{tab:nt_to_aa_mapping}
\begin{tabularx}{\linewidth}{@{}lccllX@{}}
\toprule
\textbf{Subtype} & \textbf{Nucleotide} & \textbf{Segment} & \textbf{AA Pos.} & \textbf{Site/Region} & \textbf{Biological Context} \\
\midrule
H3N2 & 7377 & HA & 155/156 & Site B & Immunodominant antigenic site \\
H3N2 & 7412 & HA & 166 & Site B/D & Antibody contact surface \\
H3N2 & 7460 & HA & 182 & Site D & Neutralising antibody target \\
H3N2 & 7992 & HA & 360 & HA2 & Fusion peptide region \\
H3N2 & 9986 & NP & 437 & NP & Conserved internal protein \\
H3N2 & 11066 & NA & 274 & NA & Neuraminidase active site \\
H3N2 & 6746 & PA & 611 & PA & Polymerase subunit \\
H3N2 & 4561 & PB1 & 740 & PB1 & Polymerase subunit \\
\midrule
H1N1 & 7423 & HA & 170 & Site Sa & Hypervariable antigenic site \\
H1N1 & 7547 & HA & 211 & Site Sb & Antigenic drift hotspot \\
H1N1 & 7563 & HA & 216 & Site Sb & Antigenic drift hotspot \\
H1N1 & 7778 & HA & 288 & HA & Structural residue \\
H1N1 & 8349 & HA & 478 & HA2 & Stalk domain \\
H1N1 & 117  & PB2 & 39 & PB2 & Host-adaptation region \\
H1N1 & 1554 & PB2 & 518 & PB2 & RNA polymerase subunit \\
\bottomrule
\end{tabularx}
\end{table}

\subsection{Biological interpretation of SHAP-identified positions}

To assess the biological plausibility of the genomic positions identified by the CNN via SHAP analysis, the top-ranked non-HA positions were cross-referenced against experimentally confirmed immune epitopes catalogued in the Immune Epitope Database (IEDB) \cite{vita2025iedb}, and the top-ranked HA positions were mapped against the five classical H3N2 antigenic sites (A--E) \cite{wiley1981structural, wilson1990structural}. Positions were considered biologically relevant when they fell within a confirmed epitope sequence or within one residue of an epitope boundary, reflecting the biological relevance of flanking residues for proteasomal cleavage and antigen processing \cite{seifert2004hepatitis, legall2007portable}.

\subparagraph{Non-HA positions.}
Among the top-ranked non-HA SHAP positions, four exhibited correspondence with confirmed IEDB epitopes spanning the NP, PB2, and M segments.

NP position 91 was identified with a mean SHAP impact of $-0.004464$ across multiple test sequences. This position corresponds precisely to the first residue of the HLA-A3-restricted cytotoxic T-lymphocyte (CTL) epitope KTGGPIYKR (NP residues 91--99) \cite{collins1999extensive}, a well-characterised CD8$^+$ T cell target in H3N2 influenza A \cite{vita2025iedb}. CTL-mediated immune pressure at this region of NP is well established, with NP being one of the most conserved influenza virus proteins and a major target in vaccine development \cite{shuklina2024inserting}.

NP position 417 was identified with a SHAP impact of $-0.002634$, placing it immediately upstream of the extensively characterised $\text{NP}_{418-426}$ CTL epitope (canonical sequence LPFDKSTIM) \cite{boon2002sequence, wahl2009tcell}. This region is among the most hypervariable H3N2 CTL epitopes documented in the IEDB, where sequence variation is directly linked to differential T-cell recognition and immune escape \cite{wahl2009tcell}. Position 417 constitutes the residue immediately flanking the N-terminus; flanking residues at this position are known to influence proteasomal cleavage and downstream epitope presentation \cite{seifert2004hepatitis, legall2007portable}.

PB2 position 746 was identified with a SHAP impact of $-0.006$ and falls directly within the IEDB-confirmed T-cell epitope SSILTDSQTATK (PB2 residues 741--752) \cite{habel2022hla}. This is significant as the antiviral pimodivir inhibits the early stages of influenza A replication by targeting the cap-binding function of the PB2 subunit \cite{patel2021pimodivir}.

Matrix protein position 5 was identified with a positive SHAP impact of $0.004300$, consistent with a fitness-promoting role. This position falls within the N-terminal region of the M1 protein (residues 1--15), which contains the IEDB-confirmed epitope MSLLTEVETYVLSIV \cite{lee2008memory}, a documented antigenic target in H3N2 strains.

\subparagraph{HA positions.}
Among the six top-ranked HA SHAP positions, four correspond to known antigenic regions of the H3N2 haemagglutinin.

HA position 159 maps precisely to Antigenic Site B (core residues 155--160), the immunodominant site in circulating H3N2 strains and the primary target of neutralising antibodies \cite{wu2020major}. The identification of $\text{HA}_{159}$ as the highest-ranked HA position is consistent with its role in determining antibody-mediated fitness differences between H3N2 clades.

HA position 144 maps to Antigenic Site A \cite{shah_seasonal_2024}. Site A positions contribute to both antigenic drift and changes in receptor binding affinity \cite{wiley1981structural}.

HA position 171 serves as a highly conserved anchor for broad-spectrum neutralization \cite{iba2014conserved}. Antibodies targeting this site achieve efficacy by binding to the protein backbone, allowing them to remain effective despite mutations in neighboring side chains \cite{iba2014conserved}. The model’s ability to identify this position repeatedly indicates that it can identify the structural constraints that define viral fitness and limit the path of future antigenic drift.

HA position 175 falls adjacent to the core residues of Antigenic Site D (key positions 174 and 176) \cite{wiley1981structural, wilson1990structural}. Positive selection at position 175 is predicted to involve N-glycosylation, which facilitates evasion of neutralizing antibodies \cite{kirkpatrick2018influenza}.

\subparagraph{Summary.}
Across four viral segments (NP, PB2, M, and HA), 
the SHAP analysis identified positions that 
correspond to experimentally validated immune 
epitopes or established antigenic sites without 
any prior biological annotation being provided 
to the model. The concordance between CNN-derived 
feature importance and independently characterised 
immunological targets that span both CTL epitopes 
in internal proteins and neutralising antibody 
targets on the HA surface provides biological 
interpretation of the model's learned sequence-fitness 
mapping and supports the interpretability of the 
SHAP-based attribution framework applied here.

\paragraph{(H1N1 A(H1N1)pdm09)}
Top-ranked H1N1 SHAP positions were mapped against 
two complementary sources: experimentally confirmed 
epitopes in the IEDB \cite{vita2025iedb}, and the 
antigenic site map of A(H1N1)pdm09 HA established 
using monoclonal antibody escape mutants against 
A/Narita/1/2009 \cite{matsuzaki_epitope_2014}. Given 
the antigenic divergence between pre-2009 seasonal 
H1N1 and A(H1N1)pdm09 \cite{matsuzaki_epitope_2014}, 
only IEDB entries derived from confirmed pdm09 
strains (A/California/04/2009 or later) were 
considered applicable. IEDB entries from pre-2009 
strains (e.g., A/New Caledonia/20/1999, 
A/Puerto Rico/8/1934) were excluded from the 
primary analysis on the grounds of antigenic 
non-equivalence, though their coverage of the 
same positional regions is noted as supporting 
context where relevant.

Among the five top-ranked HA positions, four map to sites consistent with known antigenic biology. HA position 174, 
the most recurrently identified SHAP position 
in the H1N1 analysis (appearing in 9 of the top 
20 HA entries with both positive and negative 
contributions), appears 
within the pdm09 HA epitope 
FYKNLIWLVKKGNSYPKLSK (HA residues 161--180, 
A/California/04/2009) \cite{yang_cd4_2013}, and 
maps exactly to Antigenic Site Sa in the pdm09 
monoclonal antibody escape mutant study 
\cite{matsuzaki_epitope_2014}. Site Sa is the most 
antigenically variable site in pdm09 HA and a 
primary target of neutralising antibodies.

HA position 206 maps exactly to Antigenic Site 
Sb in the pdm09 epitope map \cite{matsuzaki_epitope_2014}, 
and is covered by three IEDB epitopes in the 
HA 190--221 region. Although these IEDB entries 
derive from A/New Caledonia/20/1999, position 
206 is structurally conserved between pre-2009 
and pdm09 HA at this site, and the Matsuzaki 
pdm09 experimental confirmation is the primary 
consistency with known biology. Site Sb has been documented to 
undergo accelerating antigenic change in 
post-2017 pdm09 strains \cite{Xing2021_H1N1_AntigenicDrift}, 
consistent with the fitness-associated SHAP 
signal detected here.

HA position 141 maps exactly to Antigenic Site Sa \cite{matsuzaki_epitope_2014}. The presence of both $\text{HA}_{141}$ and $\text{HA}_{174}$ as top-ranked SHAP positions,both within Site Sa, reinforces the 
model's convergent identification of this 
immunodominant site as a fitness-relevant region 
in H1N1 pdm09.

HA position 448 falls within the pdm09 HA 
epitope ELLVLLENERTLDYHDSNVK (HA residues 441--460, 
A/California/04/2009) \cite{yang_cd4_2013}, 
providing a direct pdm09-specific IEDB match. 
This region is located in the lower HA1 domain 
and does not correspond to the classical 
antigenic sites Sa--Cb, suggesting the model 
has identified a non-canonical but 
experimentally confirmed epitope region in 
this area of the HA molecule.

HA position 106 is covered by two IEDB epitopes 
(residues 99--123) from A/New Caledonia/20/1999 
but lacks a pdm09-specific IEDB entry or 
antigenic site assignment. It is therefore 
reported as a candidate position pending 
pdm09-specific experimental characterisation.

Among non-HA positions, PA\_779 and PB2\_421 
did not correspond to any pdm09-applicable 
IEDB epitope. 

The convergent identification of the two 
dominant pdm09 antigenic sites (Sa and Sb) 
from raw sequence data, without biological 
annotation, and the additional confirmation 
of HA\_174 and HA\_448 by pdm09-specific IEDB 
entries, provides meaningful biological 
interpretation of the H1N1 CNN's learned 
sequence-fitness associations.

\clearpage
\begin{landscape}
\setlength{\tabcolsep}{4pt}
\renewcommand{\arraystretch}{0.95}
\begin{table}[H]
\centering
\caption{Biological interpretation of top-ranked H3N2 SHAP positions 
against IEDB-confirmed immune epitopes and classical HA antigenic 
sites. Positions are reported in amino acid coordinates relative 
to each viral segment. SHAP impact values represent the mean 
contribution across sequences in which the position was 
top-ranked. Epitope types: CTL = cytotoxic T-lymphocyte epitope; 
Ab = antibody/neutralising epitope; B = B-cell epitope.}
\label{tab:shap_validation_h3n2}
\footnotesize
\begin{tabularx}{\linewidth}{@{}lccp{2.8cm}Xp{3.0cm}@{}}
\toprule
\textbf{Segment} & 
\textbf{Position} & 
\textbf{SHAP} & 
\textbf{Match type} & 
\textbf{IEDB / Antigenic site} & 
\textbf{Reference} \\
\midrule
\multicolumn{6}{@{}l}{\textit{Non-HA internal proteins}} \\
\midrule
NP  & 91  & $-0.004$ & Exact        
    & CTL epitope KTGGPIYKR (NP 91--99)
    & \cite{collins1999extensive} \\
NP  & 417 & $-0.003$ & Flanking ($+1$) 
    & CTL epitope NP$_{418-426}$ (LPFDKSTIM)
    & \cite{boon2002sequence} \\
PB2 & 746 & $-0.006$ & Exact        
    & T-cell epitope SSILTDSQTATK (PB2 741--752)   
    & \cite{habel2022hla} \\
M   & 5   & $+0.004$ & Within window 
    & B-cell epitope MSLLTEVETYVLSIV (M1 1--15)                
    & \cite{lee2008memory} \\
\midrule
\multicolumn{6}{@{}l}{\textit{Haemagglutinin (HA) -- antigenic site mapping}} \\
\midrule
HA & 144 & $+0.005$ & Exact 
   & Antigenic Site A (residues 121--168); A loop, RBS periphery 
   & \cite{wiley1981structural} \\
HA & 159 & $+0.010$ & Exact 
   & Antigenic Site B (residues 155--160); immunodominant, 
     neutralising Ab target 
   & \cite{wiley1981structural} \\
HA & 171 & $\pm0.007$--$0.011$ & Documented escape 
   & Lateral-patch escape position in drifted H3N2; outside 
     classic sites A--E 
   & \cite{iba2014conserved} \\
HA & 175 & $-0.006$ & Adjacent ($\pm1$) 
   & Adjacent to Antigenic Site D (key residues 174, 176); 
     antibody contact surface 
   & \cite{wiley1981structural} \\
\bottomrule
\end{tabularx}
\end{table}

\vspace{0.75em}

\begin{table}[H]
\centering
\caption{Biological interpretation of top-ranked H1N1 A(H1N1)pdm09 SHAP positions.}
\label{tab:shap_validation_h1n1}
\footnotesize
\begin{tabularx}{\linewidth}{@{}lccXp{5.0cm}@{}}
\toprule
\textbf{Seg.} & 
\textbf{Pos.} & 
\textbf{SHAP} & 
\textbf{Biological Interpretation} & 
\textbf{Reference} \\
\midrule
\multicolumn{5}{@{}l}{\textit{Haemagglutinin}} \\
\midrule
HA & 174 & $\pm0.002$--$0.005$ 
   & pdm09 IEDB (HA 161--180, CA/04/2009) + AS Sa 
   & \cite{yang_cd4_2013, matsuzaki_epitope_2014} \\
HA & 141 & $-0.002$--$-0.003$ 
   & AS Sa (pdm09 escape mutants) 
   & \cite{matsuzaki_epitope_2014} \\
HA & 206 & $\pm0.002$--$0.010$ 
   & AS Sb (pdm09 escape mutants) 
   & \cite{matsuzaki_epitope_2014} \\
HA & 448 & $+0.003$--$0.004$ 
   & pdm09 HA epitope (HA 441--460, CA/04/2009) 
   & \cite{yang_cd4_2013} \\
\midrule
\multicolumn{5}{@{}l}{\textit{Non-HA}} \\
\midrule
PA  & 779 & $+0.002$ 
    & No applicable IEDB match 
    & --- \\
PB2 & 421 & $-0.002$ 
    & No applicable pdm09 IEDB match 
    & --- \\
\bottomrule
\end{tabularx}
\end{table}

\end{landscape}
\clearpage

\section{Discussion}

In this study, we applied the Differential Population Growth Rate 
(DPGR) framework to seasonal influenza A and showed that it can 
recover interpretable clade-level differences in relative 
transmission fitness across locations, seasons, and subtypes. 
Across both H3N2 and H1N1, the inferred pairwise growth 
advantages were consistent with major patterns of lineage 
turnover observed in contemporary influenza surveillance, 
supporting the broader idea that viral sequence dynamics contain 
recoverable signatures of competitive success 
\cite{petrova_evolution_2018, luksza_predictive_2014, 
du_evolution_2017, huddleston_genotypes_2020}. By combining 
DPGR-based inference with sequence-trained convolutional 
neural networks, conformal prediction, and SHAP-based 
interpretation, this study extends beyond retrospective 
description to a more integrated framework for estimating, 
predicting, and biologically contextualising influenza fitness.

One of the central contributions of this work is the use of 
DPGR as a simple and interpretable metric for comparing the 
growth of co-circulating clades. Influenza forecasting studies 
have previously used population-genetic, epidemiologic, and 
genotype--phenotype integration frameworks to estimate viral 
fitness and anticipate future strain success 
\cite{luksza_predictive_2014, du_evolution_2017, 
huddleston_genotypes_2020, lou_predictive_2024, 
huddleston_timely_2025}. In that context, DPGR offers a 
complementary advantage: it is directly estimated from 
observed changes in the relative abundance of variant pairs 
during windows of co-circulation, making it comparatively 
transparent and easy to interpret in surveillance settings. 
The cross-regional consistency observed here, particularly 
for recent H3N2 clades such as subclade K, strengthens the 
biological plausibility of the inferred signals and suggests 
that DPGR can capture competitive dynamics that are 
reproducible across distinct geographic settings rather than 
arising solely from local sampling noise.

At the same time, the present results underscore an important 
conceptual distinction: DPGR captures relative competitive 
growth, but it is not itself a direct measure of antigenic 
distance or vaccine effectiveness. Influenza fitness is shaped 
by multiple interacting processes, including antigenic novelty, 
replication competence, genetic-background effects, and host 
population immunity 
\cite{petrova_evolution_2018, huddleston_genotypes_2020, 
shah_seasonal_2024}. The 2022--2023 H3N2 season illustrates 
this point clearly. In that season, DPGR identified a growth 
advantage for 3C.2a1b.2a.2a.3a.1 relative to 3C.2a1b.2a.2b, 
yet this did not imply substantial vaccine failure because the 
dominant lineage remained antigenically related to the vaccine 
strain. The same pattern emerged in the H1N1 2018--2019 season, 
in which DPGR detected the competitive displacement of the 
vaccine-matched 6B.1 clade by emerging 6B.1A subclades across 
multiple continental regions, yet overall vaccine effectiveness 
against A(H1N1)pdm09 remained 46\% and was substantially 
higher (62\%) in children \cite{doyle2019interim}. In both 
cases, the DPGR fitness signal preceded but did not immediately 
translate into antigenic failure, likely because cross-reactive 
antibodies were preserved within the same broad clade lineage. 
This distinction is important for public-health interpretation: 
a positive DPGR signal should be understood as evidence of 
within-season competitive advantage, while its implications for 
vaccine performance require additional serologic, antigenic, 
and epidemiologic evidence.

These findings nonetheless suggest practical value for DPGR as 
an early quantitative surveillance signal. Timely genomic 
surveillance and rapid data sharing are already central to 
influenza monitoring and vaccine strain review 
\cite{shu_gisaid_2017, elbe_data_2017}. Recent work has further 
shown that forecast accuracy improves when evolutionary models 
are updated using contemporary genomic data and seasonally 
relevant strain information \cite{huddleston_timely_2025}. 
Within that broader surveillance ecosystem, DPGR could help 
prioritise newly expanding clades for deeper antigenic 
characterisation, laboratory follow-up, or inclusion in 
comparative forecasting pipelines. An important practical 
extension of the sequence-based CNN framework developed here 
is that, given a newly deposited genome sequence from GISAID, 
a predicted DPGR fitness score with a calibrated conformal 
prediction interval can be assigned without waiting for 
sufficient co-circulation data to accumulate for a 
regression-based DPGR estimate. The baseline comparisons presented in Table~\ref{tab:baselines} show that a clade-only lookup achieves R\textsuperscript{2} = 0.753 for H3N2 and R\textsuperscript{2} = 0.955 for H1N1, confirming that clade identity carries substantial fitness signal; the CNN adds value precisely in the scenario where clade assignment is unavailable, recovering continent-level variation from raw sequence and enabling fitness estimation for newly deposited sequences before they have been classified. This prospective applicability
is particularly relevant given the approximately three-month
average lag between sample collection and sequence submission 
to public databases \cite{huddleston_timely_2025}, during which 
time a sequence-based fitness estimate could inform early 
surveillance prioritisation. In this sense, DPGR is best 
viewed not as a replacement for WHO or CDC surveillance 
frameworks but as a complementary signal that links sequence 
abundance dynamics to an interpretable estimate of relative 
transmission fitness.

The deep-learning results extend this interpretation by 
showing that complete influenza genomes carry substantial 
information about DPGR-derived fitness. The strong predictive 
performance achieved for both subtypes indicates that sequence 
patterns alone are informative for the relative-fitness 
phenotype studied here, complementing prior work on influenza 
evolutionary forecasting and machine-learning-based antigenic 
prediction \cite{huddleston_genotypes_2020, lou_predictive_2024, 
shah_seasonal_2024}. Importantly, the use of conformal 
prediction adds a practical layer of uncertainty 
quantification that is especially relevant for surveillance 
applications, where decision-making benefits from calibrated 
ranges rather than point estimates alone. The broader 
prediction intervals observed for H3N2 are consistent with 
its more heterogeneous and rapidly shifting fitness landscape, 
whereas the much tighter H1N1 intervals reflect the narrower 
distribution of observed fitness values in that subtype. 
Together, these results support the feasibility of 
sequence-based fitness screening while also highlighting the 
need to interpret model confidence in light of 
subtype-specific evolutionary structure.

The SHAP analyses provide an additional layer of biological 
interpretation and strengthen confidence that the CNNs learned 
meaningful sequence--fitness relationships rather than only 
statistical correlates. For both subtypes, the dominant 
contribution of haemagglutinin is biologically coherent with 
the central role of HA in antigenic drift, immune escape, and 
lineage replacement \cite{petrova_evolution_2018}. At the same 
time, the models also identified contributions from internal 
genes, particularly NP and polymerase segments, suggesting 
that influenza fitness is not purely an HA-driven property. 
This finding is reinforced by the biological interpretation 
analyses, in which several SHAP-highlighted positions mapped 
to experimentally characterised epitopes or known antigenic 
sites in both H3N2 and A(H1N1)pdm09 
\cite{yang_cd4_2013, matsuzaki_epitope_2014}.

The contrasting SHAP profiles of H3N2 and H1N1 further 
illuminate subtype-specific evolutionary regimes. For H3N2, 
SHAP contributions shift progressively from suppressive at 
low fitness to strongly positive at high fitness, centred on 
HA antigenic sites consistent with the rapid antigenic 
cluster transitions that characterise H3N2 evolution 
\cite{petrova_evolution_2018, smith2004mapping}. For H1N1, 
SHAP contributions are uniformly suppressive across all 
fitness tiers, with no fitness-increasing features appearing 
among the top-ranked positions for any representative 
sequence. This pattern is consistent with the documented 
antigenic stability of A(H1N1)pdm09 in the post-pandemic 
period: following its emergence in 2009, the virus underwent 
an extended period of antigenic stasis before new antigenically 
distinct variants emerged during the 2015--2016 season when 
clade 6B.1 appeared \cite{su2015phylodynamics, 
muawan2025h1n1}. This trajectory is markedly different from 
H3N2, which undergoes major antigenic cluster transitions 
every two to five years \cite{smith2004mapping}. The uniformly 
suppressive SHAP profile for H1N1 is consistent with stronger 
purifying selection operating on the H1N1 genome, in which 
the vast majority of genomic variation relative to the 
reference represents departures from the stable post-pandemic 
optimum rather than fitness-enhancing innovations. Notably, 
antigenic cluster transitions in A(H1N1)pdm09 have been 
shown to be largely concentrated on epitopes Sa and Sb 
\cite{cheng2024h1n1antigenic}, which is consistent with the 
top-ranked HA positions identified by SHAP in the present 
study (HA positions 174 and 141 in Site Sa; position 206 in 
Site Sb). That the CNN models independently recapitulated 
this known biological asymmetry from raw sequence-fitness 
associations alone, without any prior annotation of 
evolutionary rates or selective pressures, provides additional 
support for the biological validity of the learned 
representations. The resulting picture is consistent with 
current understanding that influenza success emerges from a 
combination of antigenic and non-antigenic determinants, and 
that models trained on whole genomes may therefore capture 
biologically relevant interactions that are not obvious from 
single-segment analyses alone.

Several limitations should be considered when interpreting 
these results. First, the underlying sequence data are subject 
to the well-known biases of influenza genomic surveillance, 
including uneven sequencing effort across regions, variation 
in sampling intensity over time, and the fact that sequence 
counts do not directly measure true infection incidence 
\parencite{shu_gisaid_2017, elbe_data_2017}. Second, the DPGR 
framework depends on windows of sufficient co-circulation 
and on the assumption that short-term changes in log abundance 
ratios provide a reasonable approximation to relative growth 
advantage. Although the sliding-window procedure used here 
was designed to improve statistical robustness, the inferred 
values remain sensitive to sampling sparsity, window 
selection, and the availability of suitable intermediate 
clades for additive comparisons. Third, CNN analyzes 
were trained on restricted geographic subsets with complete 
eight-segment genomes, which improves data quality but may 
limit generalizability. The H1N1 results also revealed a 
strong class imbalance, with fewer than 0.5\% of test 
sequences exhibiting normalized DPGR values above 0.75, 
likely contributing to underestimation in the upper tail of 
the H1N1 fitness distribution. This is a known challenge for 
regression models trained on skewed target distributions 
\cite{branco2019imbalanced}, and prospective collection of 
high-fitness sequences as they emerge would be expected to 
improve calibration at the upper end of the fitness range in 
future model iterations.

These limitations suggest several directions for future work. 
Prospective validation on truly unseen seasons will be 
important for determining how early DPGR and sequence-based 
predictions can identify emerging dominant clades in practice. 
Integrating DPGR with additional data layers, including 
antigenic assays, vaccine-effectiveness estimates, and 
incidence-based surveillance, may also help separate 
growth advantage from immune escape and improve biological 
interpretability 
\cite{du_evolution_2017, huddleston_genotypes_2020, 
huddleston_timely_2025}. Methodologically, it would be 
valuable to compare the present CNN framework with 
alternative architectures and multimodal models that jointly 
incorporate genome sequence, clade metadata, and surveillance 
context. Extending the approach to other influenza lineages 
and to finer spatial scales may further clarify how regional 
ecological conditions interact with viral genotype to shape 
apparent fitness.

Overall, this study supports DPGR as a useful and 
interpretable measure of clade-level relative transmission 
fitness in seasonal influenza A. The agreement between 
DPGR-based inference and independent surveillance patterns, 
together with the strong sequence-based predictive 
performance and biologically plausible SHAP attributions, 
suggests that influenza fitness can be studied within a 
unified framework linking population growth, genomic 
prediction, and mechanistic interpretation. Such an approach 
may help bridge the gap between large-scale genomic 
surveillance and the practical need to identify which 
influenza clades are most likely to expand, persist, or 
warrant closer public-health attention.

\section{Conclusion}
In conclusion, the integration of Differential Population Growth Rate (DPGR) with deep learning provides a robust and interpretable framework for seasonal influenza surveillance. Our results demonstrate that DPGR successfully recovers the competitive hierarchy of co-circulating clades across multiple geographic regions and flu seasons, with inferred fitness advantages consistently aligning with real-world lineage replacement patterns observed by the WHO and CDC. By transforming raw genomic sequences into quantitative fitness predictions, the subtype-specific convolutional neural networks achieve high accuracy ($R^2 > 0.95$) and provide statistically grounded uncertainty estimates through conformal prediction.

Furthermore, the SHAP-based interpretability analysis suggests the models learned to prioritize biologically relevant genomic features, specifically the immunodominant antigenic sites of haemagglutinin and critical epitopes in internal proteins. The identification of subclade K as a fitter variant across several continents during the 2025-2026 season exemplifies the practical utility of this framework for early detection of emerging high-risk lineages. These findings suggest that DPGR-derived signals can serve as a critical component of the global surveillance toolkit, bridging the gap between observed epidemiological trends and the genomic basis of viral success.

By providing a unified path from sequence abundance to biological interpretation, this framework supports more informed public health decision-making, particularly in the context of vaccine strain selection and the prioritization of variants for laboratory characterization. Future efforts will focus on incorporating real-time surveillance feeds to further improve the timeliness of these predictions, ensuring that global public health responses remain ahead of the virus's rapid evolutionary trajectory.

\section{Acknowledgments}
We acknowledge the support of the USA NSF awards 2525493 and 2200138 and internal support of the Old Dominion University. A patent is pending based on a portion of this work.
%We gratefully acknowledge the authors, and the originating and submitting laboratories for the sequence data obtained from the GISAID EpiFlu database. Specifically, we acknowledge the Crick Worldwide Influenza Centre (originating laboratory) and the Centers for Disease Control and Prevention (submitting laboratory) for the sequence A/Hong Kong/4801/2014 (EPI\_ISL\_189814).

\section*{Bibliography}
\printbibliography[heading=none]

\clearpage
\appendix
\section{Supplementary CNN Figures and DPGR Regression Panels}
\label{app:dpgr_appendix}
This appendix collects the CNN training curves and detailed seasonal
regression panels moved from the main Results section to shorten the
primary manuscript.

\subsection{CNN Training Curves}

\begin{figure}[H]
  \centering
  \captionsetup[subfigure]{font=tiny, labelfont=tiny}
  \begin{subfigure}[b]{0.48\textwidth}
    \centering
    \includegraphics[width=\linewidth, height=5cm]{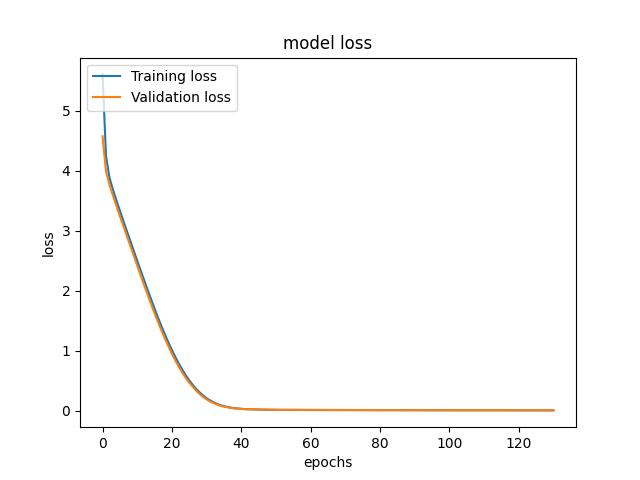}
    \caption{H3N2 training and validation loss over
    131 epochs. The sharp initial decline stabilises
    by epoch 40, with training and validation curves
    converging closely.}
  \end{subfigure}\hfill
  \begin{subfigure}[b]{0.48\textwidth}
    \centering
    \includegraphics[width=\linewidth, height=5cm]{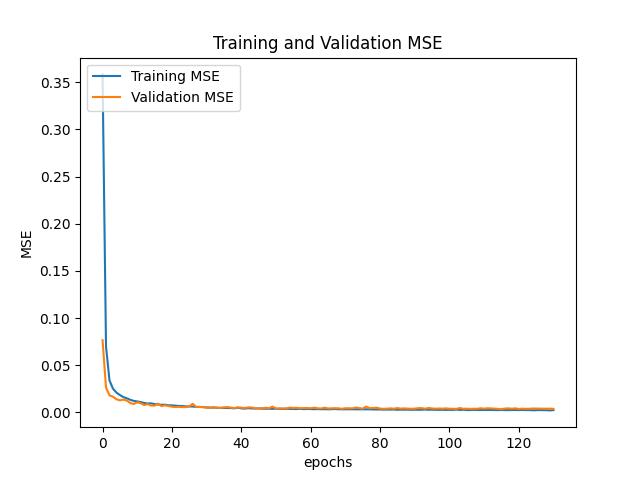}
    \caption{H3N2 training and validation MSE over
    131 epochs, reaching a final validation
    MSE of 0.0038 with no evidence of overfitting.}
  \end{subfigure}
  \vspace{1em}
  \begin{subfigure}[b]{0.48\textwidth}
    \centering
    \includegraphics[width=\linewidth, height=5cm]{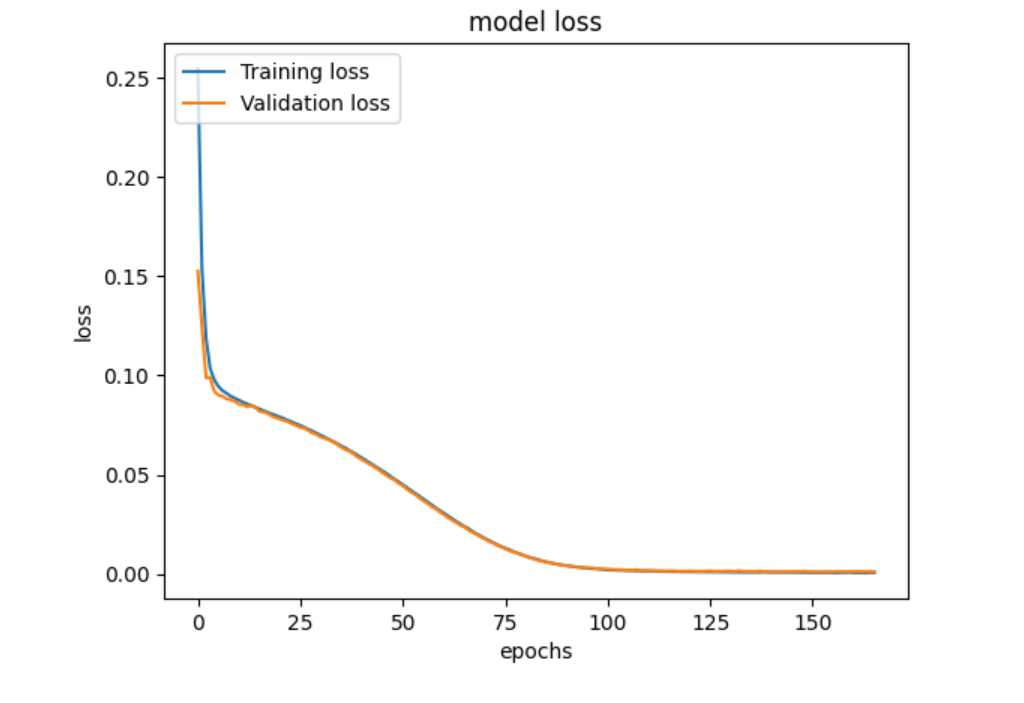}
    \caption{H1N1 training and validation loss over
    166 epochs. Convergence is smooth, with
    training and validation curves tracking
    closely throughout.}
  \end{subfigure}\hfill
  \begin{subfigure}[b]{0.48\textwidth}
    \centering
    \includegraphics[width=\linewidth, height=5cm]{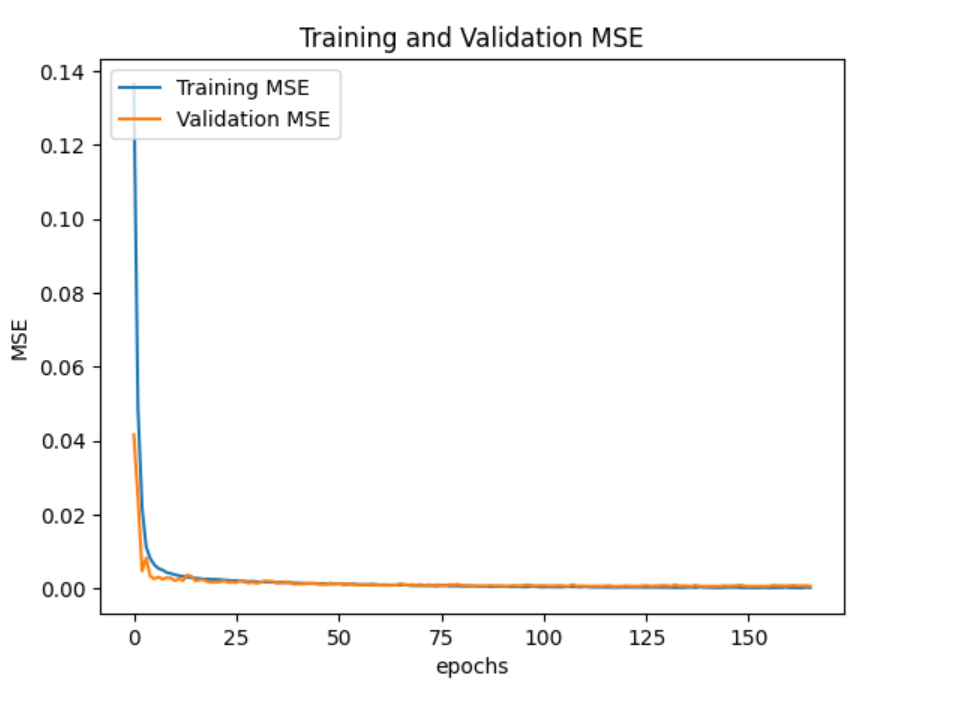}
    \caption{H1N1 training and validation MSE over
    166 epochs, reaching a final validation
    MSE of 0.0007 with strong generalisation
    and no overfitting.}
  \end{subfigure}
  \caption{Training and validation loss and MSE curves
  for the A/H3N2 (panels a--b) and A/H1N1
  (panels c--d) convolutional neural networks.
  Both models converge smoothly without overfitting,
  with H1N1 achieving lower final validation MSE
  (0.0007) than H3N2 (0.0038), consistent with
  the narrower fitness distribution of the H1N1
  training set.}
  \label{fig:cnn_training}
\end{figure}

\subsection{A/H3N2}

\subsubsection{Continent-level clades A/H3N2 2017-2018 flu season}

\paragraph{2017--2018 season.}
The dominant competitive signal at the continental level was the emergence 
of clade 3C.3a1 as the fitter variant over all co-circulating 3C.2a 
subclades in North America. DPGR analysis showed 3C.3a1 outcompeting 
3C.2a1b.1 (DPGR $= -0.0181$, window: Dec~2017--Feb~2018), 3C.2a1 
(DPGR $= -0.0320$, window: Dec~2017--Feb~2018), and 3C.2a2 
(DPGR $= -0.0240$, window: Dec~2017--Feb~2018), with all negative slopes 
confirming 3C.3a1 as the faster-growing denominator in each pair. Within 
the 3C.2a lineage, 3C.2a2 showed a fitness advantage over 3C.2a1 in 
North America (DPGR $= +0.0152$, window: Oct~2017--Jan~2018). In Oceania, 
3C.2a1 outcompeted 3C.2a1b.1 (DPGR $= +0.0262$, window: Sep--Oct~2017), 
indicating that 3C.2a1b.1 had not yet achieved dominance in that region 
during this period.

\begin{figure}[H]
  \centering
  \captionsetup[subfigure]{font=tiny, labelfont=tiny}

  \begin{subfigure}[b]{0.32\textwidth}
    \centering
    \includegraphics[width=\linewidth]{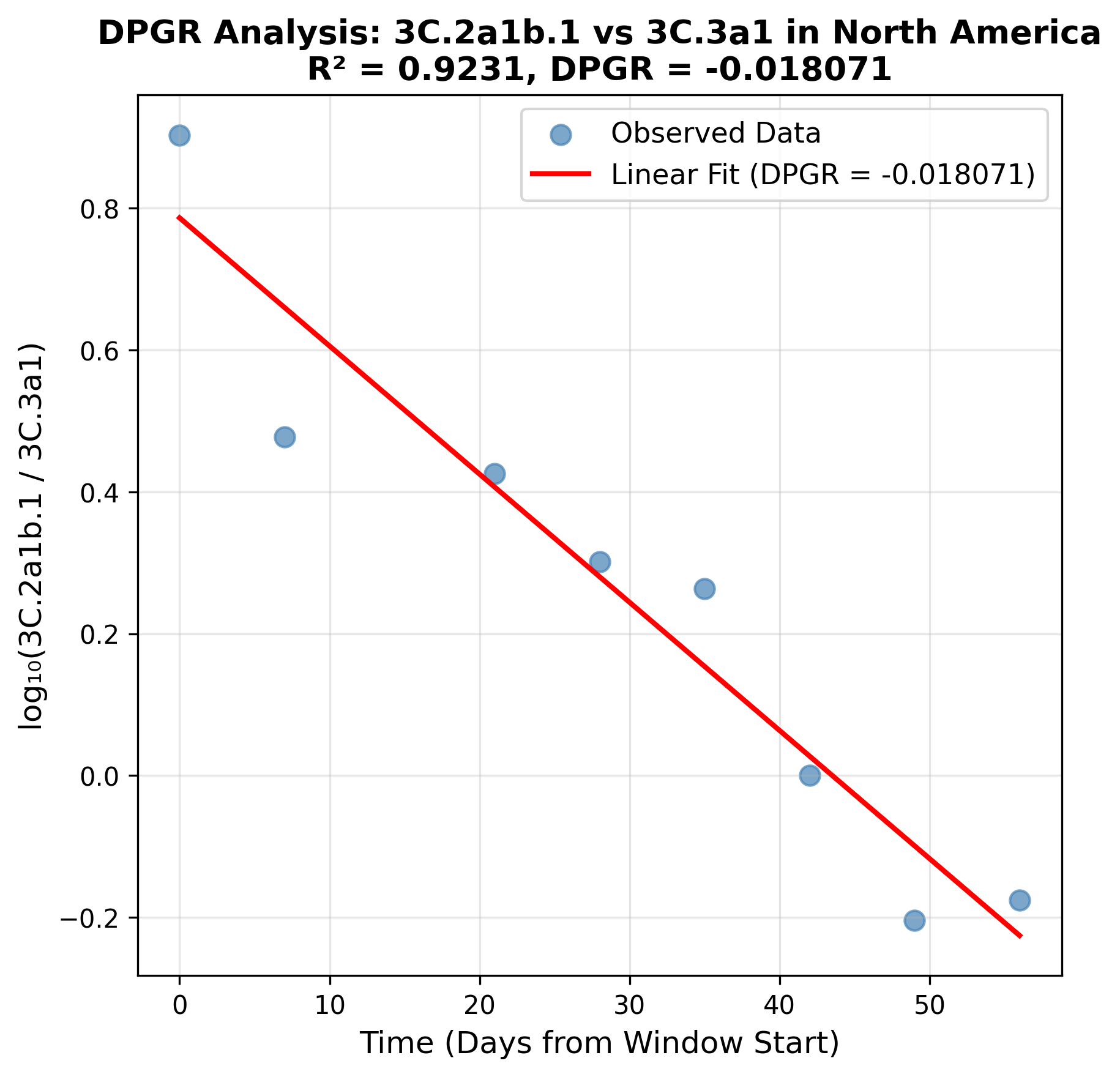}
    \caption{2017-12-18 - 2018-02-12}
  \end{subfigure}\hfill
  \begin{subfigure}[b]{0.32\textwidth}
    \centering
    \includegraphics[width=\linewidth]{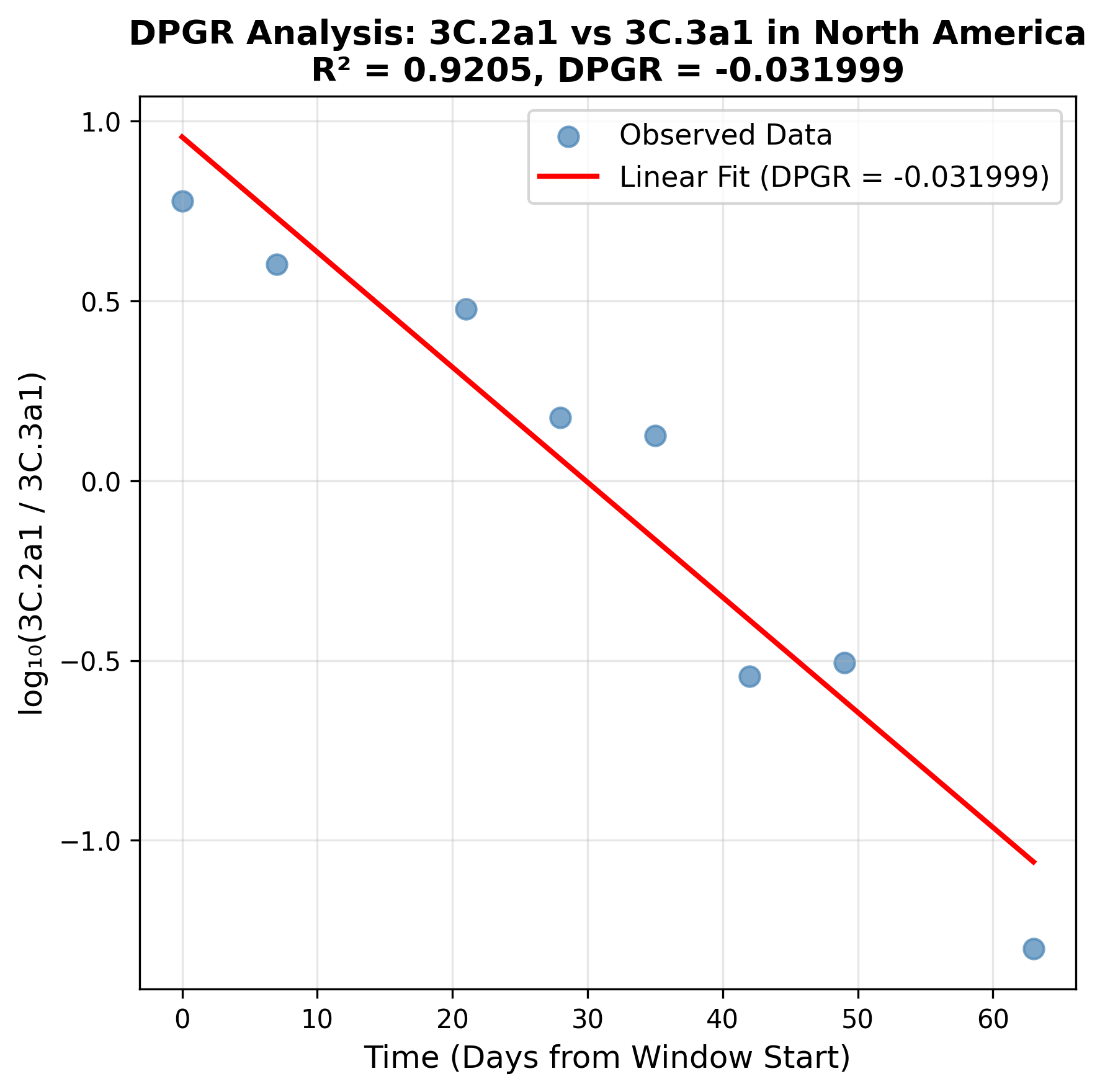}
    \caption{2017-12-18 - 2018-02-19}
  \end{subfigure}\hfill
  \begin{subfigure}[b]{0.32\textwidth}
    \centering
    \includegraphics[width=\linewidth]{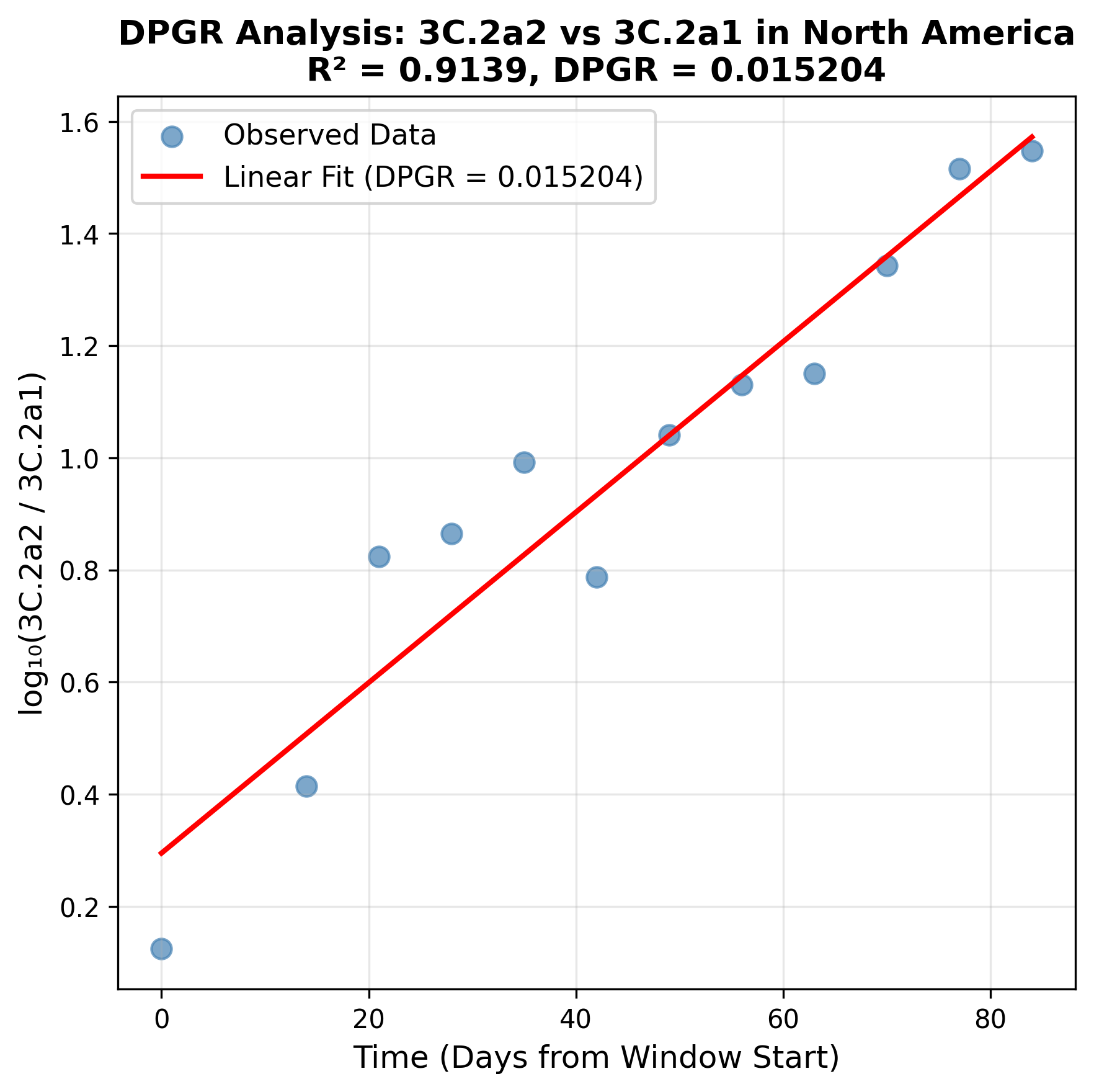}
    \caption{2017-12-18 - 2018-02-19}
  \end{subfigure}

  \vspace{1em}

  \centering
  \begin{subfigure}[b]{0.32\textwidth}
    \centering
    \includegraphics[width=\linewidth]{regression_3C_2a2_vs_3C_2a1_North_America.png}
    \caption{2017-10-09 - 2018-01-01}
  \end{subfigure}
  \hspace{2em}
  \begin{subfigure}[b]{0.32\textwidth}
    \centering
    \includegraphics[width=\linewidth]{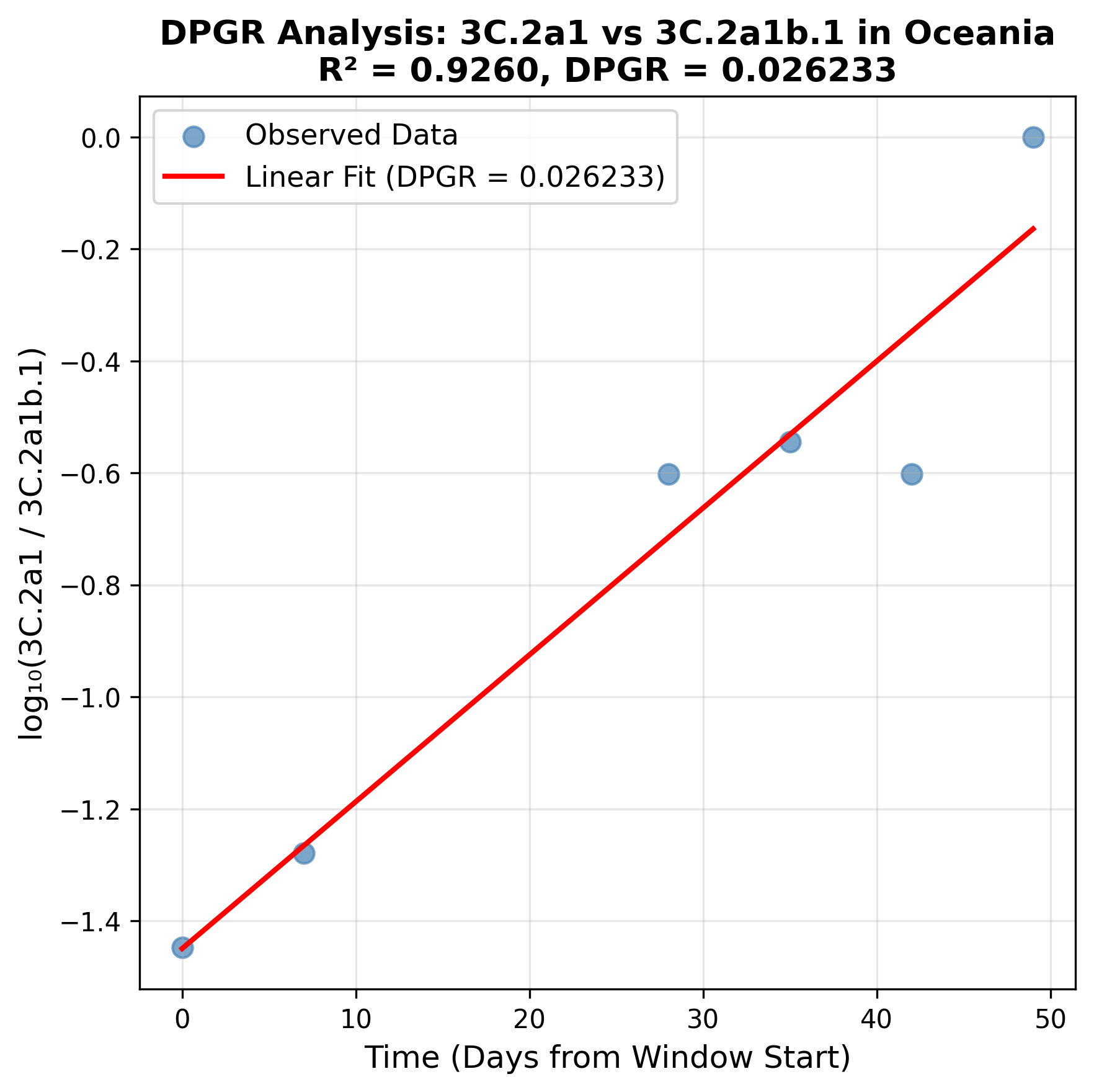}
    \caption{2017-09-04 - 2017-10-23}
  \end{subfigure}

  \caption{Supplementary H3N2 continent-level regression panels for the 2017--2018 season.}
  \label{fig:appendix-h3n2-2017-2018}
\end{figure}

\subsubsection{Continent-level clades A/H3N2 2016-2017 flu season}
At the continental level, DPGR analysis identified active competitive 
dynamics among 3C.2a subclades across multiple regions. In Europe, clade 
3C.2a1 demonstrated a fitness advantage over the vaccine-reference clade 
3C.2a (DPGR $= +0.0211$, window: Jan--Mar~2017), indicating faster growth 
of 3C.2a1 relative to the parental vaccine lineage. Simultaneously, in 
Europe, clade 3C.2a1b.1 showed a fitness advantage over 3C.2a1 
(DPGR $= -0.0162$, window: Jan--Mar~2017), and in Asia the same 
competitive relationship was observed (DPGR $= -0.0249$, window: 
Dec~2016--Jan~2017), reflecting a competitive hierarchy within the 3C.2a 
lineage where successive subclades were displacing one another. In South 
America, the older vaccine strain 3C.2a retained a fitness advantage over 
3C.2a1 (DPGR $= -0.0357$, window: Jan--Feb~2017), indicating that the 
transition to 3C.2a1 dominance had not yet occurred in that region during 
this period.

\begin{figure}[H]
  \centering
  \captionsetup[subfigure]{font=tiny, labelfont=tiny}

  \begin{subfigure}{0.24\textwidth}
    \centering
    \includegraphics[width=\linewidth]{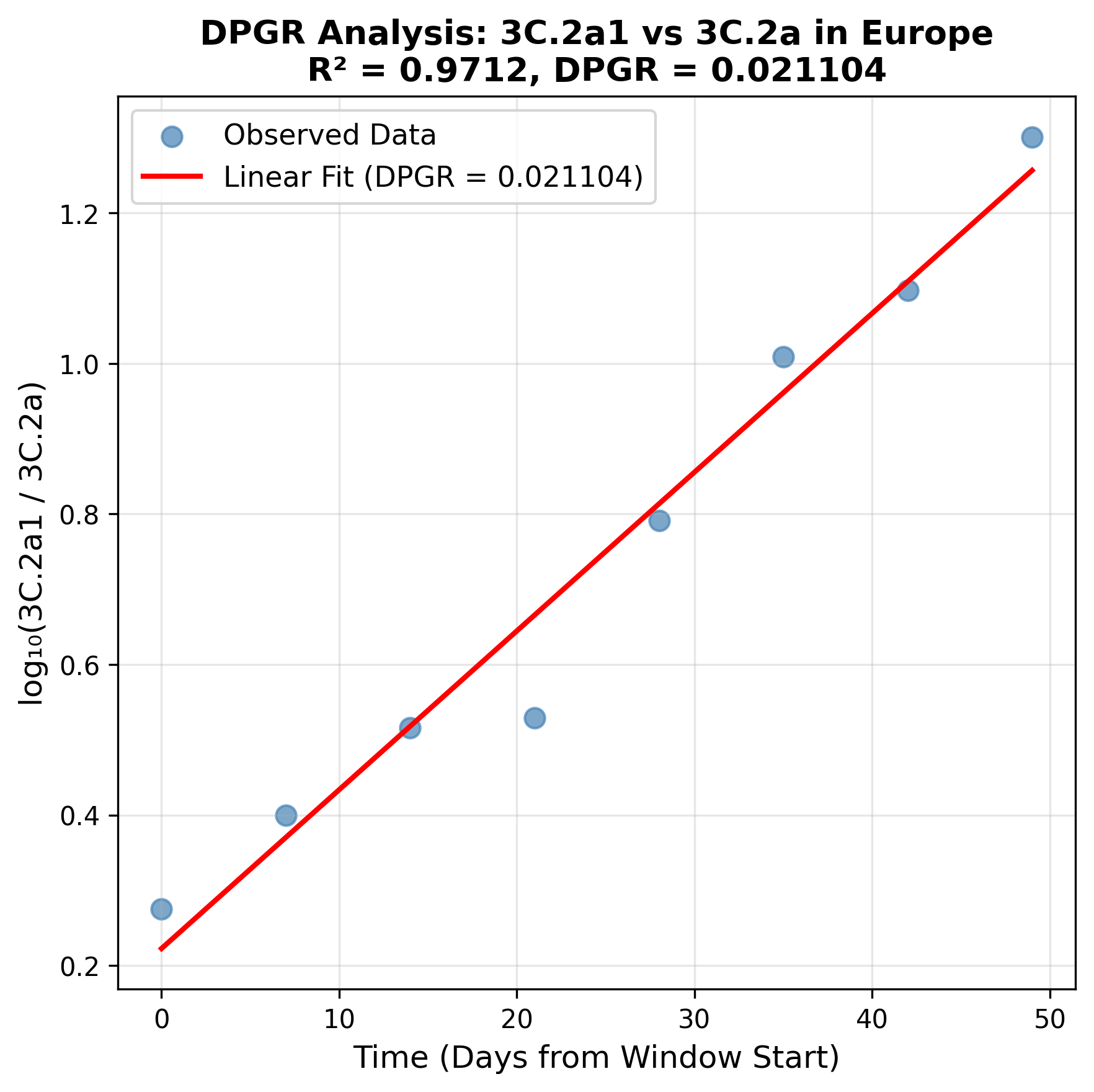}
    \caption{2017-01-23 - 2017-03-13}
  \end{subfigure}\hfill
  \begin{subfigure}{0.24\textwidth}
    \centering
    \includegraphics[width=\linewidth]{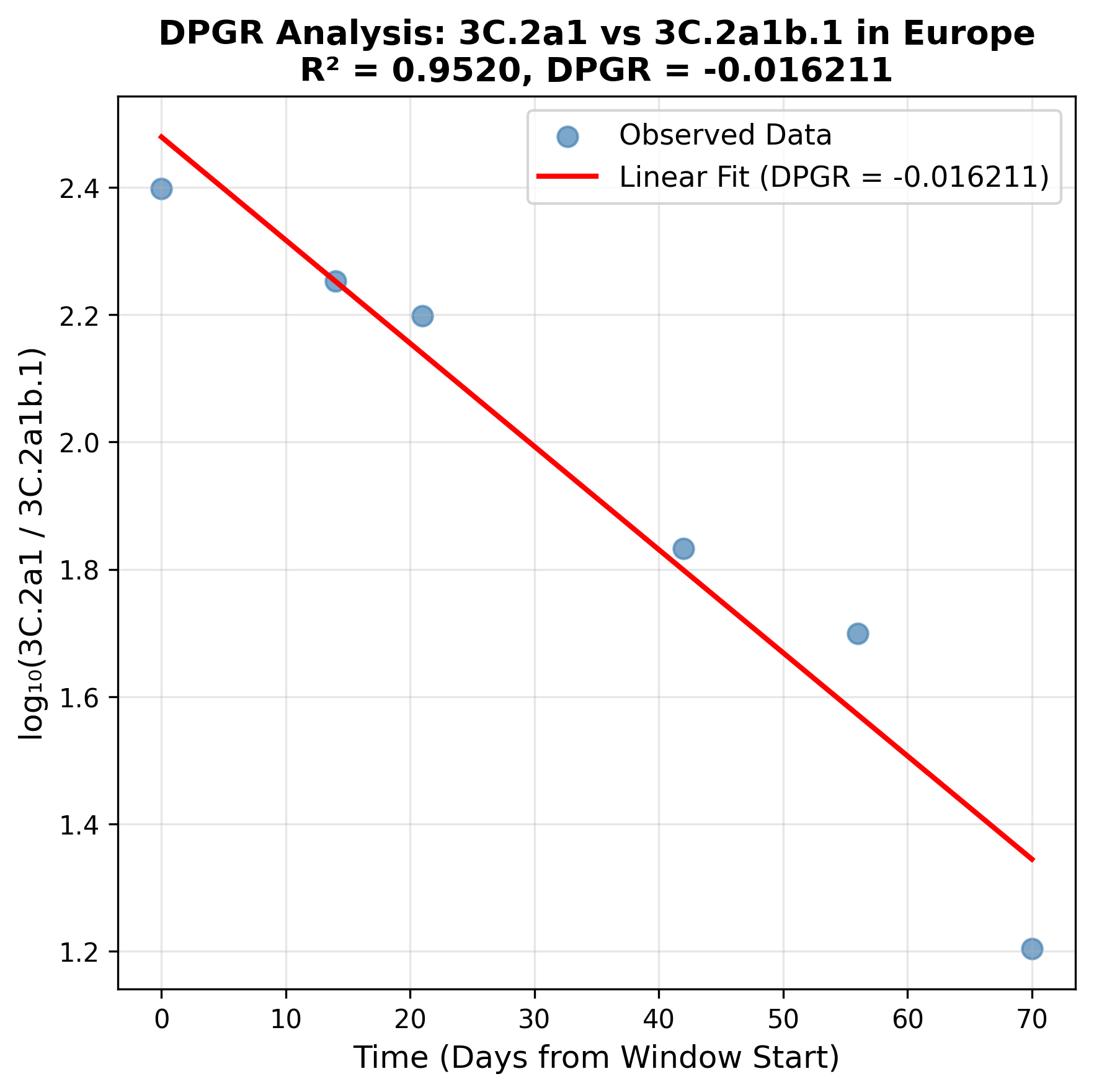}
    \caption{2017-01-09 - 2017-03-20}
  \end{subfigure}\hfill
  \begin{subfigure}{0.24\textwidth}
    \centering
    \includegraphics[width=\linewidth]{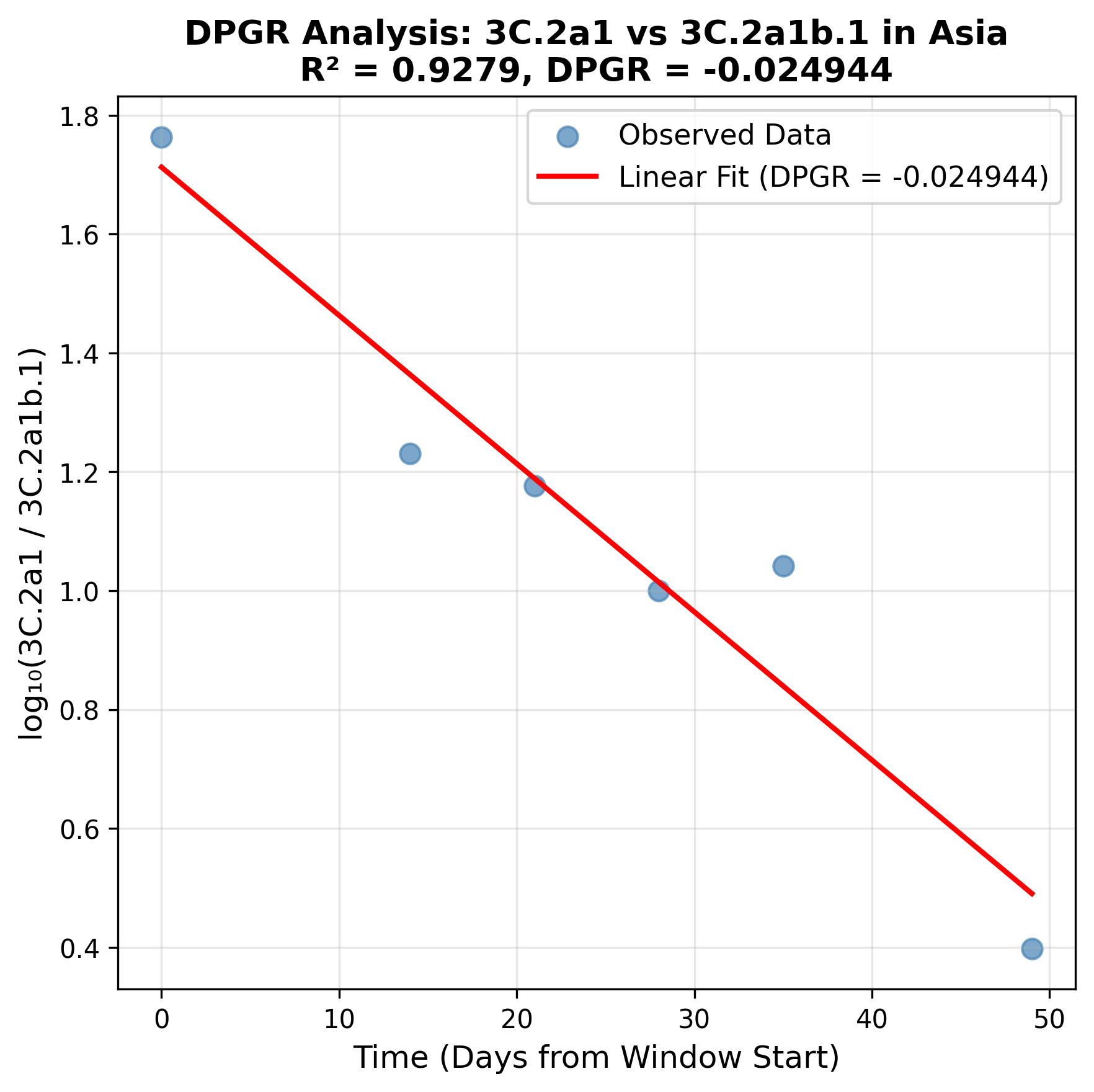}
    \caption{2016-12-12 - 2017-01-30}
  \end{subfigure}\hfill
  \begin{subfigure}{0.24\textwidth}
    \centering
    \includegraphics[width=\linewidth]{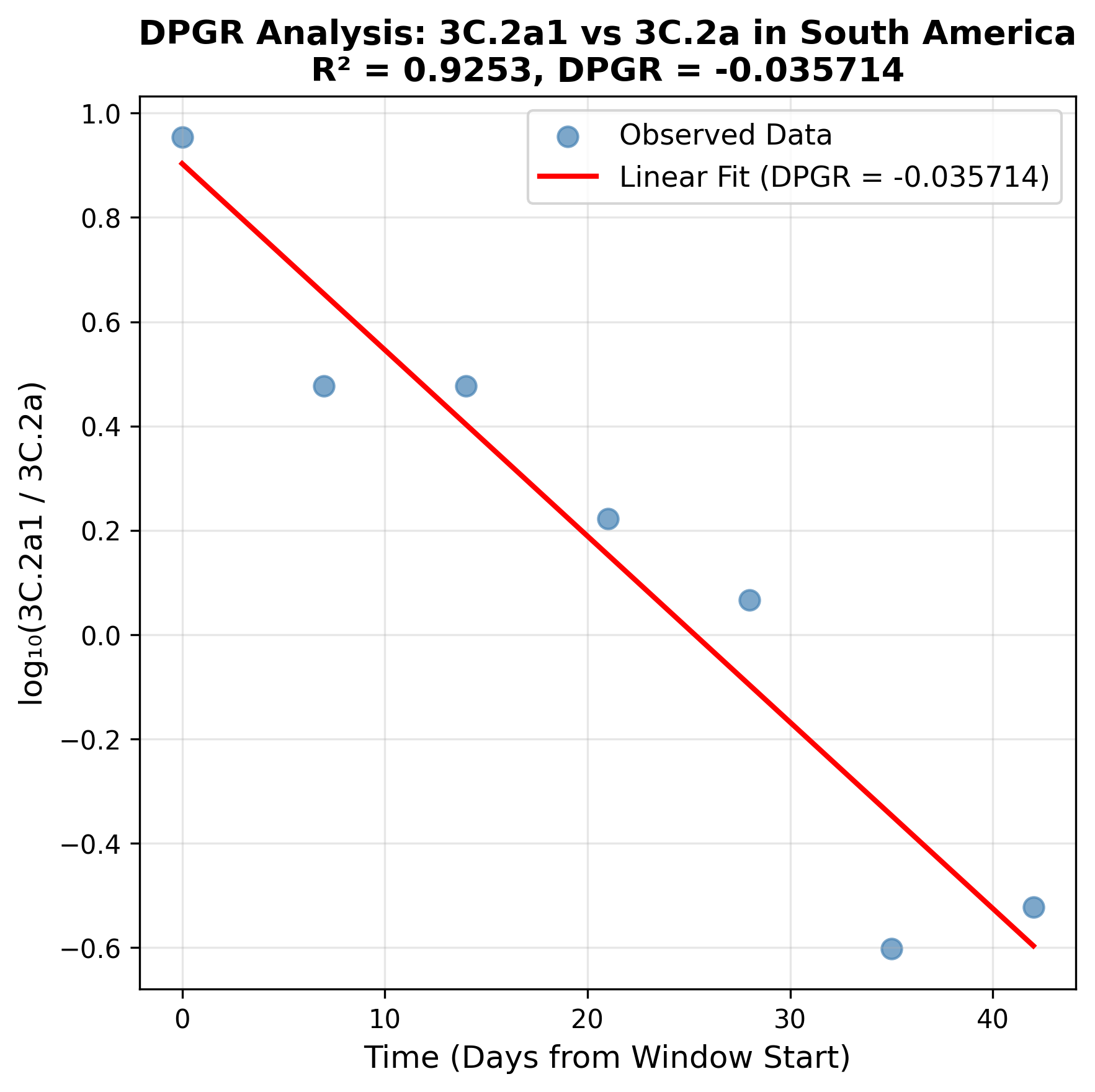}
    \caption{2017-01-02 - 2017-02-13}
  \end{subfigure}
  
  \caption{Supplementary H3N2 continent-level regression panels for the 2016--2017 season.}
  \label{fig:appendix-h3n2-2016-2017}
\end{figure}

\subsubsection{USA clades A/H3N2 across flu seasons}
 
\paragraph{2025--2026 season.}
In the United States, DPGR analysis identified subclade~K 
(3C.2a1b.2a.2a.3a.1/K) as growing faster than the vaccine strain 
3C.2a1b.2a.2a.3a.1 (DPGR $= +0.0265$, window: Nov--Dec~2025), 
consistent with the signal observed across all other monitored 
regions. The positive slope confirms subclade~K as the faster-growing 
numerator variant in the United States during the 2025--2026 season.
 
\paragraph{2022--2023 season.}
In the United States, DPGR analysis identified 3C.2a1b.2a.2a.3a.1 
as the faster-growing subclade relative to 3C.2a1b.2a.2b 
(DPGR $= -0.0225$, window: Dec~2022--Mar~2023), consistent with 
the continental-level pattern. The negative slope confirms 
3C.2a1b.2a.2a.3a.1 as the denominator growing faster, indicating 
its emergence as the dominant 3C.2a1b.2a.2 subclade in the 
United States by early 2023.
 
\paragraph{2018--2019 season.}
In the United States, DPGR analysis confirmed 3C.3a1 as the 
faster-growing variant relative to co-circulating 3C.2a-derived 
subclades. 3C.3a1 outgrew 3C.2a1 (DPGR $= -0.0338$, window: 
Oct--Dec~2018) and 3C.2a1b.1 (DPGR $= -0.0242$, window: 
Nov~2018--Jan~2019), with both negative slopes reflecting 3C.3a1 as 
the faster-growing denominator. The strong linearity of the 3C.2a1 
vs 3C.3a1 regression indicates clear competitive displacement of 
3C.2a1 by 3C.3a1 in the United States during this season.
 
\paragraph{2016--2017 season.}
In the United States, DPGR analysis identified two subclades outcompeting 
the vaccine-reference clade 3C.2a. Clade 3C.2a1 showed a fitness 
advantage over 3C.2a (DPGR $= +0.0116$, window: Nov~2016--Feb~2017), 
as did clade 3C.2a2 (DPGR $= +0.0135$, window: Dec~2016--Jan~2017). 
Both subclades were growing faster than the vaccine strain 
simultaneously, indicating active diversification within the 3C.2a 
lineage in the United States during this period.

\begin{figure}[H]
  \centering
  \captionsetup[subfigure]{font=tiny, labelfont=tiny}

  \begin{subfigure}[b]{0.23\textwidth}
    \centering
    \includegraphics[width=\linewidth, height=2.5cm]{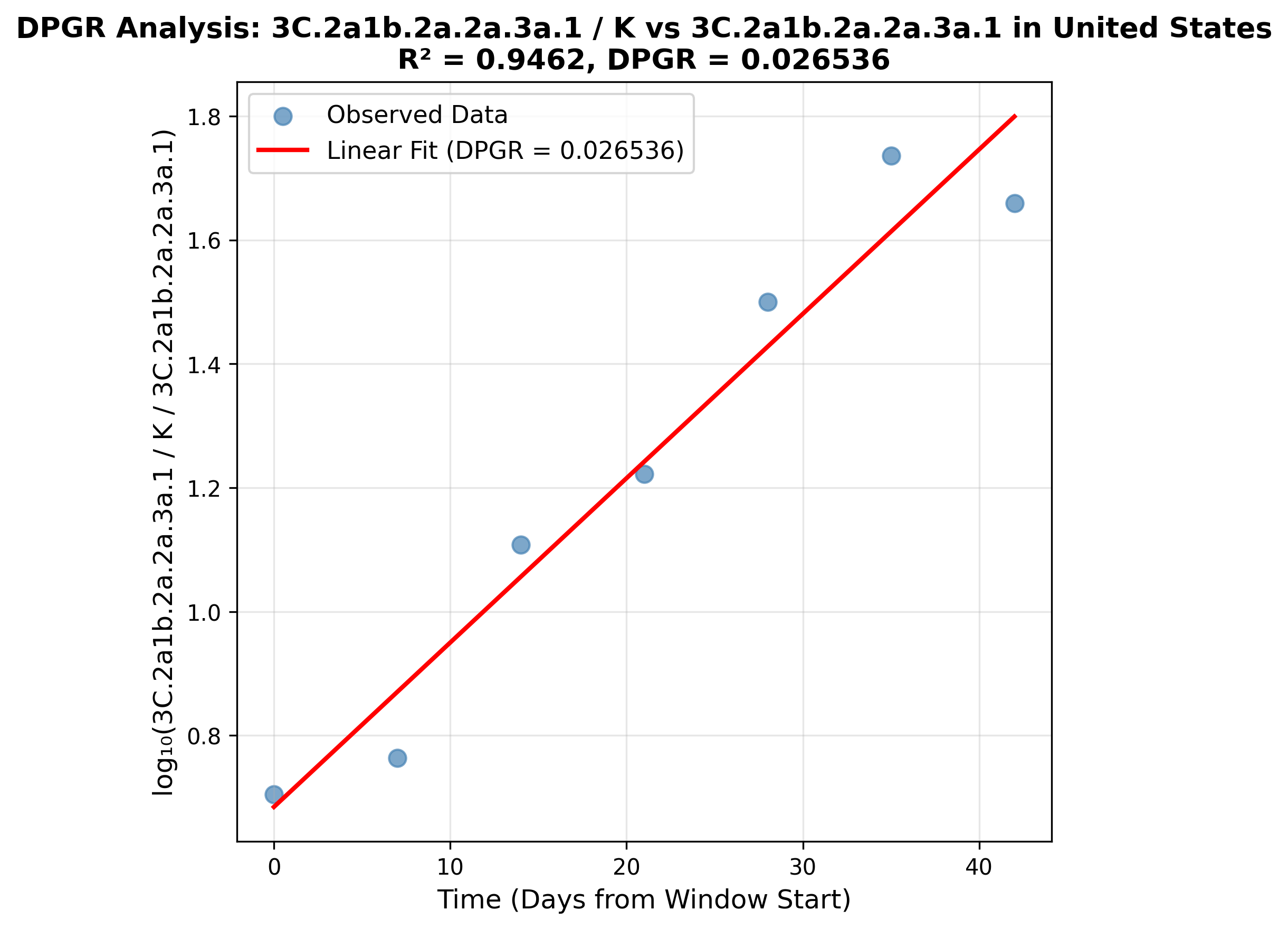}
    \caption{2025-11-03 - 2025-12-15}
  \end{subfigure}\hfill
  \begin{subfigure}[b]{0.23\textwidth}
    \centering
    \includegraphics[width=\linewidth, height=2.5cm]{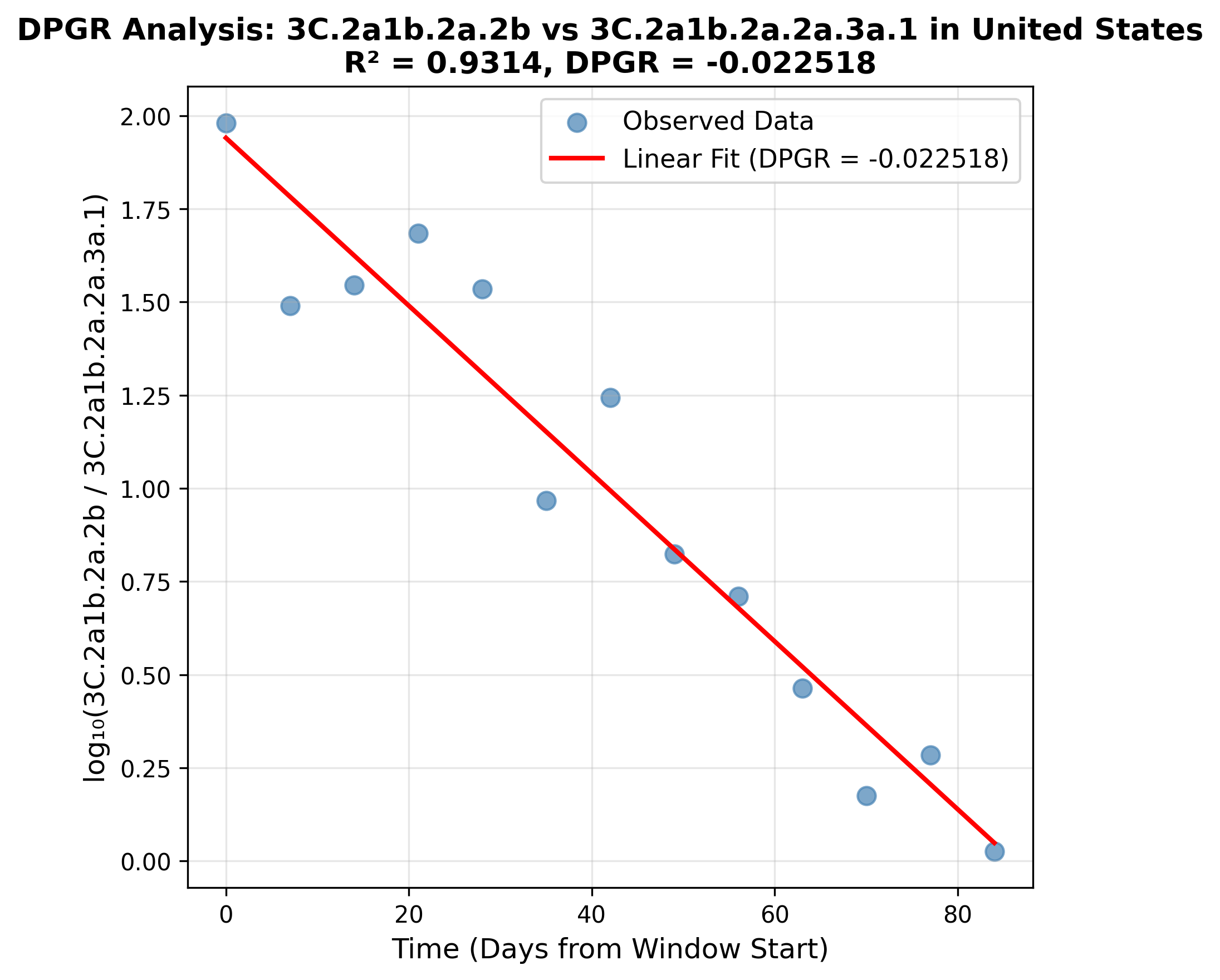}
    \caption{2022-12-19 - 2023-03-13}
  \end{subfigure}\hfill
  \begin{subfigure}[b]{0.23\textwidth}
    \centering
    \includegraphics[width=\linewidth, height=2.5cm]{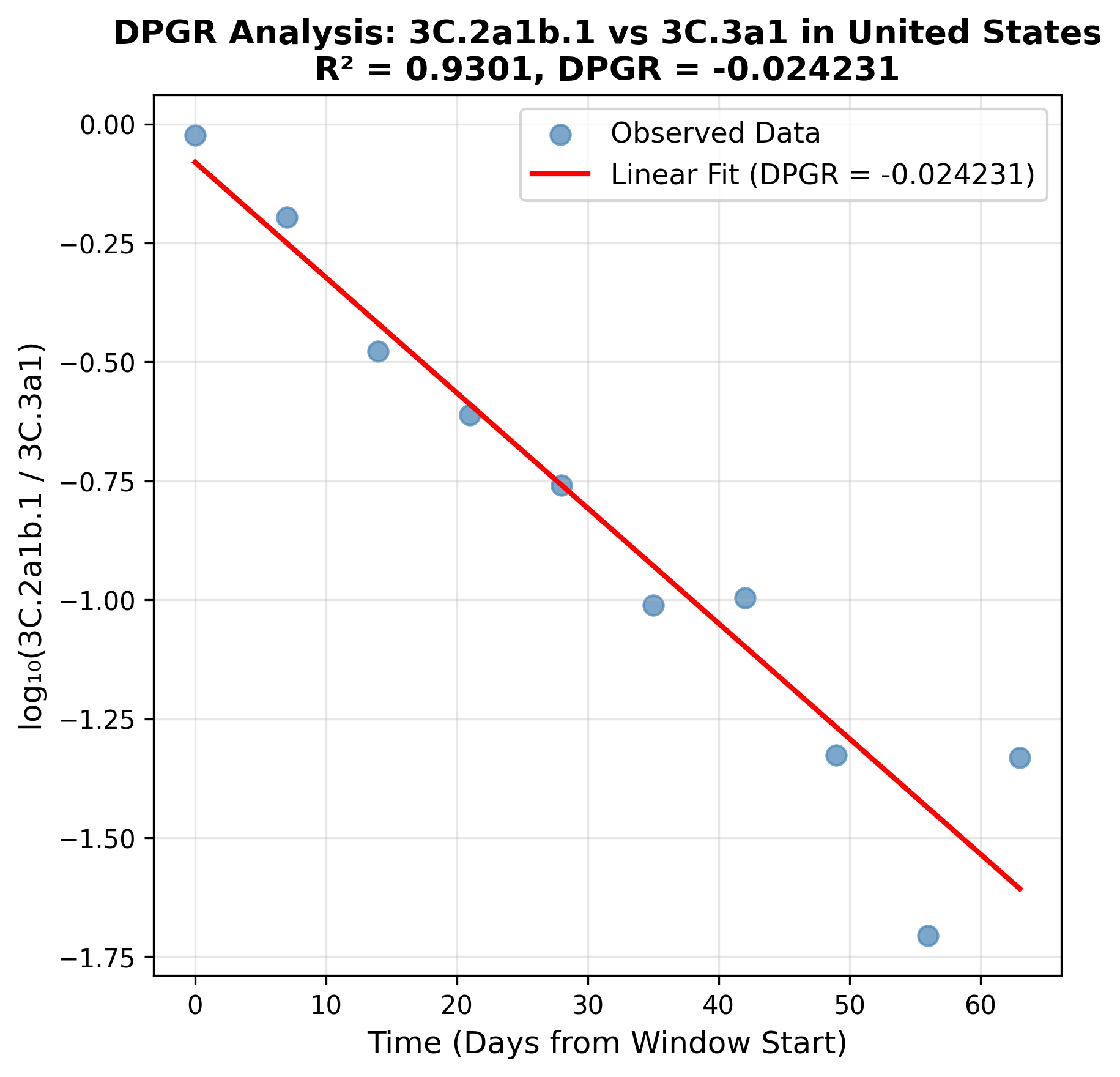}
    \caption{2018-11-26 - 2019-01-28}
  \end{subfigure}\hfill
  \begin{subfigure}[b]{0.23\textwidth}
    \centering
    \includegraphics[width=\linewidth, height=2.5cm]{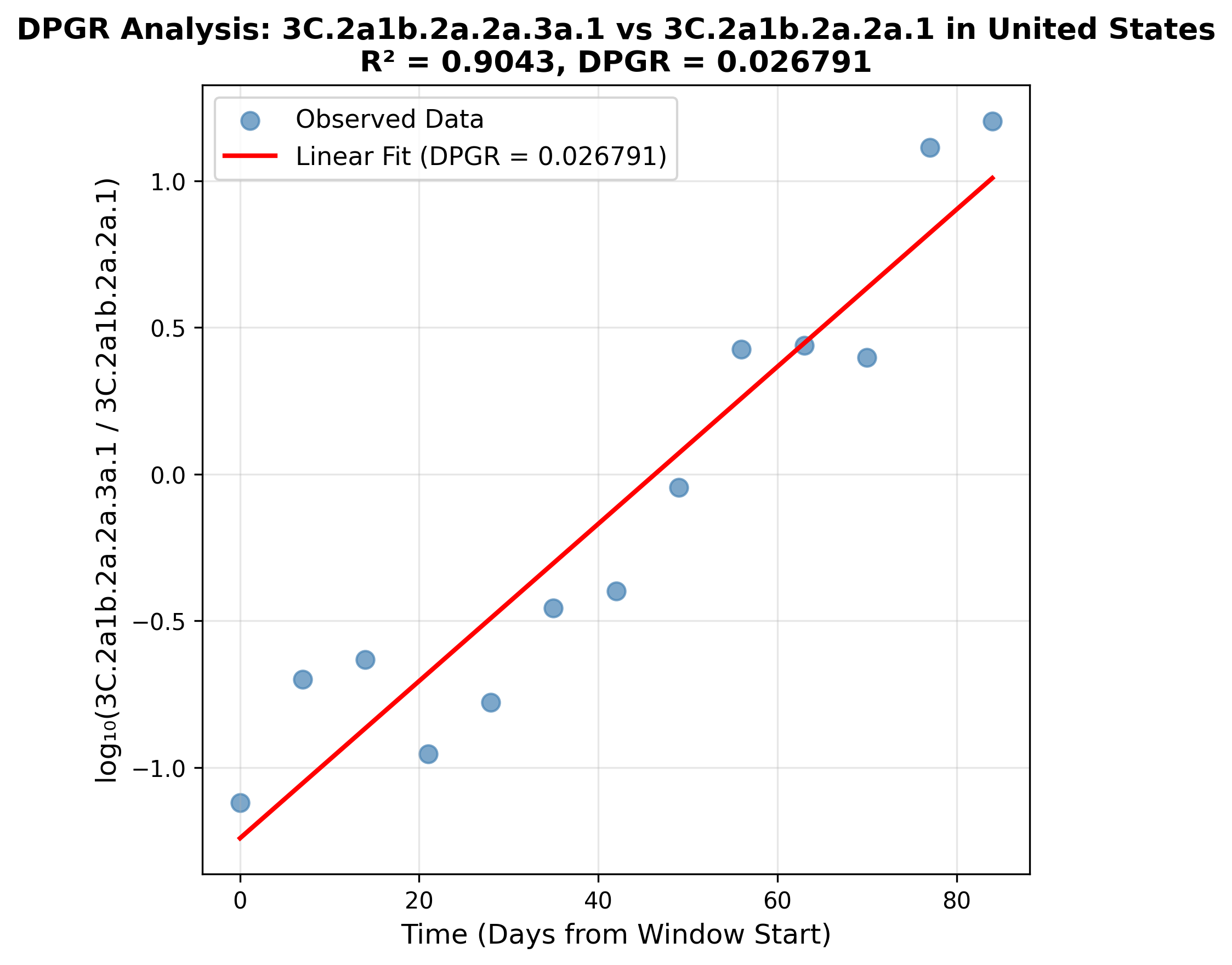}
    \caption{2022-12-19 - 2023-03-13}
  \end{subfigure}

  \vspace{1em}

  \centering
  \begin{subfigure}[b]{0.23\textwidth}
    \centering
    \includegraphics[width=\linewidth, height=2.5cm]{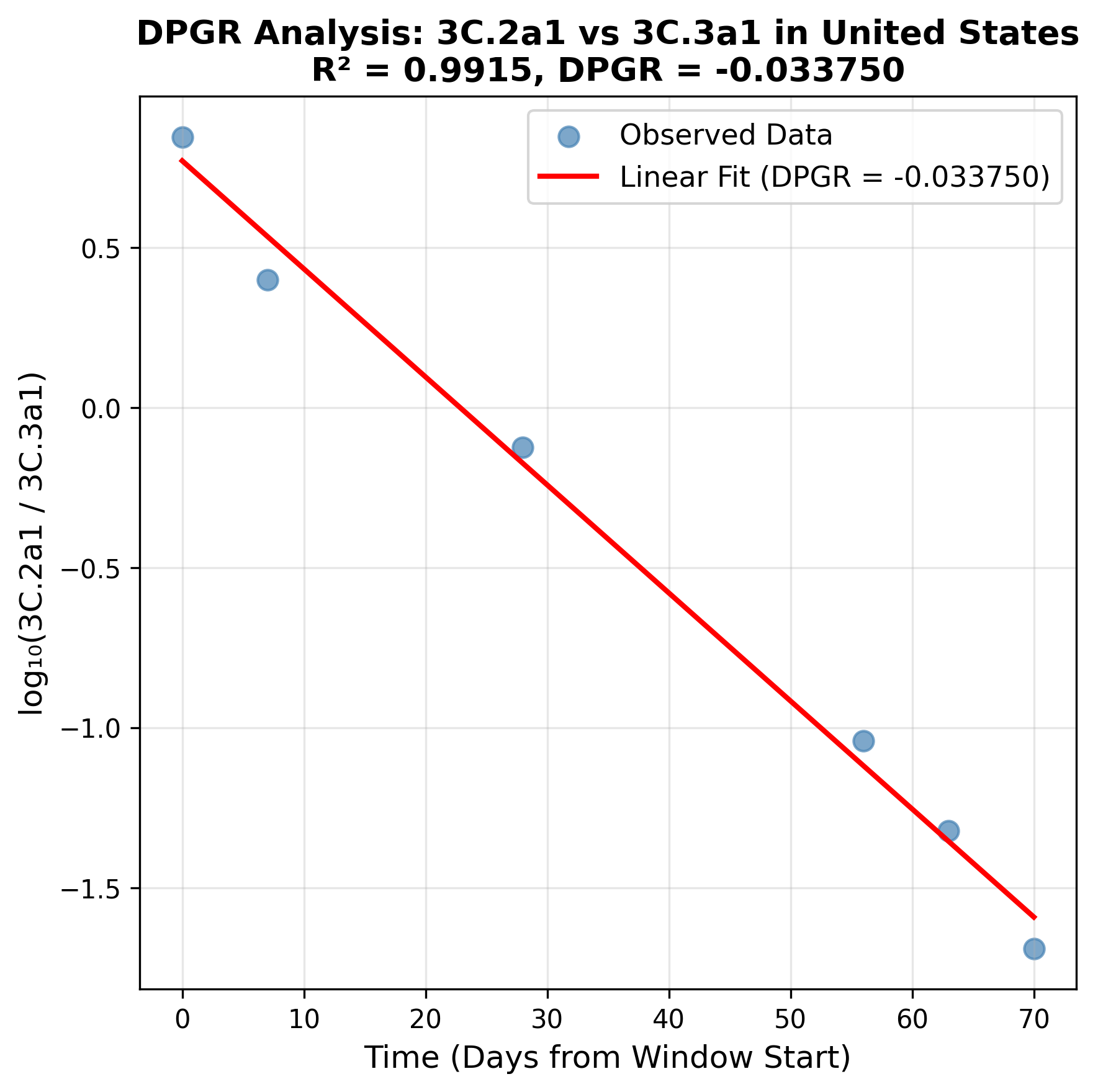}
    \caption{2018-10-08 - 2018-12-17}
  \end{subfigure}
  \hspace{1em}
  \begin{subfigure}[b]{0.23\textwidth}
    \centering
    \includegraphics[width=\linewidth, height=2.5cm]{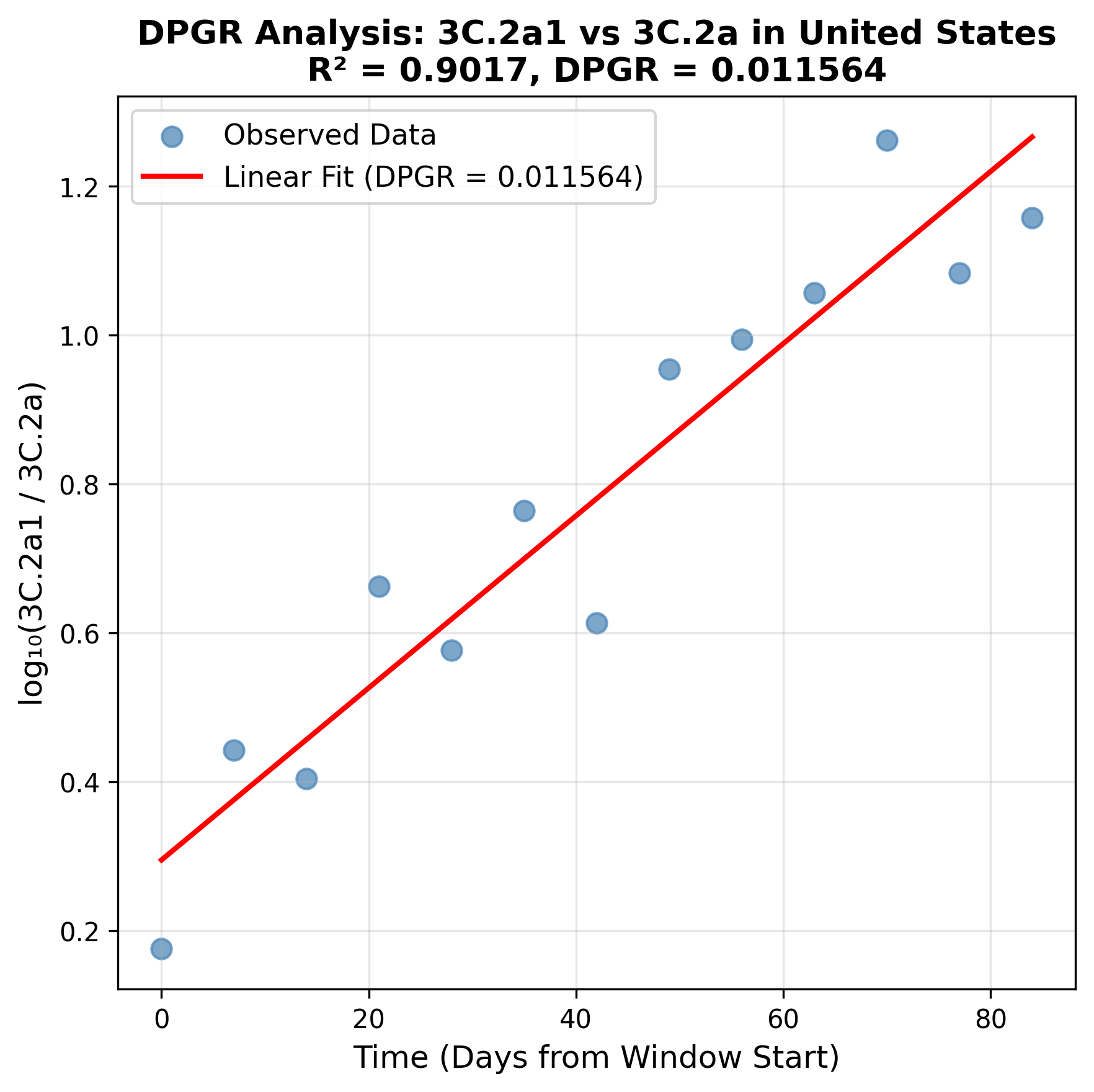}
    \caption{2016-11-21 - 2017-02-13}
  \end{subfigure}
  \hspace{1em}
  \begin{subfigure}[b]{0.23\textwidth}
    \centering
    \includegraphics[width=\linewidth, height=2.5cm]{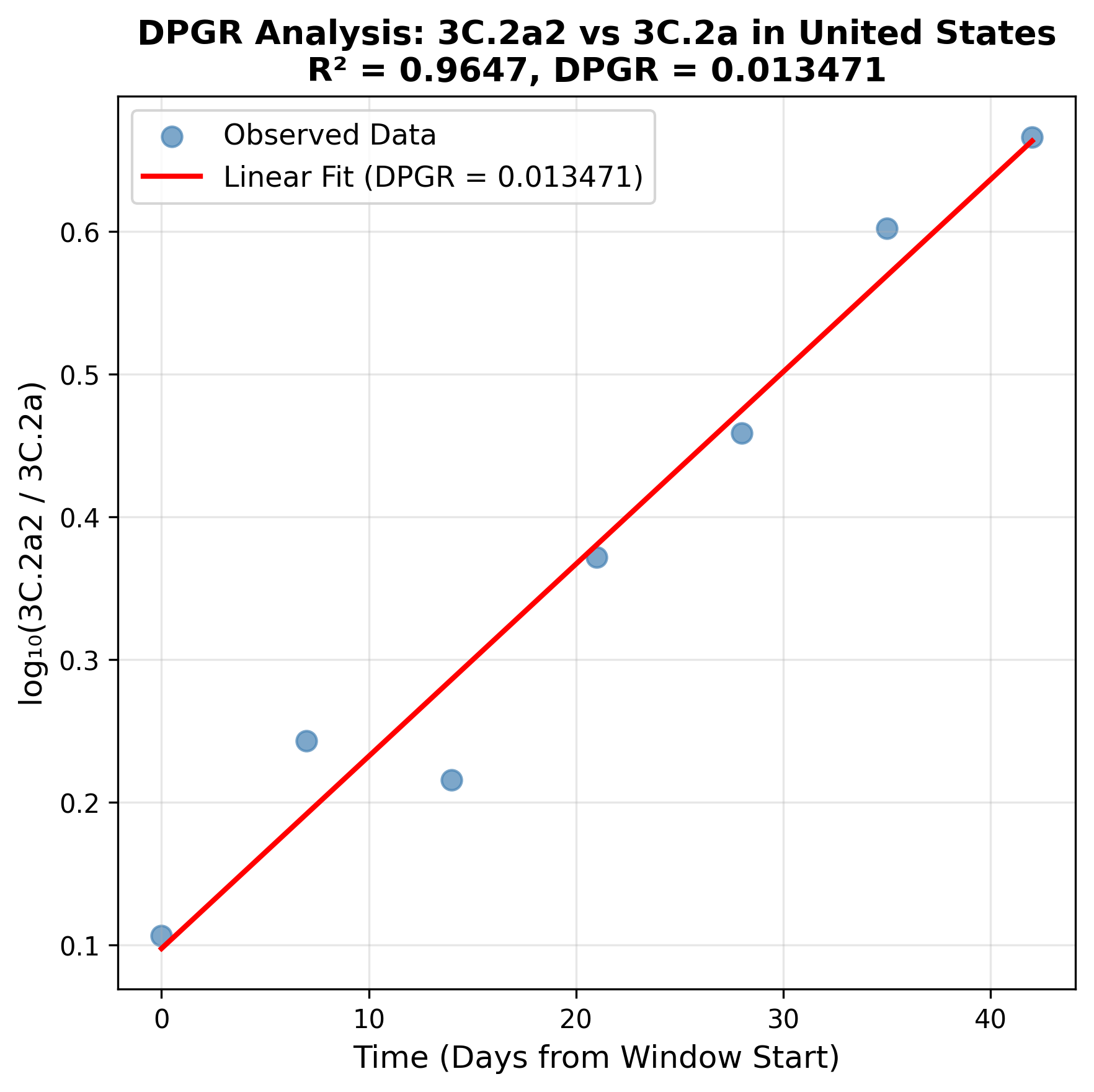}
    \caption{2016-12-19 - 2017-01-30}
  \end{subfigure}

  \caption{Supplementary H3N2 regression panels for United States flu seasons.}
  \label{fig:appendix-h3n2-us}
\end{figure}

\subsection{A/H1N1}

\subsubsection{Continental clades A/H1N1 2024-2025 flu season}
\begin{figure}[H]
  \centering
  \captionsetup[subfigure]{font=tiny, labelfont=tiny}

  \begin{subfigure}{0.24\textwidth}
    \centering
    \includegraphics[width=\linewidth]{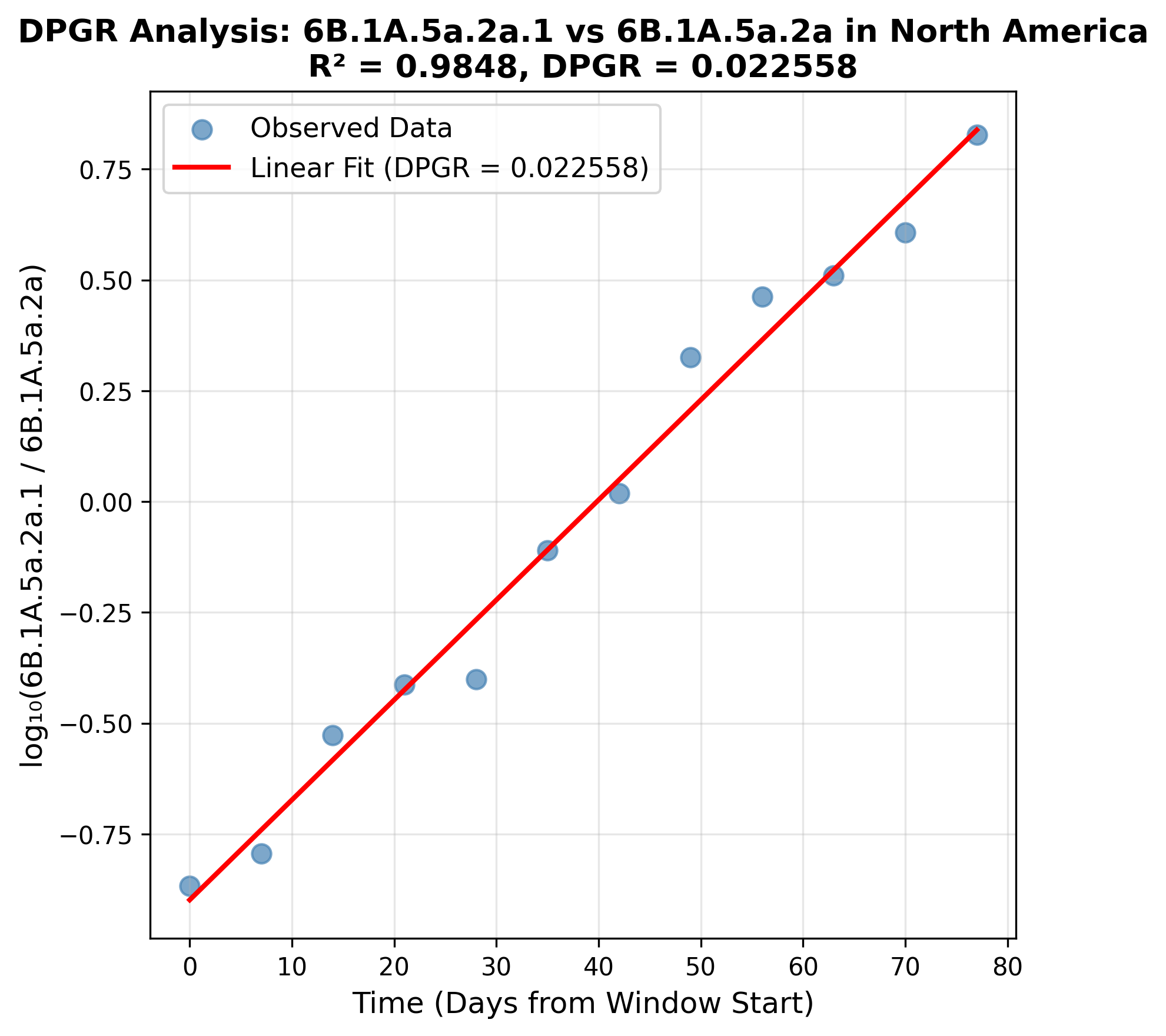}
    \caption{2024-12-30 - 2025-03-17}
  \end{subfigure}\hfill
  \begin{subfigure}{0.24\textwidth}
    \centering
    \includegraphics[width=\linewidth]{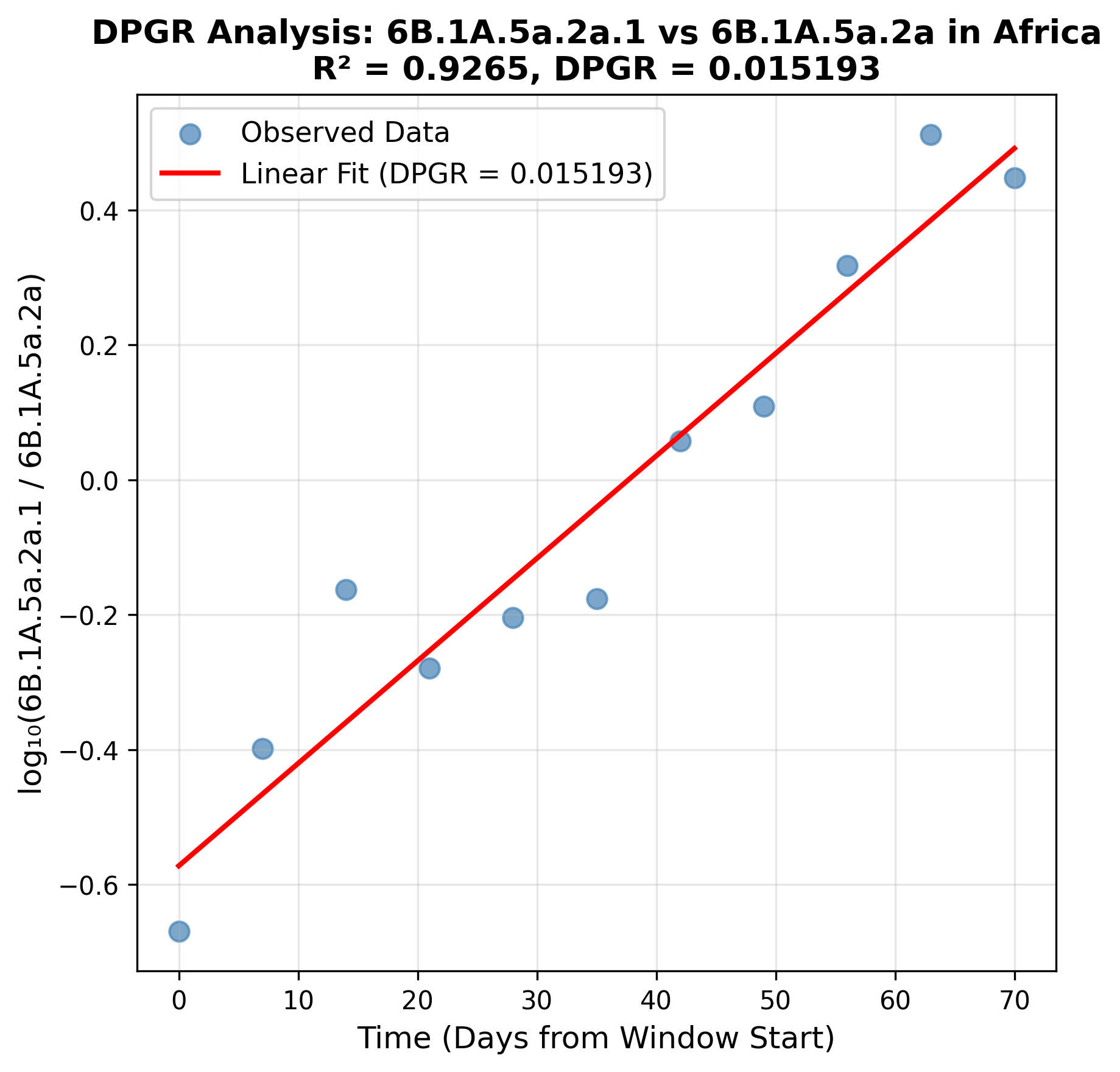}
    \caption{2025-04-28 - 2025-07-07}
  \end{subfigure}\hfill
  \begin{subfigure}{0.24\textwidth}
    \centering
    \includegraphics[width=\linewidth]{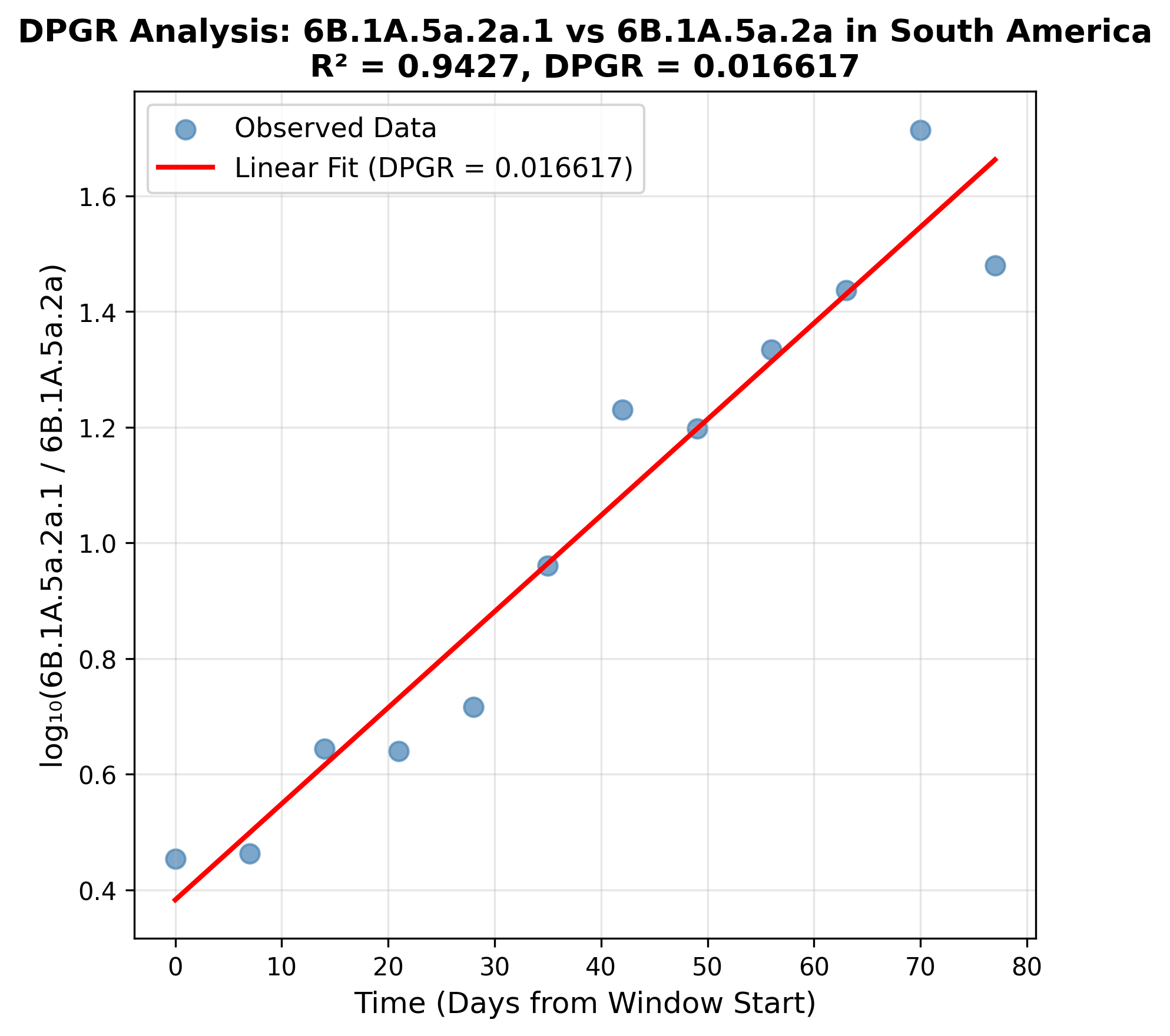}
    \caption{2025-03-03 - 2025-05-19}
  \end{subfigure}\hfill
  \begin{subfigure}{0.24\textwidth}
    \centering
    \includegraphics[width=\linewidth]{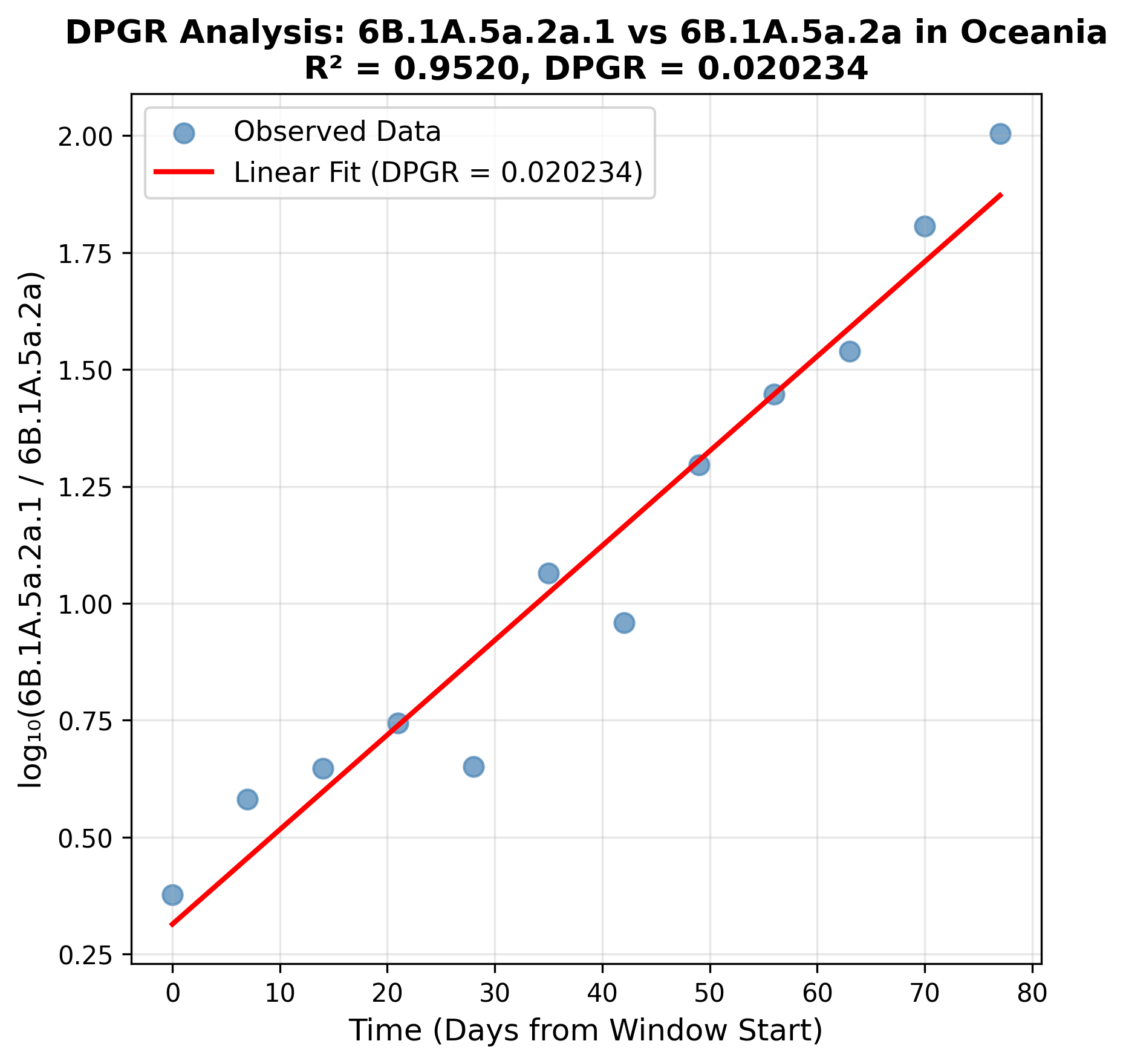}
    \caption{2025-03-24 - 2025-06-09}
  \end{subfigure}
  
  \caption{Supplementary H1N1 continent-level regression panels for the 2024--2025 season.}
  \label{fig:appendix-h1n1-2024-2025}
\end{figure}

DPGR analysis identified subclade 6B.1A.5a.2a.1 as having a 
consistent fitness advantage over its parent subclade 
6B.1A.5a.2a across all five regions with available data. 
Positive slopes were observed in North America 
(DPGR $= +0.0226$, window: Dec~2024-Mar~2025), Africa 
(DPGR $= +0.0152$, window: Apr-Jul~2025), South America 
(DPGR $= +0.0166$, window: Mar-May~2025), and Oceania 
(DPGR $= +0.0202$, window: Mar-Jun~2025). The consistency 
of positive DPGR values across geographically diverse regions 
confirms 6B.1A.5a.2a.1 as the globally dominant H1N1 subclade 
during the 2024-2025 season, consistent with its selection 
as both the egg-based (A/Victoria/4897/2022) and cell 
culture-based (A/Wisconsin/67/2022) vaccine component for 
this season.

\paragraph{2019--2020 season.}
DPGR analysis identified subclade 6B.1A.5a.2 as having a 
strong and consistent fitness advantage over multiple 
co-circulating subclades across all analysed continental 
regions. In North America, 6B.1A.5a.1 outgrew 6B.1A.5b 
(DPGR $= +0.0167$, window: Jan--Feb~2020), and 6B.1A.5a.2 
outgrew 6B.1A.5b (DPGR $= +0.0224$, window: Jan--Mar~2020). 
In Europe, 6B.1A.5b was outcompeted by 6B.1A.5a 
(DPGR $= -0.0196$, window: Jan--Mar~2020), while 
6B.1A.5a.2 grew faster than both 6B.1A.5a 
(DPGR $= +0.0226$, window: Nov~2019--Jan~2020) and 
6B.1A.5b (DPGR $= +0.0160$, window: Dec~2019--Mar~2020). 
Across all continental comparisons, 6B.1A.5a.2 consistently 
emerged as the fittest subclade, outcompeting all 
co-circulating lineages including the vaccine-matched 6B.1A.1.

\paragraph{2018-2019 season.}
DPGR analysis identified emerging 6B.1A subclades as having 
fitness advantages over older lineages across multiple 
continental regions. In North America, 6B.1A outgrew the 
parent 6B.1A.5a (DPGR $= -0.0200$, window: Oct--Dec~2018), 
indicating faster growth of 6B.1A.5a as the denominator 
variant. In Europe, 6B.1A.1 was outcompeted by 6B.1A.5a 
(DPGR $= -0.0271$, window: Oct-Dec~2018), and 6B.1A.5b 
outgrew 6B.1A (DPGR $= -0.0268$, window: Oct--Nov~2018), 
reflecting active subclade competition within the 6B.1A 
lineage. In Asia, 6B.1A outgrew the vaccine-matched 6B.1 
clade (DPGR $= -0.0148$, window: Dec~2018--Mar~2019), 
indicating displacement of the vaccine strain clade by 
emerging 6B.1A subclades.

\subsubsection{USA clades A/H1N1 across flu seasons}

\paragraph{2024--2025 season.}
In the United States, 6B.1A.5a.2a.1 demonstrated a 
fitness advantage over 6B.1A.5a.2a 
(DPGR $= +0.0209$, window: Jan--Mar~2025), consistent 
with the continental-level pattern and confirming 
6B.1A.5a.2a.1 as the dominant circulating subclade 
nationally during this period.

\paragraph{2019--2020 season.}
The United States data for 2019--2020 revealed the 
most complex competitive dynamics of any H1N1 season 
analyzed. Subclade 6B.1A.5a.2 demonstrated fitness 
advantages over all co-circulating lineages: over 
6B.1A.5a (DPGR $= +0.0516$, window: Dec~2019-Feb~2020, 
the highest DPGR value observed in the H1N1 dataset), 
over 6B.1 (DPGR $= +0.0346$, window: Nov~2019-Feb~2020), 
over 6B.1A.5a.1 (DPGR $= +0.0231$, window: Dec~2019-Feb~2020), 
and over 6B.1A.5b (DPGR $= +0.0277$, window: Nov~2019-Feb~2020). 
Subclade 6B.1A.5a.1 also outgrew 6B.1A.5b 
(DPGR $= +0.0100$, window: Oct~2019-Jan~2020) and 
6B.1A.1 (DPGR $= +0.0263$, window: Oct-Dec~2019). 
The only negative slope in the US data was 6B.1A.5b 
vs 6B.1 (DPGR $= -0.0434$, window: Feb-Mar~2020), 
indicating 6B.1 briefly outgrowing 6B.1A.5b in the 
very late season.

\paragraph{2018--2019 season.}
In the United States, 6B.1A.1 was outcompeted by 
6B.1A.5a (DPGR $= -0.0169$, window: Jan-Mar~2019), 
while 6B.1A.5b outgrew 6B.1A.1 
(DPGR $= +0.0140$, window: Jan-Mar~2019). These 
results indicate that within the 6B.1A lineage, 
the 5a and 5b subclades were simultaneously 
displacing the 6B.1A.1 vaccine-matched clade, 
with 6B.1A.5a having the stronger fitness advantage.
\begin{figure}[H]
  \centering
  \captionsetup[subfigure]{font=tiny, labelfont=tiny}

  \begin{subfigure}{0.24\textwidth}
    \centering
    \includegraphics[width=\linewidth]{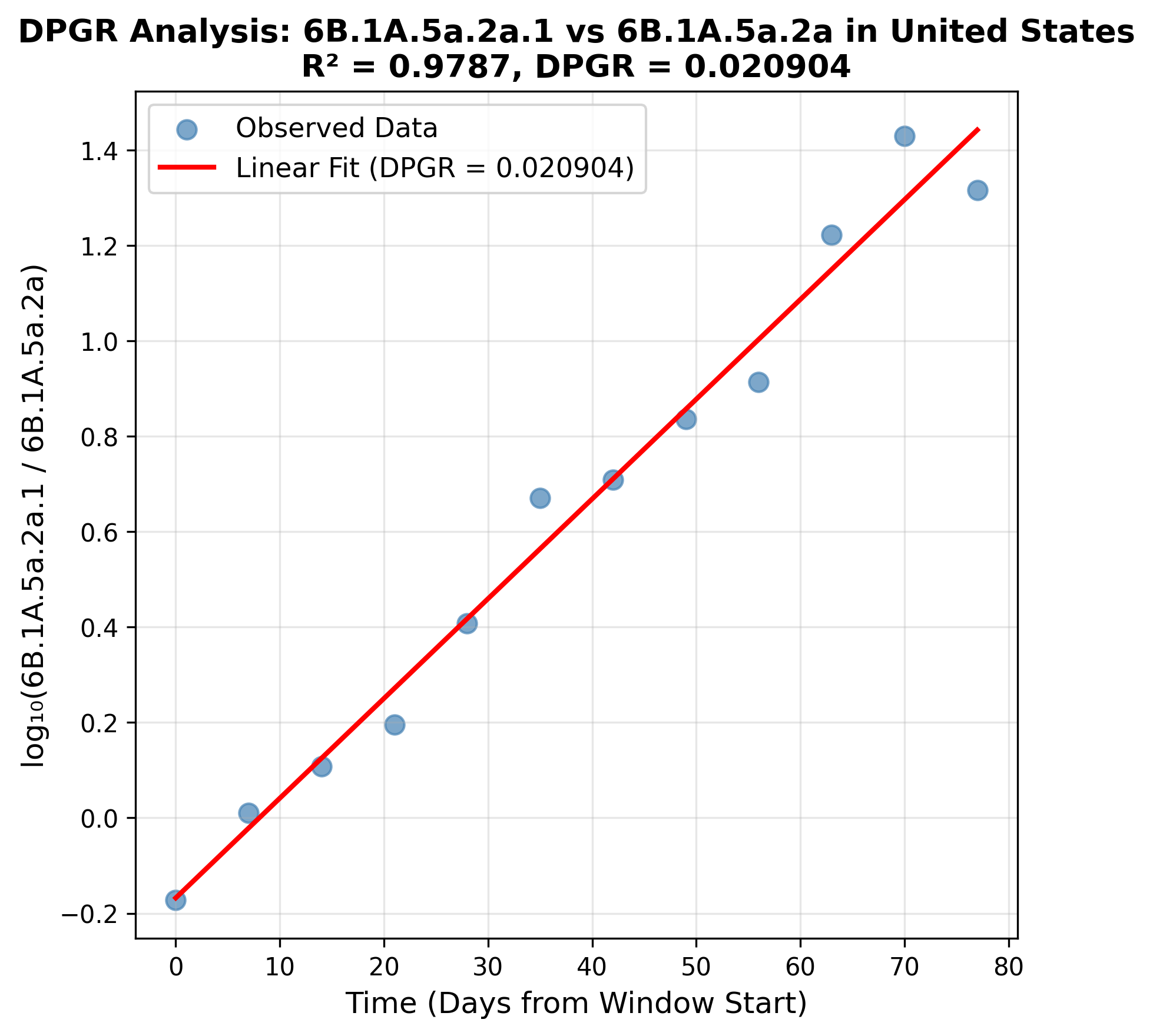}
    \caption{2025-01-06 - 2025-03-24}
  \end{subfigure}\hfill
  \begin{subfigure}{0.24\textwidth}
    \centering
    \includegraphics[width=\linewidth]{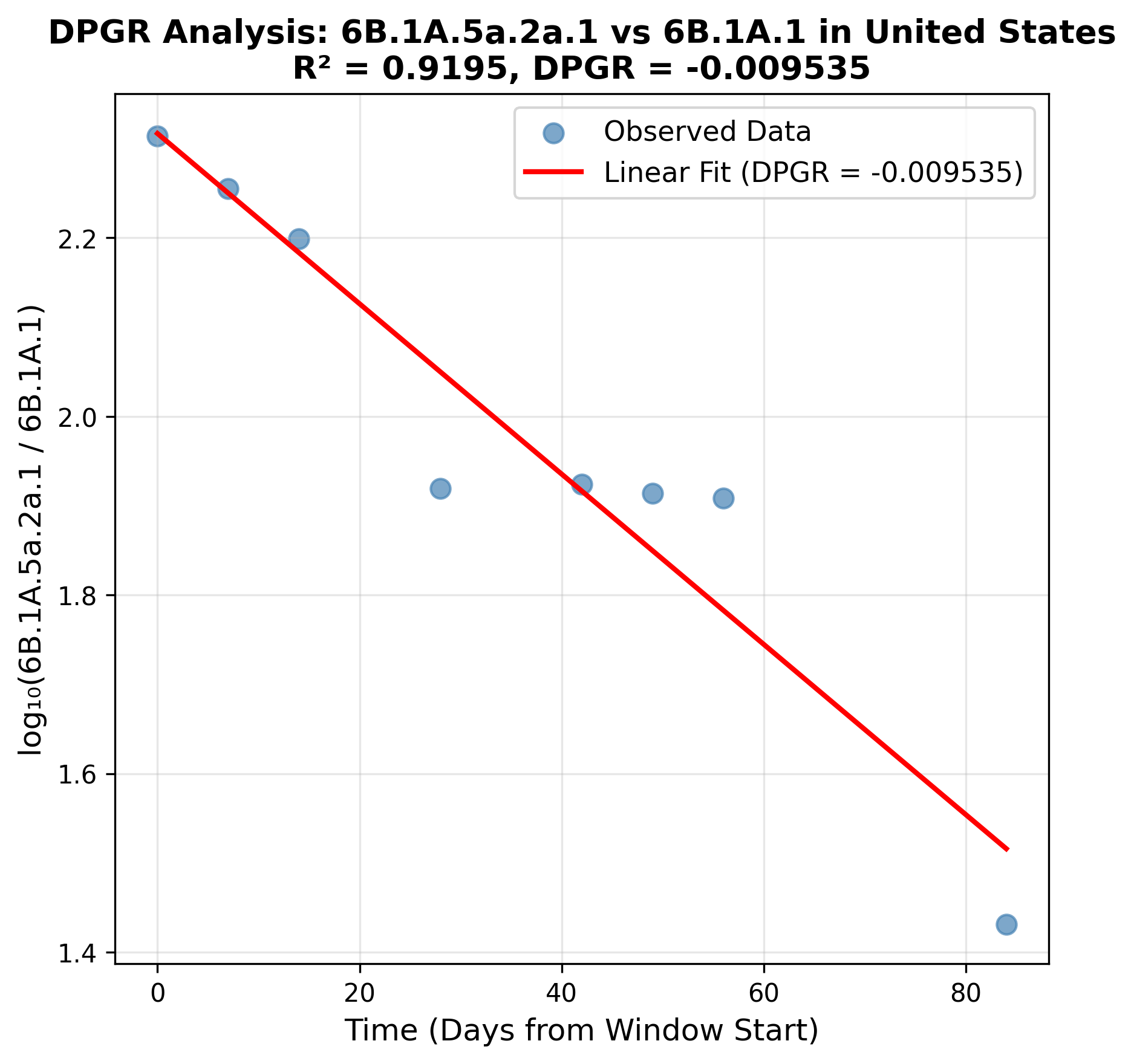}
    \caption{2022-12-12 - 2023-03-06}
  \end{subfigure}\hfill
  \begin{subfigure}{0.24\textwidth}
    \centering
    \includegraphics[width=\linewidth]{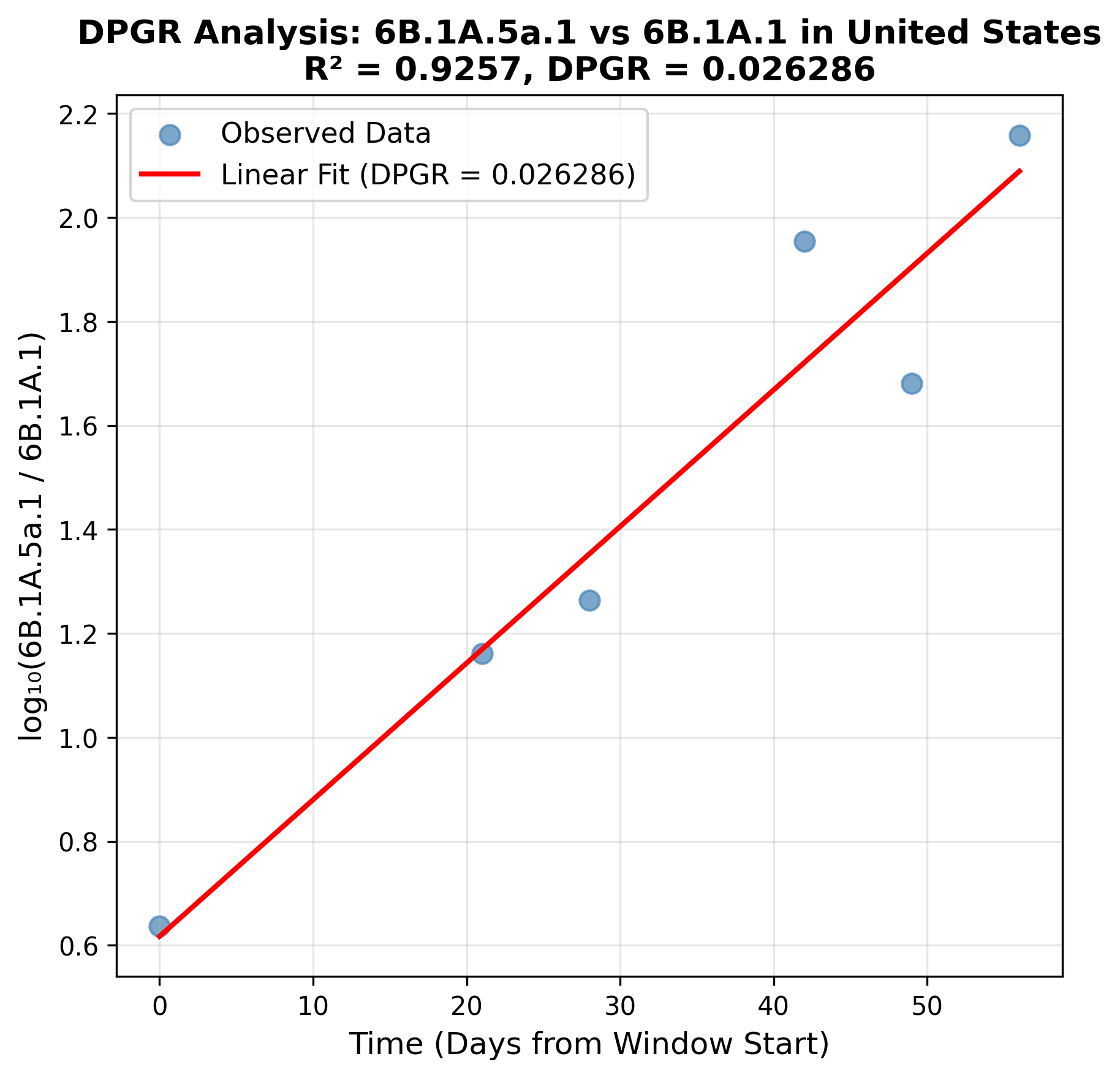}
    \caption{2019-10-21 - 2019-12-16}
  \end{subfigure}\hfill
  \begin{subfigure}{0.24\textwidth}
    \centering
    \includegraphics[width=\linewidth]{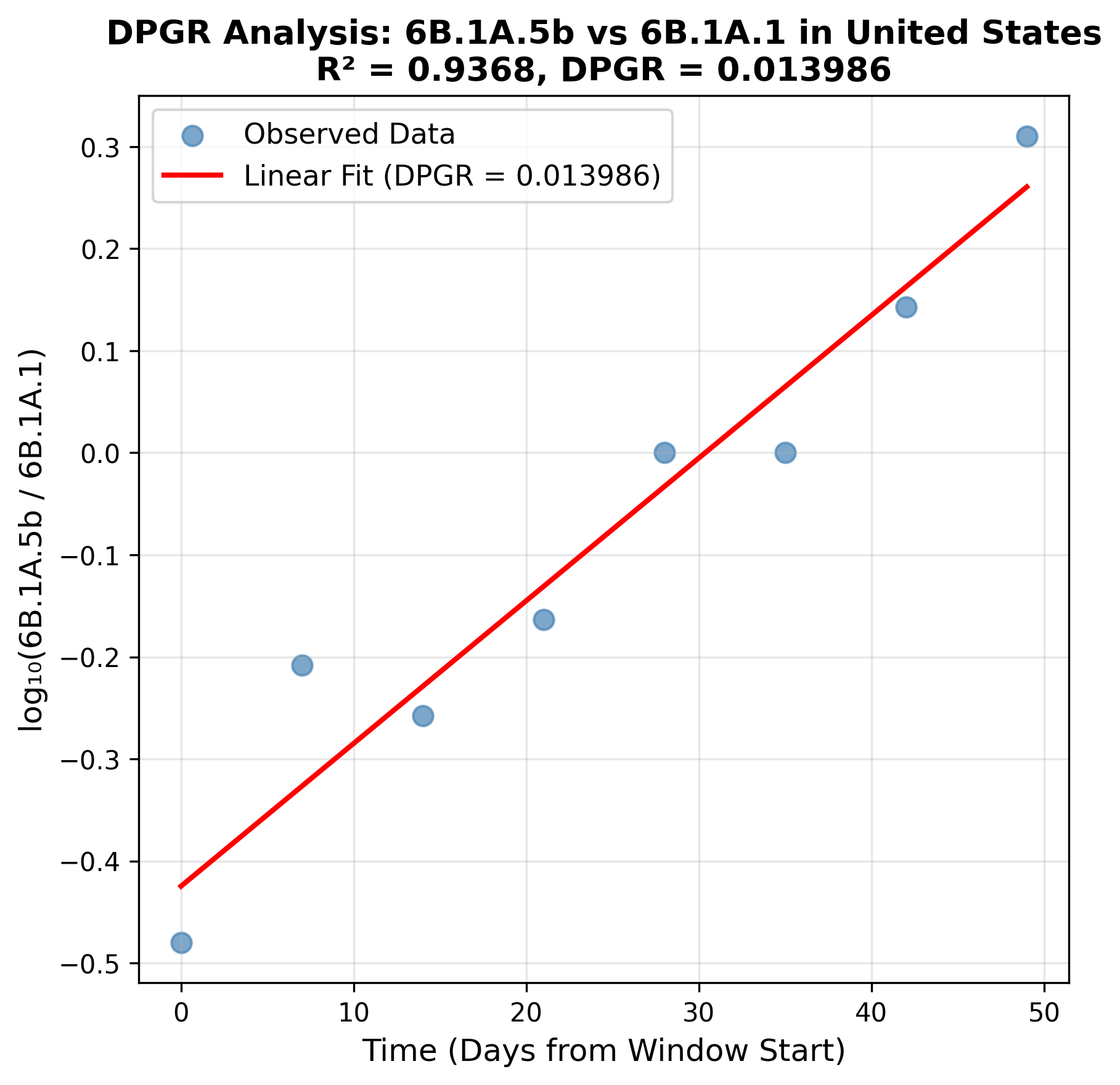}
    \caption{2019-01-28 - 2019-03-18}
  \end{subfigure}
  
  \caption{Supplementary H1N1 regression panels for United States flu seasons.}
  \label{fig:appendix-h1n1-us}
\end{figure}

\end{document}